\documentclass[journal]{IEEEtran}
\usepackage{mathtools,amssymb,lipsum}

\usepackage{cuted}
\usepackage{tikz}
\usetikzlibrary{arrows,chains,matrix,positioning,scopes,shapes}
\pgfdeclarelayer{background}
\pgfdeclarelayer{behind}
\pgfdeclarelayer{above}
\pgfdeclarelayer{glass}
\pgfsetlayers{background,behind,main,above,glass}
\makeatletter
\tikzset{join/.code=\tikzset{after node path={%
\ifx\tikzchainprevious\pgfutil@empty\else(\tikzchainprevious)%
edge[every join]#1(\tikzchaincurrent)\fi}}}
\makeatother
\tikzset{>=stealth',every on chain/.append style={join},
         every join/.style={->}}
\tikzstyle{labeled}=[execute at begin node=$\scriptstyle,
   execute at end node=$]
\makeatletter
\def\ps@headings{%
\def\@oddhead{\mbox{}\scriptsize\rightmark \hfil \thepage}%
\def\@evenhead{\scriptsize\thepage \hfil \leftmark\mbox{}}%
\def\@oddfoot{}%
\def\@evenfoot{}}
\makeatother
\IEEEoverridecommandlockouts
\newcommand{\figscale}{0.8}

\usepackage{amsmath}
\usepackage{amsfonts}
\usepackage{graphicx}
\usepackage[captionskip=-0.1cm,subrefformat=parens,labelformat=parens]{subfig}

\usepackage[skip=0.1cm]{caption}

\usepackage{tikz}
\usepackage{times}
\usepackage{cite}
\newcommand{\finaltext}[1]{{#1}}
\usepackage{lettrine}
\usepackage{amsthm}

\allowdisplaybreaks[4]
\usepackage{array}
\usepackage{amssymb}

\usepackage{stfloats}
\usepackage{footnote}
\usepackage{booktabs}
\usepackage{array}
\usepackage{algorithmic}
\usepackage{algorithm}
\usepackage{subeqnarray}
\usepackage{cases}
\usepackage{threeparttable}
\usepackage{url}
\usepackage{bm}
\usepackage{physics}
\usepackage[hidelinks]{hyperref}
\usepackage[margin=0.5in]{geometry}
\usepackage{scalerel}

\usepackage[export]{adjustbox}

\usepackage{svg}

\usepackage{enumitem} 
\newtheorem{theorem}{Theorem}
\newtheorem{assumption}[theorem]{Assumption}
\newtheorem{lemma}[theorem]{Lemma}

\newtheorem{proposition}[theorem]{Proposition}
\newtheorem{example}[theorem]{Example}
\newtheorem{remark}[theorem]{Remark}

\newtheorem{conjecture}[theorem]{Conjecture}
\newtheorem{definition}[theorem]{Definition}

\newcommand{\manualendsymbol}[1]{%
  \leavevmode\unskip\penalty9999\hbox{}\nobreak\hfill
  \quad\hbox{#1}}

\newcommand{\eproposition}{\manualendsymbol{\ensuremath{\square}}}

\newcommand{\edefinition}{\manualendsymbol{\ensuremath{\triangle}}}

\newcommand{\eexample}{\manualendsymbol{\ensuremath{\square}}}

\newcommand{\eremark}{\manualendsymbol{\ensuremath{\triangle}}}

\newcommand{\eassumption}{\manualendsymbol{\ensuremath{\square}}}

\newcommand{\vect}[1]{ \bm{#1} }

\newcommand{\set}[1]{\mathcal{#1}}

\newcommand{\beli}{\beta}
\newcommand{\vbeli}{ \vect{\beli} }

\newcommand{\setEfull}{ \mathcal{E} }
\newcommand{\setF}{ \mathcal{F} }
\newcommand{\sfN}{ \mathsf{N} }
\newcommand{\setpf}{ \partial f }

\newcommand{\sR}{\mathbb{R}}
\newcommand{\sC}{\mathbb{C}}
\newcommand{\sZ}{\mathbb{Z}}

\newcommand{\FB}{F_{ \mathrm{B} }}
\newcommand{\ZB}{Z_{ \mathrm{B} }}
\newcommand{\Herm}{\mathsf{H}}
\newcommand{\tran}{\mathsf{T}}

\newcommand{\xe}{ x_{e} }
\newcommand{\vxf}{ \vect{x}_{\setpf} }

\newcommand{\setxe}{ \mathcal{X}_{e} }
\newcommand{\setx}{ \mathcal{X} }
\newcommand{\setxf}{ \mathcal{X}_{\setpf} }
\newcommand{\vv}{ \vect{v} }
\newcommand{\vx}{ \vect{x} }

\newcommand{\vt}{ \vect{t} }

\newcommand{\vmu}{ \vect{ \mu } }

\newcommand{\wh}{w_{\mathrm{H}}}

\newcommand{\graphN}{ \mathsf{N} }

\newcommand{\setE}{ \mathcal{E} }
\newcommand{\defeq}{ \triangleq }
\newcommand{\setHerm}[1]{ \mathrm{Herm}_{#1} }
\newcommand{\setPSD}[1]{ \mathrm{PSD}_{#1} }

\newcommand{\setPD}[1]{ \mathrm{PD}_{#1} }

\newcommand{\sRp}{ \mathbb{R}_{\geq 0} }
\newcommand{\sRpp}{ \mathbb{R}_{> 0} }
\newcommand{\sZp}{ \mathbb{Z}_{\geq 0} }
\newcommand{\sZpp}{ \mathbb{Z}_{\geq 1} }
\newcommand{\etof}{ e,f }

\newcommand{\fSPAN}{ f_{\mathrm{SPA},\mathsf{N}} }
\newcommand{\LMP}{ \set{B} }
\newcommand{\ZBSPA}{ Z_{\mathrm{B,SPA}} }
\newcommand{\ZBM}{ Z_{\mathrm{B},M} }
\newcommand{\hgraphN}{ \hat{\graphN} }
\newcommand{\setcovN}[1]{ \hat{\set{N}}_{#1} }
\newcommand{\tset}[1]{\tilde{\mathcal{#1}}}
\newcommand{\matr}[1]{\vect{#1}}
\newcommand{\LCTtset}[1]{\mathring{\tilde{\mathcal{#1}}}}
\newcommand{\LCTset}[1]{\mathring{\mathcal{#1}}}
\newcommand{\LCTtv}[1]{ \mathring{ \tilde{ \vect{#1} } } }
\newcommand{\LCTv}[1]{ \mathring{ \vect{#1} } }
\newcommand{\LCT}[1]{ \mathring{#1} }
\newcommand{\LCTt}[1]{ \mathring{ \tilde{#1} } }

\newcommand{\Sigmae}{\Sigma_{(e)}}
\newcommand{\vsigma}{\vect{\sigma}}
\newcommand{\vSigma}{\vect{\Sigma}}
\newcommand{\hgraphNavg}{ \hat{\graphN}_{M,\mathrm{avg}} }

\newcommand{\ef}{ e,f }
\newcommand{\efi}{ e,f_i }
\newcommand{\efj}{ e,f_j }

\newcommand{\epfi}{ e',f_i }
\newcommand{\epfj}{ e',f_j }
\newcommand{\setpfi}{ \partial f_{i} }
\newcommand{\setpfj}{ \partial f_{j} }

\newcommand{\setpff}{ \partial f, f }

\newcommand{\xf}{ \bm{x}_{\setpf} }

\newcommand{\xfi}{ \bm{x}_{\setpf_i} }
\newcommand{\xfj}{ \bm{x}_{\setpf_j} }

\newcommand{\hLCT}[1]{ \hat{ \LCT{#1} } }
\newcommand{\hLCTgraphN}{ \hLCT{\graphN} }

\newcommand{\hgraphPsig}{ \hgraphN_{M,P_{\vsigma}} }

\newcommand{\LCTvxf}{ \LCTv{x}_{\setpf} }
\newcommand{\LCTvxfi}{ \LCTv{x}_{\setpfi} }

\newcommand{\LCTsetx}{ \LCTset{X} }
\newcommand{\LCTsetxe}{ \LCTset{X}_{e} }
\newcommand{\LCTsetxf}{ \LCTset{X}_{\setpf} }

\newcommand{\tx}{ \tilde{x} }
\newcommand{\txe}{ \tilde{x}_{e} }

\newcommand{\tvx}{ \tilde{\vect{x}}  }

\newcommand{\LCTvxl}{ \LCTtv{x}_{\mathrm{l}} }
\newcommand{\LCTvxs}{ \LCTtv{x}_{\mathrm{s}} }
\newcommand{\LCTvxlm}{ \LCTtv{x}_{\mathrm{l},m} }

\newcommand{\LCTvxsm}{ \LCTtv{x}_{\mathrm{s},m} }

\newcommand{\LCTvxlM}{ \LCTtv{x}_{\mathrm{l},[M]} }
\newcommand{\LCTvxsM}{ \LCTtv{x}_{\mathrm{s},[M]} }

\newcommand{\tzero}{\tilde{0}}

\newcommand*{\ie}{\textit{i.e.}}
\newcommand*{\eg}{\textit{e.g.}}
\newcommand*{\etal}{\textit{et al.}}

\newcommand{\gzero}{\LCT{g}_{0}}
\newcommand{\gone}{\LCT{g}_{1}}

\newcommand{\tv}[1]{ \tilde{\vect{#1}} }

\newcommand{\etofone}{ e, f_1 }
\newcommand{\onetoftwo}{ 1, f_2 }

\newcommand{\pmf}[1]{ \set{P}_{#1} }

\IEEEoverridecommandlockouts

\usepackage{stfloats}
\usepackage{balance}
\allowdisplaybreaks[4]

\begin{document}

\title{Graph-Cover-based Characterization of the Bethe Partition Function of Double-Edge Factor Graphs}

\author{Yuwen~Huang,~\IEEEmembership{Member, IEEE,}
  and Pascal~O.~Vontobel,~\IEEEmembership{Fellow, IEEE}
  \thanks{
    The work described in this paper was partially supported by the Research Grants Council of the Hong Kong Special Administrative Region, China (Project Nos. CUHK 14209317, CUHK 14207518, CUHK 14208319), by The Chinese University of Hong Kong under Direct Grant 4055204, by the Department of Education of Guangdong Province under the Grant 2026KQNCX066, and by the Hong Kong University of Science and Technology (Guangzhou) under the Grant G0101000351.
    An earlier version of this paper was presented in part at the IEEE International Symposium on Information Theory (ISIT), Los Angeles, CA, USA, June~2020~[DOI: 10.1109/ISIT44484.2020.9174508].
    \textit{(Corresponding author: Yuwen Huang.)}
  }
  \thanks{Y.~Huang is with the
    Thrust of Data Science and Analytics,
    The Hong Kong University of Science and Technology (Guangzhou),
    Guangzhou, China.
    Part of the work for this paper was done while being with the Department of Information Engineering, The Chinese University of Hong Kong, Hong Kong SAR
    (e-mail: yuwen.huang@ieee.org).}%
  \thanks{P.~O.~Vontobel is with the
    Department of Information Engineering and
    the Institute of Theoretical Computer Science and Communications,
    The Chinese University of Hong Kong, Hong Kong SAR
    (e-mail: pascal.vontobel@ieee.org).}%
}

\maketitle

\begin{abstract}

  For standard factor graphs (S-FGs), whose local functions are
    nonnegative-real valued, a popular approximation of the partition function
    is given by the Bethe approximation, which is also known as the Bethe
    partition function. Vontobel gave a combinatorial characterization of the
    Bethe partition function with the help of graph covers. The proof of this
    characterization heavily relied on the method of types.

  In this paper, we study double-edge factor graphs (DE-FGs), whose local
    functions are complex valued and satisfy certain positive-semidefinite
    constraints.  DE-FGs and their partition functions are particularly
    relevant to quantum information processing. Approximating the partition
    function of a DE-FG is more difficult than approximating the partition
    function of an S-FG because the DE-FG partition function is a sum of complex values rather than nonnegative-real values.

  We develop a Bethe approximation of the partition function of a DE-FG
    based on sum-product-algorithm (SPA) fixed points. However, one cannot
    directly apply the method of types to prove a similar graph-cover-based
    combinatorial characterization of the Bethe partition function as in the
    case of S-FGs.

  Despite this, we establish such a combinatorial characterization for a
    class of DE-FGs that satisfy an easily checkable condition. Towards
    proving this characterization, we apply a suitable loop-calculus transform
    (LCT) to DE-FGs. Originally, the LCT was introduced by Chertkov and
    Chernyak as a special linear transform for S-FGs and later extended by
    Mori. Our proposed LCT is applicable to both S-FGs and DE-FGs and
    generalizes prior versions by handling zero-valued SPA fixed-point message
    components, which are common in DE-FGs.

  Supported by numerical results, we conjecture that the
    graph-cover-based combinatorial characterization of the Bethe partition
    function holds more broadly for DE-FGs.

\end{abstract}


\section{Introduction}

\label{sec:introduction:1}

Factor graphs and, more generally, graphical models are widely used to
  represent both classical statistical models~\cite{Kschischang2001,
  Forney2001,Loeliger2004} and quantum information processing
systems~\cite{Cao2017,Alkabetz2021}. Classical statistical models are
  typically represented by standard factor graphs
  (S-FGs)~\cite{Kschischang2001, Forney2001,Loeliger2004}, whose local
  functions take on nonnegative real values. S-FGs have extensive
applications across various fields, including statistical mechanics (see,
\eg,~\cite{Mezard2009}), coding theory (see, \eg,~\cite{Richardson2008}), and
communications (see, \eg,~\cite{Wymeersch2007}). Double-edge factor graphs
(DE-FGs)~\cite{Cao2017,Cao2021} represent quantities of interest in quantum
information processing; their local functions take on complex values and
satisfy certain positive-semidefinite (PSD) constraints.  DE-FGs extend
the concept of S-FGs and are closely related to other graphical models in
quantum information processing, such as tensor
networks~\cite{Cirac2009,Coecke2010,Robeva2019,Alkabetz2021} and
ZX-calculus~\cite{Coecke2008,Coecke2011}. DE-FGs provide a rigorous framework
for analyzing tensor network contraction problems that are central to quantum
simulation~\cite{Alkabetz2021}.

A key application of factor graphs is the reformulation of inference
  problems in terms of the computation of the marginals and the partition
  function of a multivariate function whose factorization is depicted by an
S-FG~\cite{Kschischang2001,Loeliger2004} or a
DE-FG~\cite{Alkabetz2021}. However, the exact computation of the marginals and
the partition function is, in general, challenging for both S-FGs and
DE-FGs. A common approach is to apply the sum-product algorithm (SPA),
  also known as loopy belief propagation (LBP), to an S-FG or a DE-FG. The SPA
  is a heuristic algorithm that can be used to approximate marginals and the
  partition function for many classes of factor graphs. In fact, for
  cycle-free S-FGs and DE-FGs, the SPA beliefs equal the exact marginals (see,
  \eg,~\cite{Mezard2009}). For a graph with cycles, the SPA often approximates
  the marginals and the partition function accurately, although instances with
  poor approximations exist~\cite{Murphy1999,Heskes2003,Weller2014}.
\finaltext{For S-FGs, Yedidia \etal~\cite{Yedidia2005} established \finaltext{a}
correspondence between SPA fixed points with strictly positive beliefs and
interior stationary points of the Bethe free energy
function~\cite{Bethe1935}. The minimum of this function determines the Bethe
approximation of the partition function.}

The loop calculus technique provides a systematic framework for characterizing
the partition function and the Bethe partition function for
S-FGs. Initially introduced by Chertkov and Chernyak~\cite{Chertkov2006} for S-FGs with binary alphabets, the loop calculus technique
expresses the partition function as a finite series: the first term in
this series equals the Bethe partition function, while higher-order terms
correspond to loop corrections. Later on, their loop calculus technique was extended to non-binary alphabets by Chernyak and Chertkov~\cite{Chernyak2007} and by Mori~\cite{Mori2015}. (For some limitations of these extensions, see the discussion in Section~\ref{sec:LCT}, in particular Table~\ref{tab:lct_comparison}.)

Finite graph covers provide another framework for characterizing the
  partition function and the Bethe partition function of S-FGs. The relevance
  of finite graph covers for the SPA follows from a separation between local
  computation and global topology. A finite graph cover may change the global
  cycle structure, but it preserves each node's local incidence structure and
  local function~\cite{Koetter2003}. Because SPA message updates depend only
  on this local information, the SPA uses the same SPA message update rule on
  the base factor graph and its covers. This local indistinguishability
  motivates the graph-cover-based analysis of the SPA. Specifically:
\begin{itemize}

  \item For an arbitrary S-FG, Vontobel~\cite{Vontobel2013} provided a
        combinatorial characterization of the Bethe partition function in terms of
        its finite graph covers, establishing the graph-cover theorem. Let
          $\sfN$, $Z(\sfN)$, $\ZB(\sfN)$, and $\ZBM(\sfN)$ denote an arbitrary S-FG,
          its partition function, its Bethe partition function, and its degree-$M$
          Bethe partition function (defined through the $M$-covers of $\sfN$),
          respectively. The theorem states that
        \begin{align*}
          \ZBM(\sfN)                         & = Z(\sfN), \quad \text{for } M = 1, \nonumber \\
          \limsup_{ M \to \infty} \ZBM(\sfN) & = \ZB(\sfN),
        \end{align*}
        where the first identity holds because the only $1$-cover of $\sfN$ is $\sfN$ itself.

  \item Building on this result, Ruozzi~\cite{Ruozzi2012} and
          Csikv\'{a}ri~\etal~\cite{Csikvari2022} proved that the partition
          function is lower-bounded by the Bethe partition function for certain
        classes of S-FGs, thereby in particular resolving a conjecture of
          Sudderth~\etal~\cite{E.B.Sudderth2007} on log-supermodular graphical
        models.

  \item The paper~\cite{Vontobel2025} used double-cover-based methods to
          analyze the partition-function-to-Bethe-approximation ratio for two
          classes of log-supermodular factor graphs, and the papers~\cite{Wu2026}
          and~\cite{Zhou2026} analyzed factor graphs whose partition function equals
          the permanent of block-constant matrices with positive real and complex
          entries, respectively.

  \item For positive integers $M$, the
          papers~\cite{Altieri2017,Angelini2022,AngeliniPalazzi2024,Palazzi2025} used
          $M$-covers (called the $M$-layer construction in these papers) to study
          spin glasses, Ising models, and critical phenomena.

  \item Leleu \etal~\cite{Leleu2026} used graph covers to formulate
          algorithms that accelerated the convergence speed of finding optimal
          solutions to certain problems.

\end{itemize}

We now turn from S-FGs to factor graphs with complex-valued local
  functions, which arise naturally in quantum information
  processing~\cite{Loeliger2012,Loeliger2017,Loeliger2020,Mori2015a}. (For
connections of these factor graphs to other graphical notations in physics,
see~\cite[Appendix~A]{Loeliger2017}.) Cao and Vontobel~\cite{Cao2017}
  formalized such factor graphs as DE-FGs, whose local functions take complex
  values and satisfy certain PSD constraints. They also showed that the
partition function of a DE-FG is a nonnegative real number. However,
because the partition function involves summing complex-valued quantities,
approximating the partition function is, in general, much more challenging for
DE-FGs than for S-FGs. This challenge is known as the numerical sign problem
in applied mathematics and theoretical physics~(see,
\eg,~\cite{LohJr1990}). To approximate DE-FG partition functions, Cao and
  Vontobel~\cite{Cao2017} defined the Bethe partition function based on SPA
  fixed-point messages.

The SPA on DE-FGs is closely related to tensor-network methods, for which
  SPA-based approximations of marginals and the partition function turn out to
  perform well numerically for certain setups. Alkabetz and
Arad~\cite{Alkabetz2021} formalized this connection by mapping
projected-entangled-pair-state (PEPS) tensor networks to DE-FGs, thereby
establishing a direct link between the SPA on DE-FGs and tensor network
contractions.  Subsequent work has built on this foundation:
  Guo~\etal~\cite{Guo2023} proposed a block SPA for two-dimensional tensor
  networks; Tindall and Fishman~\cite{Tindall2023} related the SPA to known
tensor network gauging methods; and Tindall~\etal~\cite{Tindall2024}
demonstrated state-of-the-art accuracy of SPA-based approximations for
  the kicked Ising system~\cite{Kim2023}. Generalized and block belief
  propagation~\cite{Tindall2026,Gutman2026} further extend the SPA to richer
  contraction schedules. These results motivate a deeper theoretical study of
  the SPA on DE-FGs.

Given the combinatorial characterization of the Bethe partition function
for S-FGs via finite graph covers, it is natural to ask whether an analogous
characterization exists for DE-FGs. Numerical results on small DE-FGs suggest
that such a characterization may exist. However, establishing this
characterization for DE-FGs is significantly more challenging than for
S-FGs. The complex-valued nature of the local functions of DE-FGs prevents the
direct application of the method of types, as the following example
  illustrates.

\begin{example}
  \label{example:simple:complex:sum:1}

  Let $ \alpha $ be a complex number.
  For any positive integer $ M $, we define
  \begin{align}
    Z_{M} \defeq
    \sum\limits_{\ell = 0}^{M} c_{M,\ell},
    \label{eqn: expression of ZM}
  \end{align}
  where
  \begin{align*}
    c_{M,\ell} = \binom{M}{ \ell }
    \cdot ( 1 - \alpha )^{M - \ell}
    \cdot \alpha^{ \ell }.
  \end{align*}
  Using the binomial theorem, one can verify that $ Z_{M} = 1 $ for any
    positive integer $M$, which implies
    $\lim_{M \to \infty} \frac{1}{M} \log(Z_M) = 0$.

  When $ 0 \leq \alpha \leq 1 $, every term $ c_{M,\ell} $ is
    nonnegative, so the sum in~\eqref{eqn: expression of ZM} can be bounded as
    follows:
    \begin{align*}
      \max\limits_{\ell\in\{0,1,\ldots,M\}} c_{M,\ell}
      \leq
      Z_M
      \leq
      (M \! + \! 1) \cdot \max\limits_{\ell\in\{0,1,\ldots,M\}} c_{M,\ell}.
    \end{align*}
    A Stirling estimate of $c_{M,\ell}$ yields
    $ \max_{\ell} c_{M,\ell} = e^{o(M)} $, whence
    $\lim_{M \to \infty} \frac{1}{M} \log(Z_{M}) = 0$. Such a regime of sums
    with polynomially many nonnegative terms as here is exactly where
    Vontobel~\cite{Vontobel2013} applied the method of types to characterize
    the Bethe partition function of an S-FG via finite graph covers.

  However, when $ \alpha $ is a complex number with real part strictly
  less than zero or greater than one, the real and the imaginary parts of the
  terms in $ \{c_{M,\ell} \}_{\ell \in \{0,\ldots,M\} } $ may take on negative
  or positive values and are added constructively and destructively in the
  expression for $Z_{M}$ in~\eqref{eqn: expression of ZM}. Consequently, the
  term with the largest magnitude in the sum in~\eqref{eqn: expression of ZM}
  provides a poor estimate of $ Z_{M} $, and the limit
  $ \lim\limits_{ M \to \infty } \frac{1}{M} \log(Z_{M}) $ cannot be
  approximated using the same approach as in the case
  $ 0 \leq \alpha \leq 1 $.  \eexample
\end{example}

\subsection{Main Contributions}

We develop a novel approach to establish a finite-graph-cover-based
combinatorial characterization of the Bethe partition function for a certain
class of DE-FGs. Beyond this characterization, the proof itself is a key
contribution of this work.

We separate what this paper inherits, adapts, and
  contributes. \emph{Inherited} from the S-FG literature are finite graph
  covers, the SPA, the Bethe approximation, and classical loop
  calculus. \emph{Adapted} to DE-FGs are the SPA, its beliefs, and the
  SPA-based Bethe partition function, together with finite graph covers that
  permute the two strands of each double edge as a unit. \emph{Genuinely new}
  are an explicit loop-calculus transform (LCT) for non-binary alphabets
  admitting \finaltext{complex-valued SPA fixed-point messages with possibly zero-valued components}; the
  graph-cover conjecture for DE-FGs (Conjecture~\ref{sec:GraCov:conj:1}); and
  the graph-cover theorem for a certain class of DE-FGs
  (Theorem~\ref{sec:CheckCon:thm:1}), with its proof.

Our proof of Theorem~\ref{sec:CheckCon:thm:1} relies on three key ingredients:
\begin{enumerate}

  \item \textbf{Loop-Calculus Transform (LCT):} Given an SPA fixed point of
          a DE-FG, we develop and apply the LCT\footnote{The LCT can be seen
          as a particular instance of a holographic
          transform~\cite{AlBashabsheh2011}.}\ to the DE-FG. (If the SPA has
        multiple fixed points, we choose the one with the largest Bethe partition
        function value.) Unlike prior
        versions~\cite{Chertkov2006,Chernyak2007,Mori2015}, our proposed LCT handles
        SPA fixed-point messages with zero-valued components, which is a necessary
        extension for the analysis of DE-FGs. The transform preserves the
          partition function and gives the transformed SPA fixed-point messages an
          elementary form independent of the edge and direction.

  \item \textbf{$M$-Cover Construction:} For any positive integer $M$, we
        \finaltext{construct $M$-covers of the transformed DE-FG}. We then construct a
        factor graph whose partition function equals the arithmetic mean of the
        partition functions of all $M$-covers.

  \item \textbf{Limit Evaluation:} We let $M \to \infty$ and evaluate the limit
        of the $M$-th root of the partition function of the factor graph constructed
        in the previous step. We establish that, under a specific (easily checkable)
        condition, this limit equals the Bethe partition function of the original
        DE-FG, thereby proving the graph-cover theorem for this special class of
        DE-FGs. Numerical results suggest, interestingly, that the graph-cover
        theorem may hold more broadly. Proving the graph-cover theorem for general
        DE-FGs remains an open problem.

\end{enumerate}

The LCT introduced in this paper should also be of interest beyond this
paper's scope. The LCT generalizes the one presented in~\cite{Mori2015} and
extends its applicability to SPA fixed-point messages consisting of
complex-valued and zero-valued components.

Based on the $M$-cover construction of a DE-FG, Huang in~\cite[Chapter
  5]{Huang2024} further developed the symmetric-subspace transform. The
symmetric-subspace transform allows one to express the partition function of
the average $M$-cover of a DE-FG in terms of an integral. We anticipate that
the symmetric-subspace transform may serve as a key technique in proving the
graph-cover theorem for a broad class of DE-FGs.

\textbf{Comparison with related work}: Our research complements recent work by
Evenbly~\etal~\cite{Evenbly2024}, but the objectives of the two papers differ
fundamentally. Evenbly~\etal~\cite{Evenbly2024} applied the loop series
expansion as a practical, numerical tool to improve the accuracy of tensor
network contractions, showing significant performance gains on benchmark
models. In contrast, we employ the LCT as a theoretical instrument to prove a
formal, combinatorial characterization of the Bethe partition function for
DE-FGs (Theorem~\ref{sec:CheckCon:thm:1}).

\subsection{Basic Notations and Definitions}

Unless stated otherwise, all variable alphabets are assumed to be finite.

We define $\sZ$, $\sZp$, $\sZpp$, $\sR$, $\sRp$, $\sRpp$, and $\sC$ to be the
ring of integers, the set of nonnegative integers, the set of positive
integers, the field of real numbers, the set of nonnegative real
numbers, the set of positive real numbers, and the field of complex
  numbers, respectively. We use an overline to denote complex conjugation.


We use square brackets in two different ways. First, for any $L \in \sZpp$,
square brackets are used for representing the set
$[L] \defeq \{1, \ldots, L\}$. Second, for any statement $S$ we use Iverson's
convention, which defines $[S] \defeq 1$ if $S$ is true and $[S] \defeq 0$
otherwise.

For a finite set $\set{X}$, we define $\pmf{\set{X}}$ to be the set of probability
mass functions over $\set{X}$, \ie,
\begin{align*}
   & \pmf{\set{X}} \defeq \Biggl\{ p: \set{X} \to \sR \ \Biggl| \ p(x) \geq 0 \
  \text{for all} \,
  x \in \set{X}, \sum_{x \in \set{X}} p(x) = 1 \Biggr\}.
\end{align*}
For a positive integer $M \in \sZpp$, we define $\set{S}_M$ to be the set of all
permutations over $[M]$.

All logarithms are natural logarithms. As usually done in information theory,
we define $\log(0) \defeq -\infty$ and $0 \cdot \log(0) \defeq 0$.

We use standard linear algebra notation, similar to that
  in~\cite{Loeliger2017,Loeliger2020}, rather than the bra-ket notation of
  quantum mechanics. We use $\matr{I}$ to denote an identity matrix. We
denote the Hermitian transpose of a complex-valued matrix $\matr{A}$ by
$\matr{A}^{\Herm} = \overline{\matr{A}}^{\tran}$, where $\matr{A}^{\tran}$ and
$\overline{\matr{A}}$ denote the transpose of $\matr{A}$ and the componentwise
complex conjugate of $\matr{A}$, respectively. We associate a matrix
$\matr{A}$ with a function $A(x, y)$ such that
$ \matr{A} = \bigl( A(x, y) \bigr)_{ \! x \in \set{X}, y \in \set{Y} } $,
where $x \in \set{X}$ is the row index of $\matr{A}$ and $y \in \set{Y}$ is
the column index of $\matr{A}$. In this notation, the expression $\matr{A}(:, y)$ gives the
column $y$ of $\matr{A}$, and $\matr{A}(x, :)$ gives the row $x$ of
$\matr{A}$.

In this paper, we consider normal factor graphs (NFGs), \ie, factor graphs
where variables are associated with edges~\cite{Kschischang2001,Loeliger2004}.

The rest of this paper is structured as follows. Section~\ref{sec:SNFG}
  introduces the basics of standard normal factor graphs (S-NFGs).
  Section~\ref{sec:DENFG} introduces double-edge normal factor graphs
  (DE-NFGs). Section~\ref{sec:GraCov} defines finite graph covers for S-NFGs
  and DE-NFGs. Section~\ref{sec:LCT} introduces a variant of the
  LCT. Section~\ref{sec:symmetric-subspace transform} considers the average
  $M$-cover for S-NFGs and DE-NFGs. Section~\ref{sec:CheckCon} presents the
  graph-cover theorem for a certain class of DE-NFGs satisfying an easily
  checkable condition.  Section~\ref{sec: conclusion and open problems}
  concludes the paper and presents open problems.  Various (longer) proofs and
  additional material have been collected in the appendices.

\section{Standard Normal Factor Graphs (S-NFGs)}
\label{sec:SNFG}

In this section, we review basic concepts and properties of an
S-NFG,\footnote{Here and throughout, an S-NFG is an NFG whose local
    functions are nonnegative and real-valued. The adjective ``standard''
    serves only to distinguish this class from the complex-valued DE-NFGs
    studied in Section~\ref{sec:DENFG}. This terminology does not imply that
    factor graphs with complex-valued local functions are new; such graphs
    appear, for example, in factor-graph realizations of the fast Fourier
    transform~\cite{Kschischang2001} and in quantum-probability
    models~\cite{Loeliger2017,Loeliger2020}.} along with the associated Bethe
approximation of the partition function.  We use an example to introduce the
key concepts of an S-NFG.

\begin{figure}
  \centering
  \subfloat{
    \begin{minipage}[t]{0.4\linewidth}
      \centering
      \scalebox{0.7}{%
        \begin{tikzpicture}[node distance=2cm, remember picture]
          \input{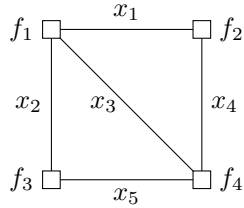}
          \node[state] (f1) at (0,0) [label=left: $f_{1}$] {};
          \node[state,right of=f1] (f2) [label=right: $f_{2}$] {};
          \node[state,below of=f1] (f3) [label=left: $f_{3}$] {};
          \node[state,right of=f3] (f4) [label=right: $f_{4}$] {};
          \begin{pgfonlayer}{background}
            \draw[-,draw]
            (f1) edge node[above]  {$x_{1}$} (f2)
            (f1) edge node[left]  {$x_{2}$} (f3)
            (f1) edge node[left]  {$x_{3}$} (f4) 
            (f2) edge node[right] {$x_{4}$} (f4)
            (f3) edge node[below] {$x_{5}$} (f4);
          \end{pgfonlayer}
        \end{tikzpicture}}
    \end{minipage}%
  }
  \caption{The S-NFG in Example \ref{sec:SNFG:exp:1}.}
  \label{sec:SNFG:fig:1}
\end{figure}

\begin{example}\label{sec:SNFG:exp:1}

  Consider the multivariate function
  $g(x_{1},\ldots,x_{5}) \defeq f_{1}(x_{1},x_{2},x_{3}) \cdot f_{2}(x_{1},x_{4})
    \cdot f_{3}(x_{2},x_{5})\cdot f_{4}(x_{3},x_{4},x_{5})$, where $g$, the
  global function, is defined to be the product of the
  local functions $f_{1}$, $f_{2}$, $f_{3}$ and $f_{4}$. The factorization of $g$ is visualized by the S-NFG $\graphN$ in
  Fig.~\ref{sec:SNFG:fig:1}, which consists of four function nodes
  $f_1, \ldots, f_4$ corresponding to local functions $f_1, \ldots, f_4$
  and five (full) edges $ 1,\ldots,5 $ corresponding to variables
  $x_1, \ldots, x_5$.
  \eexample
\end{example}

An edge that connects to two function nodes is called a full edge, whereas an
edge that is connected to only one function node is called a half
edge. Half edges are useful for many practical applications (see,
\eg,~\cite{Loeliger2004}). In this paper, it suffices to consider S-NFGs
with only full edges, as half edges can be converted into full edges by
attaching a dummy $1$-valued function node to the other end of every half edge
without changing any marginals or the partition function, and also without
changing the cycle structure of the S-NFG.

\begin{definition}
  \label{def: def of snfg}
  An S-NFG $\graphN(\setF, \setEfull, \set{X})$ consists of the following
  objects.
  \begin{enumerate}

    \item The graph $(\setF,\setEfull)$ with vertex set $\setF$ and edge set
          $\setEfull$, where $\setF$ is also known as the function-node
            set. (As mentioned above, we assume that every edge
            $e \in \setEfull$ is a full edge connecting two function nodes.)

    \item The alphabet $\set{X} \defeq \prod_{e \in \setEfull} \setxe$, where
          $\setxe$ is the finite alphabet of the variable $\xe$ associated with the
          edge $e \in \setEfull$. In this paper, without loss of generality, we
          assume that $ 0 \in \setxe$ for all $ e \in \setEfull $.

  \end{enumerate}
  For an arbitrary S-NFG $\graphN(\setF,\setEfull,\set{X})$, we assume that
  $ \set{F} = \{ f_{1},\ldots,f_{|\set{F}|} \} $, \ie, the function node set
  $ \set{F} $ consists of the elements $ f_{1},\ldots,f_{|\set{F}|}. $ We
    then define the following quantities.

  \begin{enumerate}
    \setcounter{enumi}{2}

    \item For every function node $f \in \setF$, the set $\setpf$ is defined to
          be the set of edges incident on $f$.

    \item For each edge $e\in\setEfull$ with endpoints $f_i$ and $f_j$,
            orient $e$ from $f_i$ to $f_j$ when $i < j$. This convention is only
            notational; all results are invariant under the orientation.

            Moreover, when writing an expression like $e = (f_i, f_j)$, we mean an
            edge that has endpoints $f_i$ and $f_j$. Note that we allow S-NFGs with
            parallel edges, and so an edge is not uniquely defined by its two
            endpoints.

    \item An assignment $\vx \defeq (\xe)_{e \in \setEfull} \in \set{X}$ is
          called a configuration of the S-NFG $ \sfN $.

    \item For any finite set $ \set{I} \subseteq \setEfull $, we define
          \begin{align*}
            \setx_{ \set{I} } \defeq
            \prod_{ e \in \set{I} } \setxe, \qquad
            \vx_{\set{I}} \defeq (\xe)_{e \in \set{I}}
            \in \setx_{ \set{I} }.
          \end{align*}

    \item For every function node $f \in \setF$, the local function associated
          with $f$ is, by a slight abuse of notation, also denoted by
            $f$. Here, the local function $f$ is an arbitrary mapping from $\setxf$
          to $\sRp$.

    \item The global function $g$ is defined to be the mapping
          $g:\set{X}\to \sRp, \ \vx\mapsto \prod_{f \in \setF} f(\xf)$.

    \item A configuration $\vx \in \set{X}$ satisfying $g(\vx) \neq 0$ is called
          a valid configuration.

    \item The partition function\footnote{In this paper, the partition function
            $Z(\graphN)$ of $\graphN$ is a scalar, \ie, it is not really a
            function. If $\graphN$ depends on some parameter (say, some pressure
            parameter), then $Z(\graphN)$ is a function of that parameter.} is
          defined to be $Z(\graphN) \defeq \sum_{\vx \in
              \set{X}}g(\vx)$. Thus, $Z(\graphN) \in \sRp$.\edefinition

  \end{enumerate}
\end{definition}

We consider only S-NFGs $\graphN$ for which
$Z(\graphN) \in \sRpp$.

When no ambiguity arises, we use the shorthand notations
$\sum_{\vx}$, $\sum_{\xe}$, $\sum_e$, $\sum_{f}$,
$\prod_{\vx}$, $\prod_{\xe}$, $\prod_e$, $\prod_{f}$,
$ (\cdot)_{\xe} $,
and
$ (\cdot)_{e} $ for
$\sum_{\vx \in \set{X}}$,
$\sum_{\xe\in\setxe}$,
$\sum_{e \in \setEfull}$, $\sum_{f \in \setF}$,
$\prod_{\vx \in \set{X}}$,
$\prod_{\xe \in \setxe}$, $\prod_{e \in \setEfull}$, $\prod_{f \in \setF}$,
$ (\cdot)_{\xe \in \setxe} $,
and
$ (\cdot)_{e\in \setEfull} $,
respectively.
For any set $ \set{I} \subseteq \setEfull $, we use the shorthand notations
$\sum_{\vx_{\set{I}}}$,
$\prod_{\vx_{\set{I}}}$,
and $ (\cdot)_{\vx_{\set{I}}} $
for
$\sum_{ \vx_{\set{I}} \in \setx_{\set{I}} }$,
$\prod_{ \vx_{\set{I}} \in \setx_{\set{I}} }$,
and $ (\cdot)_{ \vx_{\set{I}} \in \setx_{\set{I}} } $, respectively.
Moreover, the notation $\setpf \setminus e$ denotes
$\setpf \setminus \{ e \}$.

In the following, we give only the technical details of the SPA on an
  S-NFG; for a general discussion of the motivation for the SPA, see,
  \eg,~\cite{Kschischang2001,Loeliger2004}.

\begin{definition}\label{sec:SNFG:def:4}
  \label{def:belief at SPA fixed point for S-NFG}

  Consider an S-NFG $\graphN$. The SPA is an iterative algorithm that
    sends messages along the edges, where every message is a function of the
    variable associated with the corresponding edge. At every iteration, two
    messages are sent along every edge, one in each direction. More
    precisely, the SPA consists of the following steps.
  \begin{enumerate}

    \item (Initialization) For every $f \in \setF$ and $ e \in \setpf $, initialize
            $\mu_{\etof}^{(0)}: \setxe \to \sR_{> 0}$ as any function satisfying
          $ \sum_{\xe} \mu_{\etof}^{(0)}(\xe) = 1 $.
          (Typically, we set $\mu_{\etof}^{(0)}(\xe) \defeq 1 / |\setxe|$
          for all $\xe \in \setxe$.)

    \item (Iteration) For $t = 1, 2, 3, \ldots$, do the following calculations
          until some termination criterion is met.\footnote{The termination
            criterion is typically a combination of numerical convergence and an
            upper bound on the number of iterations.}
          \begin{enumerate}
            \item\label{SPA: update message} For every $e \in \setEfull$,
                  we define the messages from edge $ e = (f_{i}, f_{j}) $ to the function nodes
                  $ f_{j} $ and $ f_{i} $ to be the mappings
                  \begin{alignat*}{3}
                    \mu_{\efj}^{(t)}:\, \set{X}_{e} & \to \sR_{\geq 0}, \\
                    x_{e}                           & \mapsto
                    \frac{1}{\kappa_{\efj}^{(t)}}
                    \cdot
                    \sum_{\vx_{ \setpfi \setminus e }}
                    f_{i}(\xfi)
                    \cdot \!\!\!\! \prod_{e' \in \setpfi \setminus e}\!\!\!\!
                    \mu_{\epfi}^{(t-1)}(x_{e'}), \nonumber
                    \\
                    \mu_{\efi}^{(t)}:\, \set{X}_{e} & \to \sR_{\geq 0}, \\
                    x_{e}                           & \mapsto
                    \frac{1}{\kappa_{\efi}^{(t)}}
                    \cdot
                    \sum_{ \vx_{ \setpfj \setminus e } }
                    f_{j}(\xfj)
                    \cdot \!\!\!\! \prod_{e' \in \setpfj \setminus e}\!\!\!\!
                    \mu_{\epfj}^{(t-1)}(x_{e'}), \nonumber
                  \end{alignat*}
                  respectively, where the scaling factors $ \kappa_{\efj}^{(t)} $ and $ \kappa_{\efi}^{(t)} $ are defined by default to be
                  \begin{align*}
                    \kappa_{\efj}^{(t)}
                                        & \defeq
                    \sum_{\xfi}
                    f_{i}(\xfi)
                    \cdot
                    \prod_{e' \in \setpfi \setminus e}
                    \mu_{\epfi}^{(t-1)}(x_{e'}),
                    \nonumber                    \\
                    \kappa_{\efi}^{(t)} & \defeq
                    \sum_{\xfj}
                    f_{j}(\xfj)
                    \cdot
                    \prod_{e' \in \setpfj \setminus e}
                    \mu_{\epfj}^{(t-1)}(x_{e'}).
                  \end{align*}
                  The \emph{unnormalized message} is obtained from
                    the corresponding message-update formula by omitting the
                    factor $1/\kappa_{\ef}^{(t)}$. The normalized message is
                    defined only when $\kappa_{\ef}^{(t)}>0$.
                  With this choice of scaling factors, each message sums
                    to one and can thus be viewed as a probability mass
                    function over $\set{X}_e$; the scaling also keeps the
                    message values bounded over the iterations. Importantly,
                    as pointed out in the upcoming
                    Definition~\ref{sec:SNFG:def:2}, the main quantity of
                    interest in this paper is the value of $\ZBSPA(\graphN,\vmu)$, which is
                    invariant under any rescaling of the messages. The
                  associated message vectors are defined to be
                  $ \vmu_{\efj}^{(t)} \defeq \bigl( \mu_{\efj}^{(t)}(\xe)
                    \bigr)_{ \!  \xe } $ and
                  $ \vmu_{\efi}^{(t)} \defeq \bigl( \mu_{\efi}^{(t)}(\xe)
                    \bigr)_{ \!  \xe } $.

            \item\label{SPA: update belief edges} For every $e = (f_{i},f_{j}) \in \setEfull$, we define the belief function at edge $e$ to be the mapping
                  \begin{align}
                    \beli_e^{(t)}: \setxe \to \sR_{\geq 0}, \quad
                    \xe
                     & \mapsto
                    \frac{1}{\kappa_e^{(t)}}
                    \cdot
                    \mu_{\efi}^{(t)}(\xe) \cdot
                    \mu_{\efj}^{(t)}(\xe),
                    \label{sec:SNFG:eqn:2}
                  \end{align}
                  where the scaling factor $ \kappa_e^{(t)} $ is defined to be
                  \begin{align*}
                    \kappa_e^{(t)} \defeq \sum_{\xe}
                    \mu_{\efi}^{(t)}(\xe) \cdot \mu_{\efj}^{(t)}(\xe).
                  \end{align*}
                  The associated vector is defined to be
                  $ \vbeli_{e}^{(t)} \defeq \bigl( \beli_e^{(t)}(\xe) \bigr)_{ \!  \xe } $.

            \item\label{SPA: update belief functions} For every $f \in \setF$, we define the belief function at function node $f$ to be the mapping
                  \begin{align}
                    \beli_f^{(t)}: \setxf & \to \sR_{\geq 0}, \nonumber \\
                    \xf
                                          & \mapsto
                    \frac{1}{\kappa_f^{(t)}}
                    \cdot
                    f(\xf)
                    \cdot  \prod_{e \in \setpf}
                    \mu_{\etof}^{(t)}(\xe),
                    \label{sec:SNFG:eqn:1}
                  \end{align}
                  where the scaling factor $ \kappa_f^{(t)} $ is defined to be
                  \begin{align*}
                    \kappa_f^{(t)} \defeq
                    \sum_{ \xf }
                    f(\xf) \cdot \prod_{e \in \setpf} \mu_{\etof}^{(t)}(\xe).
                  \end{align*}
                  The associated vector is defined to be
                  $ \vbeli_{f}^{(t)} \defeq \bigl( \beli_f^{(t)}(\xf) \bigr)_{ \! \xf } $.
                  Each belief is defined only when its
                    normalization factor is positive. If that factor is zero,
                    the corresponding belief is undefined; see
                    Remark~\ref{sec:SNFG:remk:1}, Item~\ref{sec:SNFG:remk:1:item:2}.

            \item\label{SPA: restart rule: S-NFG}
                  If an unnormalized message is identically zero,
                    restart from Step~1 with fresh random strictly positive
                    initial messages whose entries sum to one.
          \end{enumerate}

    \item At iteration $t \in \sZpp$, the SPA messages can be collected as the SPA message vector
          \begin{align*}
            \vmu^{(t)} \defeq
            \bigl( \vmu_{\etof}^{(t)} \bigr)_{ \! e \in \setpf,\, f \in \setF}.
          \end{align*}
          The message updates in Step~\ref{SPA: update message} can be written as
          \begin{align*}
            \vmu^{(t)} = \fSPAN\bigl( \vmu^{(t-1)} \bigr),
          \end{align*}
          where $\fSPAN$ denotes the normalized message-update map.

    \item An SPA message vector
          \begin{align}
            \vmu \defeq ( \vmu_{\etof} )_{e \in \setpf,\, f \in \setF},
            \label{sec:SNFG:eqn:10}
          \end{align}
          where
          $ \vmu_{\etof} \defeq \bigl( \mu_{\etof}(\xe) \bigr)_{ \! \xe} $
          is called an SPA fixed-point message vector if every message-update normalization factor is positive and
          \begin{align}
            \vmu = \fSPAN(\vmu). \label{sec:SNFG:eqn:11}
          \end{align}
          The belief functions $\beli_e$ and $\beli_f$ induced by
          $\vmu$ are given by~\eqref{sec:SNFG:eqn:2}
          and~\eqref{sec:SNFG:eqn:1}, respectively, whenever their corresponding
          normalization factors are positive. In these formulas, replace $\mu_{\etof}^{(t)}$ by $\mu_{\etof}$ for all $f\in\setF$ and $e\in\setpf$.\edefinition

  \end{enumerate}
\end{definition}

\begin{remark}
  \label{asmp: assume messages are nonnegative}

  We make some remarks on the SPA as defined in
  Definition~\ref{def:belief at SPA fixed point for S-NFG}.
  \begin{itemize}
    \item Other specified positive normalization rules may be
          used separately for nonzero messages.

    \item Under the default sum-to-one normalization, we have $\vmu_{\etof}^{(t)}\in\pmf{\setxe}$ for every $f\in\setF$ and every $e\in\setpf$ at each $t\in\sZp$. Whenever defined, the beliefs are
          normalized to sum to one: $\vbeli_e^{(t)}\in\pmf{\setxe}$ and
          $\vbeli_f^{(t)}\in\pmf{\setxf}$ for $t\in\sZpp$.\eassumption
  \end{itemize}

\end{remark}

Yedidia~\etal~\cite[Section V]{Yedidia2005} showed that the SPA fixed-point
message vectors can be characterized by the stationary points of the Bethe
free energy function.  The following definitions related to the Bethe free
energy function are adapted from~\cite[Section V]{Yedidia2005} to S-NFGs.

\begin{definition}\!\!\cite[Section V]{Yedidia2005}\label{sec:SNFG:def:3}
  Let $\graphN$ be an S-NFG. The collection of beliefs is defined to be
  \begin{align*}
    \vbeli \defeq \Bigl( ( \vbeli_f )_{f \in \setF}, \,
                                                      ( \vbeli_e )_{e \in \setE} \Bigr),
  \end{align*}
  where $ \vbeli_f \defeq \bigl( \beli_f(\vxf) \bigr)_{\! \vxf \in \setxf } $
  for all $ f \in \setF $
  and $ \vbeli_e \defeq \bigl( \beli_e(\xe) \bigr)_{\! \xe \in \setxe } $
  for all $ e \in \setEfull $.
  The local marginal polytope $\LMP(\graphN)$
  associated with $\graphN$ is defined to be
  \begin{align*}
    \LMP(\graphN)
     & \defeq \left\{
    \vbeli
    \ \middle| \
    \begin{array}{c}
      \vbeli_f \in \pmf{\setxf} \ \ \text{for all} \ f \in \setF, \\
      \vbeli_e \in \pmf{\setxe} \ \ \text{for all} \ e \in \setE, \\[0.10cm]
      \sum\limits_{\vx_{\setpf \setminus e}}
      \beli_f(\xf)
      = \beli_e(\xe)                                              \\
      \text{for all} \,
      f \in \setF, e \in \setpf, \xe \in \setxe
    \end{array}
    \right\}. \tag*{\ensuremath{\triangle}}
  \end{align*}
\end{definition}

\begin{definition}\!\!\cite[Section V]{Yedidia2005}
  \label{sec:SNFG:def:1}
  Let $\graphN$ be an S-NFG. The Bethe free energy function
  is defined to be the mapping
  \begin{align*}
    \FB: \LMP(\graphN) & \to \sR \cup \{+\infty\}, \quad
    \vbeli \mapsto  U_{\mathrm{B}}(\vbeli) -  H_{\mathrm{B}}(\vbeli),
  \end{align*}
  where
  \begin{align*}
    U_{\mathrm{B}}: \LMP(\graphN) & \to \sR \cup \{+\infty\},\quad
    \vbeli \mapsto
    \sum_{f} U_{\mathrm{B},f}(\vbeli_{f}),
    \nonumber                                                      \\
    H_{\mathrm{B}}: \LMP(\graphN) & \to \sR,\quad
    \vbeli \mapsto \sum_{f}
    H_{\mathrm{B},f}(\vbeli_{f})
    - \sum_{e} H_{\mathrm{B},e}(\vbeli_{e}),
  \end{align*}
  with
  \begin{align*}
     & U_{\mathrm{B},f}: \pmf{\setxf}
    \to \sR \cup \{+\infty\},              \,
    \vbeli_{f}
    \mapsto
    -\!\!\sum_{\xf}
    \beli_f(\xf)
    \cdot
    \log\bigl( f(\xf) \bigr), \nonumber             \\
     & H_{\mathrm{B},f}: \pmf{\setxf} \to \sR,\quad
    \vbeli_{f} \mapsto
    -\sum_{\xf}
    \beli_f(\xf)
    \cdot
    \log\bigl( \beli_f(\xf) \bigr), \nonumber       \\
     & H_{\mathrm{B},e}: \pmf{\setxe} \to \sR,\quad
    \vbeli_{e} \mapsto -\sum_{\xe}
    \beli_{e}(\xe)
    \cdot
    \log\bigl( \beli_{e}(\xe) \bigr).
  \end{align*}
  If $\beli_f(\xf)>0$ and $f(\xf)=0$, the corresponding term in
  $U_{\mathrm{B},f}(\vbeli_f)$ is defined to be $+\infty$.
  Here, the function $ U_{\mathrm{B}} $ is the Bethe average energy function and $ H_{\mathrm{B}} $ is the Bethe entropy function.
  (For NFGs with full and half edges, the sum in $ \sum_{e} H_{\mathrm{B},e}(\vbeli_{e}) $ is only over all full
  edges.)  The Bethe approximation of the partition function, henceforth
  called the Bethe partition function, is defined to be
  \begin{align}
    \ZB(\graphN)
     & \defeq
    \exp
    \left(
    -
    \min_{\vbeli \in \LMP(\graphN)}
    \FB(\vbeli)
    \right).
    \label{sec:SNFG:eqn:4}
  \end{align}
  This definition is often viewed as the primal formulation of the Bethe partition function.
  \edefinition
\end{definition}

For a cycle-free S-NFG $\graphN$, we have $\ZB(\graphN) = Z(\graphN)$, \ie,
the Bethe approximation of the partition function is exact. For an S-NFG
$\graphN$ with cycles, the Bethe partition function $\ZB(\graphN)$ can be
\finaltext{smaller than, equal to, or larger than} the partition function $Z(\graphN)$.
\begin{enumerate}

  \item In~\cite[Corollary 4.2]{Ruozzi2012}, Ruozzi proved that the Bethe
        partition function lower-bounds the partition function for binary,
        log-supermodular graphical models. He later extended this result to other
        models that are not necessarily binary or
        log-supermodular~\cite{Ruozzi2013}, including ferromagnetic Potts
          models and models related to counting weighted graph homomorphisms.

  \item A separate line of research has established
        $\ZB(\graphN) \leq Z(\graphN)$ for specific S-NFGs.
        In~\cite{Vontobel2013a}, Vontobel studied the S-NFG whose partition function
        equals the permanent of a nonnegative matrix. Building on this
          result, Gurvits~\cite[Theorem 2.2]{Gurvits2011} proved that the
        associated Bethe partition function lower-bounds the partition
        function; this bound was extended to S-NFGs \finaltext{whose local generating polynomials are multi-affine and real stable}~\cite{Straszak2019,Anari2021,Huang2024a}. Degree-$M$ Bethe and
          Sinkhorn permanents yield further bounds on the
          permanent~\cite{Huang2023}.

\end{enumerate}

For an S-NFG, Vontobel~\cite{Vontobel2013} gave a combinatorial
characterization of the analytically defined Bethe partition function
$\ZB(\graphN)$ in~\eqref{sec:SNFG:eqn:4}: it is the limit superior, as
$M\to\infty$, of the $M$-th root of the average partition function over all
$M$-covers. The proof uses the method of types to characterize valid cover
configurations. Section~\ref{sec:GraCov} presents the required details.

There is an alternative definition of the Bethe partition function based on SPA fixed-point
message vectors.

\begin{definition}
  \label{sec:SNFG:def:2}
  Consider an S-NFG $\graphN$ and let $\vmu $ be an SPA message
  vector\footnote{The SPA message vector $\vmu$ need not be an SPA
    fixed-point message vector.} for $\graphN$ as defined
  in~\eqref{sec:SNFG:eqn:10}. Define the $\vmu$-based Bethe partition function
  to be
  \begin{align}
    \ZBSPA(\graphN,\vmu)
     & \defeq
    \frac{\prod\limits_f Z_f(\vmu)
    }{\prod\limits_e Z_e(\vmu)}, \qquad
    Z_e(\vmu) > 0, \, e \in \setEfull,
    \label{sec:SNFG:eqn:5}
  \end{align}
  where
  \begin{alignat}{2}
    Z_f(\vmu)
     & \defeq
    \sum_{\xf}
    f(\xf)
    \cdot
    \prod_{e \in \setpf}
    \mu_{\etof}(\xe),
    \
     &        & \quad f \in \setF,
    \label{eqn: def of zf for snfg}                   \\
    Z_e(\vmu)
     & \defeq
    \sum_{\xe}
    \mu_{\efi}(\xe)
    \cdot \mu_{\efj}(\xe),
     &        & \quad e= (f_{i},f_{j}) \in \setEfull.
    \label{sec:SNFG:eqn:6}
  \end{alignat}
  (For NFGs with full and half edges, the product in the denominator on the
  right-hand side of~\eqref{sec:SNFG:eqn:5} is only over all full
  edges. Moreover, note that the expression for $\ZBSPA(\graphN,\vmu)$ is
  scaling invariant, \ie, if any of the messages $\mu_{\ef}$ is multiplied by
  a nonzero (complex) number, then the value of $\ZBSPA(\graphN,\vmu)$
    remains unchanged because the scaling of a message gives an equal
    multiplicative contribution to the numerator and the denominator of the
    expression for $\ZBSPA(\graphN,\vmu)$.) The SPA-based Bethe partition
  function $\ZBSPA^{*}(\graphN)$ is then defined to be
  \begin{align*}
    \ZBSPA^{*}(\graphN)
     & \defeq \max_{\vmu}
    \ZBSPA(\graphN,\vmu),
  \end{align*}
  where the maximum is over all SPA fixed-point message vectors of $\graphN$,
  \ie, over all SPA message vectors satisfying~\eqref{sec:SNFG:eqn:11}. The
  above definition is also called the pseudo-dual formulation of the Bethe
  partition function.  In contrast to the primal formulation defined
  in~\eqref{sec:SNFG:eqn:4}, the pseudo-dual formulation $ \ZBSPA^{*} $ is
  defined based on an SPA fixed point. For a discussion about the primal and
  the pseudo-dual formulations, see Remark~\ref{sec:SNFG:remk:1}.
  \edefinition
\end{definition}

The Bethe free energy function $ \FB $ and the $\vmu$-based Bethe partition
function $ \ZBSPA $, as used in the definitions of the primal and the
pseudo-dual formulations, are connected through the SPA fixed-point message
vector $ \vmu $.

\begin{proposition}\!\cite[Theorem 2]{Yedidia2005} For every interior
  stationary point $ \vbeli{} \in \LMP(\graphN) $ of the Bethe free energy
  function $ \FB $, there exists an SPA fixed-point message vector $ \vmu $
  such that it consists of strictly positive elements only and
  \begin{align*}
    \exp \bigl( -\FB(\vbeli) \bigr) = \ZBSPA(\graphN,\vmu).
  \end{align*}
  \finaltext{Conversely, every SPA fixed-point message vector $\vmu$ whose
  induced beliefs are strictly positive corresponds to an interior stationary
  point of $\FB$.}  \eproposition
\end{proposition}

Building on this result, Yedidia~\etal~\cite[Conjecture 1]{Yedidia2005}
  conjectured that all SPA fixed points, including those with at least one
  zero-valued entry, correspond to stationary points of the Bethe free energy
  function. They further conjectured in~\cite[Conjecture 2]{Yedidia2005} that
  boundary local minima of the Bethe free energy function correspond to SPA fixed
  points.

\begin{remark}\label{sec:SNFG:remk:1}
  We make some remarks on the primal and the pseudo-dual formulations of the
  Bethe partition function, as defined in Definitions~\ref{sec:SNFG:def:1}
  and~\ref{sec:SNFG:def:2}, respectively.
  \begin{enumerate}

    \item For S-NFGs, the primal and pseudo-dual formulations of the Bethe
          partition function are typically well defined and exactly equal. For
          example, when all local functions in an S-NFG take only positive real
            values, \ie,
          \begin{align*}
            f(\vxf) > 0, \qquad \vxf \in \setxf,\, f \in \setF,
          \end{align*}
          then Yedidia~\etal~\cite[Propositions 5 and 6]{Yedidia2005} proved that
          \begin{align*}
            \ZB(\graphN) = \ZBSPA^{*}(\graphN) \in \sR_{>0}.
          \end{align*}

          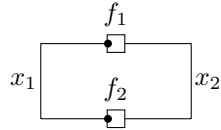
\begin{figure}[t]
            \centering
            \begin{minipage}[t]{0.4\linewidth}
              \centering
              \scalebox{\figscale}{%
                \begin{tikzpicture}[node distance=2cm, remember picture]
                  \tikzstyle{state}=[shape=rectangle,fill=white, draw, minimum width=0.2cm, minimum height = 0.2cm,outer sep=-0.3pt]
                  \tikzset{dot/.style={circle,fill=#1,inner sep=0,minimum size=3pt}}
                  \node[state] (f2) at (0,0) [label=above: $f_{2}$] {};
                  \node[state] (f1) at (0,1) [label=above: $f_{1}$] {};
                  \node [dot=black] at (f2.west) {};
                  \node [dot=black] at (f1.west) {};
                  \node (x1) at (-0.8,0.5) [label=left: $x_{1}$] {};
                  \node (x2) at (0.8,0.5) [label=right: $x_{2}$] {};
                  \begin{pgfonlayer}{background}
                    \draw[-,draw]
                    (1,0) -- (-1,0) (1,0) -- (1,1) (-1,0) -- (-1,1)
                    (-1,1) -- (1,1);
                  \end{pgfonlayer}
                \end{tikzpicture}}
            \end{minipage}%
            \caption{The S-NFG in Item~\ref{sec:SNFG:remk:1:item:2} in Remark~\ref{sec:SNFG:remk:1}.\label{sec:SNFG:fig:2}}
          \end{figure}

    \item\label{sec:SNFG:remk:1:item:2} There exist S-NFGs for which the
          primal formulation of the Bethe partition function is well defined,
          whereas the pseudo-dual formulation is not well defined.  Consider
          the S-NFG in Fig.~\ref{sec:SNFG:fig:2}, where the variables
          associated with edges $1$ and $ 2 $ are $ x_{1} \in \{0,1\} $ and
          $ x_{2} \in \{0,1\} $, respectively, and where the dots denote the
          row index of the matrices $ \matr{f}_{1} $ and $ \matr{f}_{2} $
          associated with the function nodes $ f_{1} $ and $ f_{2} $,
          respectively.  In particular, these matrices are defined to be
          \begin{align*}
            \matr{f}_{1} & \defeq
            \bigl( f_{1}(x_{1},x_{2}) \bigr)_{ \! x_{1} ,x_{2} \in \{0,1\} },\,\,
            \matr{f}_{2}\defeq
            \bigl( f_{2}(x_{1},x_{2}) \bigr)_{ \!x_{1} ,x_{2} \in \{0,1\} },
          \end{align*}
          with rows indexed by $ x_{1} $ and columns indexed by $ x_{2}
          $. Running the SPA on this S-NFG is equivalent to applying the power
          method to the matrix $ \matr{f}_{1} \cdot \matr{f}_{2}^{\tran}
          $. For the special case
          \begin{align*}
            \matr{f}_{1}
            = \begin{pmatrix}
                1 & 1 \\ 0 & 1
              \end{pmatrix},
            \qquad
            \matr{f}_{2}
            = \begin{pmatrix}
                1 & 0 \\ 0 & 1
              \end{pmatrix},
          \end{align*}
          the SPA fixed-point messages are given by
          \begin{align*}
            \vmu_{1,f_{1}}= \vmu_{2,f_{2}} = \begin{pmatrix} 0 \\ 1 \end{pmatrix}, \quad
            \vmu_{2,f_{1}}= \vmu_{1,f_{2}} = \begin{pmatrix} 1 \\ 0 \end{pmatrix}, \quad
          \end{align*}
          which results in
          \begin{align*}
            Z_{1}(\vmu)
            = Z_{2}(\vmu) = 0,
          \end{align*}
          where $ Z_{1} $ is associated with edge $ 1 $ and $ Z_{2} $ is
          associated with edge $ 2 $, as defined in~\eqref{sec:SNFG:eqn:6}. As
          for the pseudo-dual formulation, we cannot evaluate
          $ \ZBSPA(\graphN,\vmu) $ for this S-NFG $ \graphN $.  In the
            primal formulation, the Bethe free energy function satisfies
          \finaltext{$\min_{\vbeli\in\LMP(\graphN)}\FB(\vbeli)=0$}, which
          implies $ \ZB(\graphN) = 1 $, \ie, the primal formulation is well
          defined. (Note that the partition function is
          $ Z(\sfN) = \operatorname{tr}(\matr{f}_{1} \cdot
            \matr{f}_{2}^{\tran}) = 2 $.)\eremark

  \end{enumerate}
\end{remark}

\section{Double-Edge Normal Factor Graphs (DE-NFGs)}
\label{sec:DENFG}

The general definition of a DE-NFG~\cite{Cao2017} permits four edge
  types: full and half single edges and full and half double edges. We
  consider only full double edges. This restriction entails no loss of
  generality: every other edge type can be converted into a full double edge
  without changing the partition function or corresponding marginals. The
  conversion is detailed after Definition~\ref{sec:DENFG:def:4}, once the
  paired-variable and Choi-matrix notation is available.

We start by presenting an example DE-NFG.

\begin{example}
  \label{sec:DENFG:exp:1}

  \begin{figure}[t]
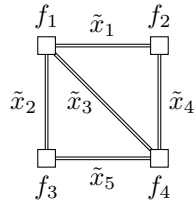

  \centering
  \begin{minipage}[t]{\linewidth}
    \centering
    \scalebox{0.8}{%
      \begin{tikzpicture}[node distance=2cm, on grid,auto]
        \input{figures/length.tex}
        \input{figures/head_files_figs.tex}
        \node[state] (f1) at (0,0) [label=above: $f_1$] {};
        \node[state] (f2) at (1.5,0) [label=above: $f_2$] {};
        \node[state] (f3) at (0,-1.5) [label=below: $f_3$] {};
        \node[state] (f4) at (1.5,-1.5) [label=below: $f_4$] {};
        \begin{pgfonlayer}{background}
          \draw[double]
          (f1) -- node[above] {$ \tilde{x}_{1} $} (f2)
          (f1) -- node[left]  {$ \tx_{2} $} (f3) 
          (f1) -- node[left, xshift=-0.0cm]  {$ \tx_{3} $} (f4)
          (f2) -- node[right]  {$ \tx_{4} $} (f4)
          (f3) -- node[below] {$ \tx_{5} $} (f4);
        \end{pgfonlayer}
      \end{tikzpicture}}
  \end{minipage}%
  \caption{The DE-NFG in Example~\ref{sec:DENFG:exp:1}.\label{sec:DENFG:fig:4}}
\end{figure}
  The DE-NFG in Fig.~\ref{sec:DENFG:fig:4} depicts the following factorization
  \begin{align*}
    \hspace{-0.25 cm} & g(x_{1},\ldots,x_{5},x_{1}',\ldots,x_{5}')
    \defeq f_{1}(x_{1},x_{2},x_{3},x_{1}',x_{2}',x_{3}')                                                                                                                  \\
                      & \cdot \! f_{2}(x_{1},x_{4},x_{1}',x_{4}') \! \cdot \! f_{3}(x_{2},x_{5},x_{2}',x_{5}') \! \cdot \! f_{4}(x_{3},x_{4},x_{5},x_{3}',x_{4}',x_{5}').
  \end{align*}
  Each double edge $e \in [5]$ is associated with the variable
  $\tx_e = (x_e, x'_e)$, where $x_e$ and $x'_e$ take values in the same
  alphabet $\set{X}_e$. Moreover, if $e$ is incident on a function node
  $f$, then $\tx_e$ is an argument of the local function $f$, \ie, both
  $x_e$ and $x'_e$ are arguments of $f$.
  \eexample
\end{example}

\begin{definition}\label{sec:DENFG:def:4}
  Let $(\setF,\setEfull)$ be a graph, and use the incidence-set notation
    and edge-orientation convention of Definition~\ref{def: def of snfg}. A
    DE-NFG $\graphN(\setF,\setEfull,\tset{X})$ on this graph is specified as
    follows.
  \begin{enumerate}

    \item\label{sec:DENFG:def:4:item:1} For each edge $e\in\setEfull$, let
            $\set{X}_e$ be a finite alphabet containing $0$. The edge carries the
            paired variable
          \begin{align*}
            \tx_e \defeq (x_e,x_e') \in
            \tset{X}_e \defeq \set{X}_e\times\set{X}_e.
          \end{align*}
          Define $\tzero\defeq(0,0)$, $\tset{X}\defeq\prod_{e\in\setEfull}\tset{X}_e$, and $\tvx\defeq(\tx_e)_{e\in\setEfull}\in\tset{X}$. The vector $\tvx$ is a configuration of $\graphN$.

    \item\label{sec:DENFG:def:4:item:3}
          For $\set{I}\subseteq\setEfull$, we define
          \begin{align*}
            \set{X}_{\set{I}}
             & \defeq \prod_{e\in\set{I}}\set{X}_e,       \qquad
            \tset{X}_{\set{I}}
            \defeq \prod_{e\in\set{I}}\tset{X}_e
            = \set{X}_{\set{I}}\times\set{X}_{\set{I}},          \\
            \vx_{\set{I}}
             & \defeq (x_e)_{e\in\set{I}}\in\set{X}_{\set{I}},   \\
            \vx_{\set{I}}'
             & \defeq (x_e')_{e\in\set{I}}\in\set{X}_{\set{I}},  \\
            \tvx_{\set{I}}
             & \defeq(\tx_e)_{e\in\set{I}}
            =(\vx_{\set{I}},\vx_{\set{I}}')
            \in\tset{X}_{\set{I}}.
          \end{align*}
          For a mapping $h:\tset{X}_{\set{I}}\to\sC$, we define the matrix
          \begin{align}
            \matr{C}_{h}
            \defeq
            \bigl(h(\vx_{\set{I}},\vx_{\set{I}}')\bigr)_{
                                                   \vx_{\set{I}},\vx_{\set{I}}'\in\set{X}_{\set{I}}}
            \in\sC^{|\set{X}_{\set{I}}|\times|\set{X}_{\set{I}}|}
            \label{sec:DENFG:eqn:6}
          \end{align}
          such that $\vx_{\set{I}}$ and $\vx_{\set{I}}'$ index the rows
            and columns of $\matr{C}_{h}$, respectively. Let
            $\setHerm{\set{X}_{\set{I}}}$, $\setPSD{\set{X}_{\set{I}}}$, and
            $\setPD{\set{X}_{\set{I}}}$ denote the corresponding sets of
            Hermitian, PSD, and positive-definite (PD) matrices, respectively.

    \item\label{sec:DENFG:def:4:item:4} Each function node
            $f\in\setF$ carries a local function $f:\tset{X}_{\setpf}\to\sC$,
            whose matrix representation $\matr{C}_f$ is called its Choi
            matrix~\cite{Wood2015}. A DE-NFG is called \textbf{strict-sense}
            if $\matr{C}_f\in\setPSD{\set{X}_{\setpf}}$ for every
            $f\in\setF$. It is called \textbf{weak-sense} if only
            $\matr{C}_f\in\setHerm{\set{X}_{\setpf}}$ is required.

          In either case, the matrix $\matr{C}_f$ is Hermitian and therefore has an
            orthonormal eigendecomposition. Let $\set{L}_f$ be an index set
            with $|\set{L}_f|=|\set{X}_{\setpf}|$, and let
            $\lambda_f:\set{L}_f\to\sR$ and
            $u_f:\set{X}_{\setpf}\times\set{L}_f\to\sC$ specify its
            eigenvalues and (normalized) eigenvectors. Then
          \begin{align}
            \!\!\!\!\! \! f\bigl(\tvx_{\setpf}\bigr)
             & =\sum_{\ell_f\in\set{L}_f}
            \lambda_f(\ell_f)
            \cdot u_f(\vx_{\setpf},\ell_f)
            \cdot  \overline{u_f(\vx_{\setpf}',\ell_f)},
            \label{sec:DENFG:eqn:5}                    \\
            \!\!\!\!\! \!  [\ell_f=\ell_f']
             & =\sum_{\vx_{\setpf}\in\set{X}_{\setpf}}
            \overline{u_f(\vx_{\setpf},\ell_f')}
            \cdot  u_f(\vx_{\setpf},\ell_f).
            \label{sec:DENFG:eqn:8}
          \end{align}
          If $\vect{u}_f(:,\ell_f)$ denotes the eigenvector with entries
            $u_f(\vx_{\setpf},\ell_f)$, then
          {
          \begin{align}
            \matr{C}_f
              &=\sum_{\ell_f\in\set{L}_f}\lambda_f(\ell_f)
                \cdot\vect{u}_f(:,\ell_f)
                \cdot\bigl(\vect{u}_f(:,\ell_f)\bigr)^{\!\Herm},
                \label{sec:DENFG:eqn:9}\\
            [\ell_f=\ell_f']
              &=\bigl(\vect{u}_f(:,\ell_f')\bigr)^{\!\Herm}
                \cdot\vect{u}_f(:,\ell_f).\nonumber
          \end{align}}

    \item The global function and the partition function of $\graphN$
            are defined to be $g:\tset{X}\to\sC$,
            $\tvx\mapsto\prod_{f\in\setF}f(\tvx_{\setpf})$ and
            $Z(\graphN)\defeq\sum_{\tvx\in\tset{X}}g(\tvx)$, respectively. For
            a strict-sense DE-NFG, it holds that $Z(\graphN)\in\sRp$, whereas
            for a weak-sense DE-NFG it only holds that $Z(\graphN)\in\sR$ (see
            Appendix~\ref{apx:property of ZN}).  \edefinition

  \end{enumerate}
\end{definition}

\emph{Reduction to full double edges.} For a general DE-NFG as
  in~\cite{Cao2017}, applying the following two conversions edge by edge
  yields an equivalent DE-NFG containing only full double edges.
  \begin{itemize}

    \item In order to convert a half single edge into a full single edge or to
          convert a half double edge into a full double edge, one attaches a
          function node whose local function is identically one.

    \item For each full single edge $e$ carrying $y_e$, one introduces a full
          double edge carrying $\widetilde{y}_e=(y_e,y_e')$ and replaces each
          incident local function $f(\,\cdot\,;y_e)$ by
          $\widehat f(\,\cdot\,;y_e,y_e')\defeq f(\,\cdot\,;y_e) \cdot [y_e \! = \!
              y_e']$.  This diagonal embedding gives $\widehat f$ a block-diagonal Choi
          matrix with PSD blocks. Summing over $(y_e,y_e')$ enforces $y_e=y_e'$ and
          recovers the original sum over $y_e$; hence, the partition function is
          unchanged, and the paired-variable marginal equals the original marginal
          on the diagonal and vanishes off the diagonal.

  \end{itemize}

When no ambiguity arises, we use the same shorthand notations as for S-NFGs (introduced after Definition~\ref{def: def of snfg}), with $\xe$, $\setxe$, $\vx_{\set{I}}$, $\setx_{\set{I}}$ replaced by $\txe$, $\tset{X}_{e}$, $\tvx_{\set{I}}$, $\tset{X}_{\set{I}}$, respectively; moreover, $\setpf \setminus e$ abbreviates $\setpf \setminus \{e\}$.

\begin{assumption}\label{sec:DENFG:asum:2}
  Unless stated otherwise, a DE-NFG denotes a strict-sense DE-NFG, \ie, we consider $ \lambda_{f}(\ell_{f}) \in \sR_{\geq 0} $ for all $ \ell_{f} \in \set{L}_{f} $ and $ f \in \setF $.
  Furthermore, we consider only strict-sense DE-NFGs
  $\graphN$ for which $Z(\graphN) \in \sRpp$.
  \eassumption
\end{assumption}

One of the motivations for considering DE-NFGs is that many NFGs
with complex-valued local functions in quantum information processing
can be transformed into DE-NFGs. (See~\cite{Cao2017} for further examples and details on this transformation.)

\begin{figure}[t]
  \centering
  \begin{minipage}[t]{\linewidth}
      \centering
      \adjustbox{max width=\linewidth}{%
      \scalebox{\figscale}{%
\begin{tikzpicture}[node distance=0.6cm, on grid,auto]
        \input{figures/head_files_figs.tex}
        \node[state] (rho) at (0,0) [] {$\rho$};
        \node[state_long] (U) [above right =0.7cm and 2cm of rho] {$U$};
        \node[state_long] (UH) [below right =0.7cm and 2cm of rho] {$U^{H}$};
        \node[state_long] (B) [right = 2.4cm of U] {$B$};
        \node[state_long] (BH) [right = 2.4cm of UH] {$B^{H}$};
        \node[state] (eqn) [below right=0.7cm and 2cm of B] {$I$};
        \draw
        (rho) |-node[above right, pos=.6] {$x_{1}$} (U)--node[above] {$x_{2}$} (B)
        -| node[above left, pos=.4] {$x_{3}$} (eqn)|-node[above left, pos=.6]
        {$x_{3}'$}(BH)
        --node[above] {$x_{2}'$} (UH)-| node[above right, pos=.4] {$x_{1}'$} (rho);
        \node [dot=black] at (rho.north) {};
        \node [dot=black] at (U.east) {};
        \node [dot=black] at (UH.west) {};
        \node [dot=black] at (B.east) {};
        \node [dot=black] at (BH.west) {};
        \node [dot=black] at (eqn.south) {};
      \end{tikzpicture}}}
  \end{minipage}

  \vspace{0.15cm}
  \begin{minipage}[t]{\linewidth}
      \centering
      \adjustbox{max width=\linewidth}{%
      \scalebox{\figscale}{%
\begin{tikzpicture}[node distance=0.8cm, on grid,auto]
        \input{figures/head_files_figs.tex}
        \input{figures/length.tex}
        \begin{pgfonlayer}{main}
          \node[state] (rho) at (0,0) [] {$\rho$};
          \node[state_long] (U) [right =2cm of rho] {$\tilde{U}$};
          \node[state_long] (B) [right = 2.4cm of U] {$\tilde{B}$};
          \node[state] (eqn) [right= 2cm of B] {$I$};
        \end{pgfonlayer}
        \begin{pgfonlayer}{background}
          \draw[double]
          (rho) -- node[above] {$\tx_{1}$} (U)
          -- node[above] {$\tx_{2}$} (B)
          -- node[above] {$\tx_{3}$} (eqn);
        \end{pgfonlayer}
        \node [dot=black] at (2.35,0) {};
        \node [dot=black] at (4.75,0) {};
        \node[] (f3) at (0,0) [] {};
      \end{tikzpicture}}}
  \end{minipage}
  \vspace{0.2cm}
  \caption[]{{Top}: the NFG in Example \ref{sec:DENFG:exp:2}. {Bottom}: the DE-NFG in Example \ref{sec:DENFG:exp:2}.}\label{sec:DENFG:fig:2}
\end{figure}

\begin{example}[\!\!\cite{Loeliger2012,Loeliger2017}]\label{sec:DENFG:exp:2}
  Consider the NFG at the top of Fig.~\ref{sec:DENFG:fig:2}. Each
    function node represents a matrix and carries a dot on one of its two
    incident edges. The variable on the edge with the dot indexes the rows of
    the matrix; the variable on the other edge indexes the columns. The graph
    describes a quantum system with two consecutive unitary evolutions.
  \begin{itemize}
    \item At the first stage, a quantum system is prepared in a mixed
          state represented by a complex-valued PSD density matrix $\matr{\rho}$
          with
          $ \matr{\rho} \defeq \bigl( \rho(x_{1},x_{1}') \bigr)_{\! x_{1},x_{1}' \in
              \set{X}_{1} } \in \setPSD{ \set{X}_{1} } $ with trace one, where the row
          is indexed by $ x_{1} $ and the column is indexed by $ x_{1}' $.

    \item The system then undergoes two consecutive unitary
            evolutions, which are represented by the pair of function nodes
          $U$ and $B$, respectively. Both the matrix
          $ \bigl( U(x_{2},x_{1}) \bigr)_{\! x_{2} \in \set{X}_{2},x_{1} \in
              \set{X}_{1}} $ with row indices $ x_{2} $ and column indices
          $ x_{1} $ and the matrix
          $ \bigl( B(x_{3},x_{2}) \bigr)_{\! x_{3} \in \set{X}_{3},x_{2} \in
              \set{X}_{2}} $ with row indices $ x_{3} $ and column indices
          $ x_{2} $ are unitary matrices.
  \end{itemize}
  We make certain modifications to transform this NFG into a DE-NFG.
  \begin{itemize}
    \item The variables $ x_{e} \in \set{X}_{e} $ and $ x_{e}' \in \set{X}_{e} $
          are merged into the variable
          $ \tx_{e} = ( x_{e}, x_{e}' ) \in \tset{X}_{e} $ for each $ e \in [3] $.

    \item New local functions are defined based on the rearranged variables.  As
          an example, the entries in the matrix $\tilde{\matr{U}}$ are defined to be
          \begin{align*}
            \tilde{U}(\tvx_{2},\tvx_{1}) \defeq U(x_{2},x_{1})
            \cdot \overline{U(x_{2}',x_{1}')}, \qquad \tvx_{2} \in \tset{X}_{2},\,
            \tvx_{1} \in \tset{X}_{1}.
          \end{align*}
          \finaltext{One can verify that the Choi matrix $\matr{C}_{\tilde U}$,
          with rows indexed by $(x_{2},x_{1})$ and columns indexed by
          $(x_{2}',x_{1}')$, is PSD.}
  \end{itemize}
  The resulting DE-NFG is shown in Fig.~\ref{sec:DENFG:fig:2} (bottom).
  \eexample
\end{example}

\begin{definition}\label{sec:DENFG:def:2}

  The SPA for a DE-NFG $\sfN$ uses the message and belief updates of
    Definition~\ref{sec:SNFG:def:4}, with $\xe$ and $\setxe$ replaced by
    $\txe$ and $\tset{X}_e$, respectively, and with the message, belief, and
    local-function codomains replaced by $\sC$. The initialization and zero-scaling-factor rule of the DE-NFG SPA must also preserve the Choi-matrix
    structure, as specified below. We call a function
    $h:\tset{X}_{\set{I}}\to\sC$ \emph{PD-normalized} if
  \begin{align*}
    \matr{C}_h\in\setPD{\set{X}_{\set{I}}}
      \quad\text{and}\quad
      \sum_{\tvx_{\set{I}}}h(\tvx_{\set{I}})=1.
  \end{align*}
  A function $h$ is \emph{PSD-normalized} if
    $\matr{C}_h\in\setPSD{\set{X}_{\set{I}}}$ and either its entries sum
    to one, or its entries sum to zero and $\Tr(\matr{C}_h)=1$.
  \begin{enumerate}
    \item (Initialization) For every $f \in \setF$ and $e \in \setpf$,
            define $\mu_{\ef}^{(0)}: \tset{X}_e \to \sC$ arbitrarily such that it is
            PD-normalized. (For example, choose $\mu_{\ef}^{(0)}(\txe)$ such that
            $\matr{C}_{\mu_{\ef}^{(0)}} \defeq (1 \! - \! \epsilon) \cdot
              \matr{I}_{|\set{X}_e| \times |\set{X}_e|} / |\set{X}_e| + \epsilon \cdot
              \bm{1}_{|\set{X}_e|}\cdot\bm{1}_{|\set{X}_e|}^{\Herm} / |\set{X}_e|^2$ with
            $0 \leq \epsilon < 1$.)

    \item\label{details of SPA in DE-NFG} (Iteration) For $t = 1, 2, 3, \ldots$,
          the following calculations are performed until a termination criterion is
          met.\footnote{The termination criterion is typically a combination of
            numerical convergence and an upper bound on the number of iterations.}
          \begin{enumerate}

            \item
                  \label{SPA: update message: DE-NFG}

                  Compute each unnormalized message
                    $\widehat\mu_{\ef}^{(t)}$ by the message-update rule of
                    Definition~\ref{sec:SNFG:def:4}, before division and with
                    $\xe$ replaced by $\txe$. For this message, define
                    $\widehat{\matr{C}}_\mu\defeq
                    \matr{C}_{\widehat\mu_{\ef}^{(t)}}$ and
                    $s\defeq\bm{1}^{\Herm}\cdot\widehat{\matr{C}}_\mu\cdot\bm{1}$.
                    Choose its normalization factor as follows:
                  {
                  \begin{align*}
                    \kappa_{\ef}^{(t)}\defeq
                    \begin{cases}
                      s & s>0,\\
                      \Tr(\widehat{\matr{C}}_\mu)
                        & s=0,\ \widehat{\matr{C}}_\mu\ne0,\\
                      0 & \widehat{\matr{C}}_\mu=0.
                    \end{cases}
                  \end{align*}}%
                  When $\kappa_{\ef}^{(t)}>0$, set
                    $\mu_{\ef}^{(t)}=\widehat\mu_{\ef}^{(t)}/\kappa_{\ef}^{(t)}$.
                  The associated vectors are defined to be
                  $ \vmu_{\efj}^{(t)} \defeq \bigl( \mu_{\efj}^{(t)}(\txe) \bigr)_{ \!  \txe } $
                  and $ \vmu_{\efi}^{(t)} \defeq \bigl( \mu_{\efi}^{(t)}(\txe) \bigr)_{ \!  \txe } $.

            \item
                  \label{SPA: update belief edges: DE-NFG}\label{SPA: update belief functions: DE-NFG}
For every edge $e=(f_i,f_j)\in\setEfull$ and every function node
                  $f\in\setF$, the belief functions $\beli_e^{(t)}$ and $\beli_f^{(t)}$,
                  their normalization factors, and their associated vectors are defined
                  by the belief-update steps of Definition~\ref{sec:SNFG:def:4}, with
                  $\xe$ and $\setxe$ replaced by $\txe$ and $\tset{X}_e$,
                  respectively. Each belief is defined only when its normalization
                  factor is positive.

            \item\label{SPA: restart rule: DE-NFG}
                  If an unnormalized message is identically zero,
                    restart from Step~1 with a fresh random PD-normalized
                    initialization.%
                  \footnote{A nonzero PSD message with zero total
                    entry sum can still be normalized by its positive trace.}

          \end{enumerate}
    \item For every $t \in \sZp$, the SPA messages form the vector
            $\vmu^{(t)} \defeq \bigl( \vmu_{\etof}^{(t)} \bigr)_{\! e \in
                \setpf,\, f \in \setF}$. The message updates in
          Step~\ref{SPA: update message: DE-NFG} can be written as
          $\vmu^{(t)}=\fSPAN\bigl(\vmu^{(t-1)}\bigr)$, where $\fSPAN$ denotes
          the normalized message-update map.

    \item An SPA message vector consisting of PSD-normalized
            messages
          $ \vmu \defeq ( \vmu_{\etof} )_{f \in \setF,\,e \in \setpf} $ with
          $ \vmu_{\etof} \defeq \bigl( \mu_{\etof}(\txe) \bigr)_{ \! \txe } $
          is called an SPA fixed-point message vector if every message-update normalization factor chosen in
            Step~\ref{SPA: update message: DE-NFG} is positive and
          \begin{align}
            \vmu = \fSPAN(\vmu).  \label{sec:DENFG:eqn:15}
          \end{align}
          The belief functions induced by $\vmu$ are defined as
          in Step~\ref{SPA: update belief edges: DE-NFG} whenever their
          corresponding normalization factors are positive, with $\mu_{\etof}^{(t)}$ replaced by $\mu_{\etof}$ for all $f\in\setF$ and $e\in\setpf$.\edefinition

  \end{enumerate}
\end{definition}

\begin{remark}
  \label{asmp: assume messages are PSD}

  We make some remarks on the SPA as defined in
  Definition~\ref{sec:DENFG:def:2}.
  \begin{itemize}
    \item We normalize each message by its total entry sum
          when that sum is positive. For a nonzero PSD message with zero total
          entry sum, we use unit-trace normalization, which is natural in
          quantum mechanics. Whenever $\ZBSPA(\graphN,\vmu)$ is defined, its
          value is unchanged by positive rescaling of any message, because
          each scaling factor cancels between the numerator and denominator.
          The LCT identities in Section~\ref{sec:LCT} below and the graph-cover
          analysis in Sections~\ref{sec:symmetric-subspace transform}--\ref{sec:CheckCon} remain valid
          under such rescalings. The maximization uses the normalized-message
          class specified in Definition~\ref{sec:DENFG:def:2}.

    \item Positive semidefiniteness of the SPA messages is not an
            additional assumption. Lemma~\ref{sec:DENFG:lem:1} proves that the
            PD-normalized initialization, the update rules, and the PD-normalized
            restart rule preserve positive semidefiniteness at every iteration.\eremark
  \end{itemize}

\end{remark}

Algebraically, one may collapse each double edge to a single edge
  carrying the paired variable $\txe$ over the enlarged alphabet
  $\tset{X}_e$. The resulting ordinary NFG has the same factorization and
  finite covers, although it is generally not an S-NFG because its local
  functions may be complex-valued (Definition~\ref{def: def of snfg}). The
  purpose of the double-edge formalism is to retain the matrix structure
  hidden by this collapse: for each local function or SPA message, the
  unprimed variables form the row index and the primed variables form the
  column index of its Choi matrix. Both the strict-sense condition on local
  functions and the PSD invariance of the SPA messages are statements about
  these Choi matrices. Accordingly, an $M$-cover uses one permutation for a
  double edge and thus lifts the two strands of each double edge
  together. This explicit Choi-matrix structure underlies the LCT of
  Section~\ref{sec:LCT} and the proof of Theorem~\ref{sec:CheckCon:thm:1}.

\begin{lemma}
  \label{sec:DENFG:lem:1}
  Consider running the SPA of Definition~\ref{sec:DENFG:def:2} on a
    DE-NFG $\graphN$. Its messages and beliefs satisfy the following
    properties.
  \begin{enumerate}

    \item\label{sec:DENFG:lem:1:item:1}
          For every $f\in\setF$ and every $e\in\setpf$, the following hold.
          The message Choi matrix $\matr{C}_{\mu_{\ef}^{(t)}}$ is PSD at each
          $t\in\sZp$. The belief Choi matrices $\matr{C}_{\beli_e^{(t)}}$ and
          $\matr{C}_{\beli_f^{(t)}}$ are PSD at each $t\in\sZpp$ whenever
          the corresponding beliefs are defined.

    \item\label{sec:DENFG:lem:1: PSD of SPA fixed point message} Every
            SPA fixed-point message vector $ \vmu $ satisfies
            $ \matr{C}_{\mu_{\ef}} \in \setPSD{\set{X}_{e}} $ for all
            $ f \in \setF $ and $ e \in \setpf $.

    \item\label{sec:DENFG:lem:1:item:2} If $Z_f(\vmu)>0$ for every
            $f\in\setF$ and $Z_e(\vmu)>0$ for every $e\in\setEfull$, the beliefs
          induced by an SPA fixed-point message vector $\vmu$ satisfy the local
          consistency constraints
          \begin{align*}
            \sum_{ \tvx_{\setpf \setminus e} }
            \beli_f(\tvx_{\setpf})
            = \beli_{e}(\tx_{e}), \qquad
            \txe \in \tset{X}_{e}, \, e \in \setpf,\, f \in \setF.
          \end{align*}
  \end{enumerate}
\end{lemma}

\begin{proof}
  See Appendix~\ref{apx: proof of lemma17}.
\end{proof}

The following definition generalizes Definition~\ref{sec:SNFG:def:2}
from S-NFGs to DE-NFGs.

\begin{definition}\label{sec:DENFG:def:3}
  For a DE-NFG $\graphN$, let $\vmu$ be a collection of PSD-normalized
    SPA messages\footnote{The SPA message vector $\vmu$ need not be an
    SPA fixed-point message vector.}\hspace{0.25em}for $\graphN$, as defined in
    Definition~\ref{sec:DENFG:def:2}. The $\vmu$-based Bethe partition
    function $\ZBSPA(\graphN,\vmu)$ and the SPA-based Bethe partition function
    $\ZBSPA^{*}(\graphN)$ are given by Definition~\ref{sec:SNFG:def:2} with
    the edge variable $\xe$ replaced by the paired variable $\txe$. The
    following lemma will show that $Z_e(\vmu)\in\sR_{\geq0}$; hence the
    denominator condition is $Z_e(\vmu)>0$, equivalently $Z_e(\vmu)\neq0$, for
    all $e\in\setEfull$.\footnote{Generalizing the primal Bethe free
      energy function of Definition~\ref{sec:SNFG:def:1} from S-NFGs to
      DE-NFGs is not obvious, because the complex logarithm is multi-valued;
      this generalization is left for future research. Instead,
      following~\cite[Definition~8]{Cao2017}, we define the Bethe partition
      function of an arbitrary DE-NFG based on the pseudo-dual Bethe free
      energy function.}  \edefinition
\end{definition}

\begin{lemma}\label{lem: nonnegative of ZBSPA for DE-NFG}
  For a DE-NFG $\graphN$ and any collection $\vmu$ of PSD-normalized SPA
    messages, it holds that $ Z_f(\vmu) \in \sR_{\geq 0} $ for all
    $f \in \setF$ and $Z_{e}(\vmu) \in \sRp$ for all $e \in \setEfull$.
  In particular, these conclusions hold for every SPA iterate
    $\vmu^{(t)}$, $t\in\sZp$, and every SPA fixed-point message vector.
\end{lemma}

\begin{proof}
  For each $f\in\setF$, Schur's product theorem gives
  \begin{align*}
    \matr{A}_f
      \defeq \matr{C}_f\circ
      \bigotimes_{e\in\setpf}\matr{C}_{\mu_{\ef}}
      \in\setPSD{\set{X}_{\setpf}}.
  \end{align*}
  Thus
    $Z_f(\vmu)=\bm{1}^{\Herm} \cdot \matr{A}_f  \cdot \bm{1}\in\sR_{\geq0}$.
    Similarly, the matrix $\matr{A}_e\defeq
      \matr{C}_{\mu_{\efi}}\circ\matr{C}_{\mu_{\efj}}$ is PSD, and
    $Z_e(\vmu)=\bm{1}^{\Herm} \cdot \matr{A}_e \cdot \bm{1}\in\sR_{\geq0}$. The final
    assertion follows from Lemma~\ref{sec:DENFG:lem:1}.\qedhere

\end{proof}

\begin{proposition}\label{sec:DENFG:prop:1}
  Let $\graphN$ be a tree-structured DE-NFG. If $\graphN$ admits an SPA
    fixed-point message vector $\vmu$ under
    Definition~\ref{sec:DENFG:def:2}, then $\vmu$ is unique and
    $
      \ZBSPA(\graphN,\vmu)=Z(\graphN).
    $%
\end{proposition}
\begin{proof}
  The standard SPA argument for trees proves both claims; see,
    \eg,~\cite[Theorem~4.2]{J.Wainwright2008}.
\end{proof}

\begin{assumption}\label{sec:DENFG:remk:1}
  We consider only DE-NFGs $ \graphN $ for which
  there exists an SPA fixed-point message vector $ \vmu $ satisfying
  \begin{align*}
    \ZBSPA(\graphN,\vmu) = \ZBSPA^{*}(\graphN) \in \sR_{>0},
  \end{align*}
  which implies $ Z_f(\vmu),Z_{e}(\vmu) \in \sR_{> 0} $ for all
  $ f \in \setF $ and $ e \in \setEfull $.  Equivalently, we consider only
  DE-NFGs whose pseudo-dual Bethe partition function is positive. \eremark
\end{assumption}

\begin{proposition}
  \label{prop: sufficient condition for pd messages}
  Consider an SPA fixed-point message vector $ \vmu $.  A sufficient condition
  for $ \ZBSPA(\graphN,\vmu) \in \sR_{>0} $ is that for each $ f \in \setF $,
  the Choi-matrix representation $ \matr{C}_{f} $ is a positive definite
  matrix.

\end{proposition}
\begin{proof}
  See Appendix~\ref{apx: sufficient condition for pd messages}.
\end{proof}

We conclude this section with the following remarks.
\begin{enumerate}
  \item In this section, DE-NFGs have been introduced along with their
        associated SPA. The SPA fixed points define the Bethe partition function
        $\ZBSPA^{*}(\graphN)$, which approximates the partition function of a
        DE-NFG.

  \item In the case of strict-sense DE-NFGs, both the partition function
        $Z(\graphN)$ and the Bethe partition function $\ZBSPA^{*}(\graphN)$ are
        nonnegative real numbers.

  \item An S-NFG can be viewed as a special case of a DE-NFG. This observation
        indicates that the concepts, techniques, and theoretical results developed
        for DE-NFGs can be applied to S-NFGs as well.

  \item Although formally Definition~\ref{sec:SNFG:def:1} could potentially
          be extended from S-NFGs to DE-NFGs, there is a challenge in dealing with
        the multi-valued nature of the complex logarithm function. We leave this
        issue for future research.

\end{enumerate}

\section{Finite Graph Covers}
\label{sec:GraCov}

This section reviews finite graph covers, a key combinatorial tool for
  analyzing NFGs~\cite{Koetter2007,Vontobel2013}.

\begin{definition}
  Following~\cite{Stark:Terras:96:1}, consider two graphs
    $\mathsf{G}=(\set{V},\set{E})$ and
    $\hat{\mathsf{G}}=(\hat{\set{V}},\hat{\set{E}})$, where parallel edges are
    allowed. Let $\pi_V: \hat{\set{V}} \to \set{V}$ and
    $\pi_E: \hat{\set{E}} \to \set{E}$ be surjective maps that preserve
    incidence. The graph $\hat{\mathsf{G}}$ is a cover of $\mathsf{G}$ if, for
    every $\hat{v} \in \hat{\set{V}}$, the map $\pi_E$ restricts to a
    bijection from the edges incident on $\hat{v}$ to the edges incident on
    $\pi_V(\hat{v})$.  If there is a positive integer $M$ such that
    $\bigl| \pi_V^{-1}(v) \bigr| = M$ for all $v \in \set{V}$, then
    $\hat{\mathsf{G}}$ is called an $M$-cover of $\mathsf{G}$. Here, the set $\pi_V^{-1}(v)$ is the preimage of $v \in \set{V}$, \ie, we have $\pi_V^{-1}(v) \defeq \{ \hat{v} \in \hat{\set{V}} \mid \pi_V(\hat{v}) = v
      \}$. \edefinition
\end{definition}

Without loss of generality, we can denote the vertex set of an $M$-cover by
$\hat{\set{V}} \defeq \set{V} \times [M]$ and define the mapping
$\pi_V: \hat{\set{V}} \to \set{V}$ by $\pi_V\bigl( (v,m) \bigr) = v$ for all
$m \in [M]$. Consider an edge in the original graph that connects nodes $v$ and $v'$ and a corresponding edge in the $M$-cover that connects nodes $(v,m)$ and $(v',m')$. Then
\begin{align*}
  \Bigl( \pi_V \bigl( (v,m) \bigr), \pi_V \bigl( (v',m') \bigr) \Bigr) = (v,v').
\end{align*}
For every edge $e =(v,v')$ in the original graph, there exists a permutation $\sigma_e \in \set{S}_{M}$ such that for each $ m \in [M] $,
the edge $ \bigl( (v,m), (v',m') \bigr) $ in the $M$-cover satisfies $ m' = \sigma_{e}(m) $. Therefore,
for every edge $e =(v,v')$ in the original graph, the corresponding edges in the $M$-cover
can be represented by the set
\begin{align*}
  \left\{
  \Bigl( (v,m),\bigl(v',\sigma_{e}(m)\bigr) \Bigr)
  \ \middle| \
  m \in [M]
  \right\}.
\end{align*}

To simplify the endpoint-pair notation in the following construction, we
first consider graphs with at most one edge between any two vertices.
Parallel edges are distinguished by retaining the label $e$ of the
corresponding base edge.

\begin{definition}
  \label{sec:GraCov:def:1}
  Let $\graphN = (\set{F}, \set{E}, \set{X})$ be an S-NFG or a DE-NFG, and let $M \in \sZpp$. An $M$-cover of $\graphN$, denoted by $\hgraphN$, is an NFG specified by a vector of permutations $\vsigma \defeq (\sigma_e)_{e \in \set{E}} \in \set{S}_M^{|\set{E}|}$. The components of $\hgraphN = (\hat{\set{F}}, \hat{\set{E}}, \hat{\set{X}})$ are defined as follows:
  \begin{enumerate}
    \item Function Nodes: The set of function nodes is $\hat{\set{F}} \defeq \set{F} \times [M]$. Each node $(f,m) \in \hat{\set{F}}$ inherits the local function of its corresponding node $f \in \set{F}$.
    \item Edges: For each edge $e = (f_i, f_j) \in \set{E}$, the cover contains $M$ corresponding edges given by the set
          \begin{align*}
            \hat{\set{E}}_e \defeq \left\{ \Bigl( (f_i, m), \bigl(f_j, \sigma_e(m) \bigr) \Bigr)
            \ \middle| \ m \in [M] \right\}.
          \end{align*}
          The full-edge set of the cover is $\hat{\set{E}} \defeq \bigcup_{e \in \set{E}} \hat{\set{E}}_e$.
    \item Alphabets: Each edge $\hat{e} \in \hat{\set{E}}_e$ inherits the alphabet of its corresponding edge $e \in \set{E}$. For an S-NFG, the alphabet is $\set{X}_{\hat{e}} \defeq \set{X}_e$; for a DE-NFG, it is $\tset{X}_{\hat{e}} \defeq \tset{X}_e$.
  \end{enumerate}
  The set of all possible $M$-covers of $\graphN$ is denoted by $\setcovN{M}$.
  \edefinition
\end{definition}

For a given NFG $\mathsf{N}$, we consider all graphs in $\hat{\set{N}}_{M}$ to
be distinct graphs, although some graphs in the set $\hat{\set{N}}_{M}$ might
be isomorphic to each other. (See also the comments on labeled graph covers
after~\cite[Definition~19]{Vontobel2013}.) More importantly, for the finite
graph covers of a DE-NFG, the double edges are permuted together.\footnote{A
  graph-cover definition based on splitting the two edges comprising a double
  edge could be considered. However, the resulting graph covers would not be
  immediately relevant for analyzing the SPA for DE-NFGs in the way that the
  SPA is formulated.}

\begin{figure}[t]
  \begin{center}
    \begin{minipage}[t]{0.14\textwidth}
  \centering
  \scalebox{\figscale}{%
    \begin{tikzpicture}[node distance=1cm, on grid,auto]
      \input{figures/head_files_figs.tex}
      \node[state] (f11) [] {};
      \node[state,right of=f11] (f21) [] {};
      \node[state,below of=f11] (f31) [] {};
      \node[state,right of=f31] (f41) [] {};
      \path[-,draw]
      (f11) edge node[above]  {} (f21)
      (f11) edge node[left]  {} (f31) 
      (f11) edge node[left] {} (f41)
      (f21) edge node[right] {} (f41)
      (f31) edge node[below] {} (f41);
    \end{tikzpicture}}
\end{minipage}
\begin{minipage}[t]{0.09\textwidth}
  \scalebox{\figscale}{%
    \begin{tikzpicture}[node distance=1cm, on grid,auto]
      \input{figures/head_files_figs.tex}

      \begin{pgfonlayer}{main}
        \node[state] (f11) [] {};
        \node[state,right of=f11] (f21) [] {};
        \node[state,below of=f11] (f31) [] {};
        \node[state,right of=f31] (f41) [] {};
        \path[-,draw]
        (f11) edge node[above]  {} (f21)
        (f11) edge node[left]  {} (f31) 
        (f11) edge node[left] {} (f41)
        (f21) edge node[right] {} (f41)
        (f31) edge node[below] {} (f41);
      \end{pgfonlayer}

      \begin{pgfonlayer}{behind}
        \node[state] (f12) [above right=0.2cm and 0.2cm of f11] {};
        \node[state,right of=f12] (f22) [] {};
        \node[state,below of=f12] (f32) [] {};
        \node[state,right of=f32] (f42) [] {};
        \path[-,draw]
        (f12) edge node[above]  {} (f22)
        (f12) edge node[left]  {} (f32) 
        (f12) edge node[left] {} (f42)
        (f22) edge node[right] {} (f42)
        (f32) edge node[below] {} (f42);
      \end{pgfonlayer}

    \end{tikzpicture}}
\end{minipage}
\begin{minipage}[t]{0.09\textwidth}
  \scalebox{\figscale}{%
    \begin{tikzpicture}[node distance=1cm, on grid,auto]
      \input{figures/head_files_figs.tex}
      \begin{pgfonlayer}{main}
        \node[state] (f11) [] {};
        \node[state,right of=f11] (f21) [] {};
        \node[state,below of=f11] (f31) [] {};
        \node[state,right of=f31] (f41) [] {};
        \path[-,draw]
        (f11) edge node[above]  {} (f21)
        (f11) edge node[left]  {} (f31) 
        (f11) edge node[left] {} (f42)
        (f21) edge node[right] {} (f41)
        (f31) edge node[below] {} (f41);
      \end{pgfonlayer}
      \begin{pgfonlayer}{behind}
        \node[state] (f12) [above right=0.2cm and 0.2cm of f11] {};
        \node[state,right of=f12] (f22) [] {};
        \node[state,below of=f12] (f32) [] {};
        \node[state,right of=f32] (f42) [] {};
        \path[-,draw]
        (f12) edge node[above]  {} (f22)
        (f12) edge node[left]  {} (f32) 
        (f12) edge node[left] {} (f41)
        (f22) edge node[right] {} (f42)
        (f32) edge node[below] {} (f42);
      \end{pgfonlayer}
    \end{tikzpicture}}
\end{minipage}
\begin{minipage}[t]{0.09\textwidth}
  \scalebox{\figscale}{%
    \begin{tikzpicture}[node distance=1cm, on grid,auto]
      \input{figures/head_files_figs.tex}
      \begin{pgfonlayer}{main}
        \node[state] (f11) [] {};
        \node[state,right of=f11] (f21) [] {};
        \node[state,below of=f11] (f31) [] {};
        \node[state,right of=f31] (f41) [] {};
        \path[-,draw]
        (f11) edge node[above]  {} (f22)
        (f11) edge node[left]  {} (f31) 
        (f11) edge node[left] {} (f41)
        (f21) edge node[right] {} (f41);
      \end{pgfonlayer}
      \begin{pgfonlayer}{behind}
        \node[state] (f12) [above right=0.2cm and 0.2cm of f11] {};
        \node[state,right of=f12] (f22) [] {};
        \node[state,below of=f12] (f32) [] {};
        \node[state,right of=f32] (f42) [] {};
        \path[-,draw]
        (f31) edge node[below] {} (f42)
        (f12) edge node[above]  {} (f21)
        (f12) edge node[left]  {} (f32) 
        (f12) edge node[left] {} (f42)
        (f22) edge node[right] {} (f42)
        (f32) edge node[below] {} (f41);
      \end{pgfonlayer}
    \end{tikzpicture}}
\end{minipage}
    \medskip
    \caption{Left: the S-NFG $\graphN$ in Example~\ref{example:finite_graph_cover}.
      Right: samples of 2-covers
      of $\graphN$.\label{sec:GraCov:fig:9}}
    \begin{minipage}[t]{0.14\textwidth}
  \centering
  \scalebox{\figscale}{%
    \begin{tikzpicture}[node distance=1cm, on grid,auto]
      \input{figures/head_files_figs.tex}
      \begin{pgfonlayer}{glass}
        \node[state] (f11) [] {};
        \node[state,right of=f11] (f21) [] {};
        \node[state,below of=f11] (f31) [] {};
        \node[state,right of=f31] (f41) [] {};
      \end{pgfonlayer}
      \begin{pgfonlayer}{main}
        \draw[double]
        (f11) -- (f21)
        (f11) -- (f31) 
        (f11) -- (f41)
        (f21) -- (f41)
        (f31) -- (f41);
      \end{pgfonlayer}
    \end{tikzpicture}}
\end{minipage}
\begin{minipage}[t]{0.08\textwidth}
  \scalebox{\figscale}{%
    \begin{tikzpicture}[node distance=1cm, on grid,auto]
      \input{figures/head_files_figs.tex}
      \begin{pgfonlayer}{glass}
        \node[state] (f11) [] {};
        \node[state,right of=f11] (f21) [] {};
        \node[state,below of=f11] (f31) [] {};
        \node[state,right of=f31] (f41) [] {};
      \end{pgfonlayer}

      \begin{pgfonlayer}{above}
        \node[state] (f12) [above right=0.2cm and 0.2cm of f11] {};
        \node[state,right of=f12] (f22) [] {};
        \node[state,below of=f12] (f32) [] {};
        \node[state,right of=f32] (f42) [] {};
      \end{pgfonlayer}
      \begin{pgfonlayer}{background}
        \draw[double]
        (f12) -- (f22)
        (f12) -- (f32) 
        (f12) -- (f42)
        (f22) -- (f42)
        (f32) -- (f42);
      \end{pgfonlayer}
      \begin{pgfonlayer}{behind}
        \draw[double]
        (f11) -- (f21)
        (f11) -- (f31) 
        (f11) -- (f41)
        (f21) -- (f41)
        (f31) -- (f41);
      \end{pgfonlayer}
    \end{tikzpicture}}
\end{minipage}
\begin{minipage}[t]{0.08\textwidth}
  \scalebox{\figscale}{%
    \begin{tikzpicture}[node distance=1cm, on grid,auto]
      \input{figures/head_files_figs.tex}
      \begin{pgfonlayer}{glass}
        \node[state] (f11) [] {};
        \node[state,right of=f11] (f21) [] {};
        \node[state,below of=f11] (f31) [] {};
        \node[state,right of=f31] (f41) [] {};
      \end{pgfonlayer}

      \begin{pgfonlayer}{above}
        \node[state] (f12) [above right=0.2cm and 0.2cm of f11] {};
        \node[state,right of=f12] (f22) [] {};
        \node[state,below of=f12] (f32) [] {};
        \node[state,right of=f32] (f42) [] {};
      \end{pgfonlayer}
      \begin{pgfonlayer}{background}
        \draw[double]
        (f12) -- (f22)
        (f12) -- (f32) 
        (f22) -- (f42)
        (f32) -- (f42);
      \end{pgfonlayer}
      \begin{pgfonlayer}{behind}
        \draw[double]
        (f12) -- (f41);
      \end{pgfonlayer}
      \begin{pgfonlayer}{main}
        \draw[double]
        (f11) -- (f21)
        (f11) -- (f31) 
        (f11) -- (f42)
        (f31) -- (f41);
      \end{pgfonlayer}
      \begin{pgfonlayer}{above}
        \draw[double]
        (f21) -- (f41);
      \end{pgfonlayer}
    \end{tikzpicture}}
\end{minipage}
\begin{minipage}[t]{0.08\textwidth}
  \scalebox{\figscale}{%
    \begin{tikzpicture}[node distance=1cm, on grid,auto]
      \input{figures/head_files_figs.tex}
      \begin{pgfonlayer}{glass}
        \node[state] (f11) [] {};
        \node[state,right of=f11] (f21) [] {};
        \node[state,below of=f11] (f31) [] {};
        \node[state,right of=f31] (f41) [] {};
      \end{pgfonlayer}

      \begin{pgfonlayer}{above}
        \node[state] (f12) [above right=0.2cm and 0.2cm of f11] {};
        \node[state,right of=f12] (f22) [] {};
        \node[state,below of=f12] (f32) [] {};
        \node[state,right of=f32] (f42) [] {};
      \end{pgfonlayer}

      \begin{pgfonlayer}{behind}
        \draw[double]
        (f11) -- (f22)
        (f11) -- (f31) 

        (f31) -- (f42);
      \end{pgfonlayer}

      \begin{pgfonlayer}{background}
        \node[state] (f12) [above right=0.2cm and 0.2cm of f11] {};
        \node[state,right of=f12] (f22) [] {};
        \node[state,below of=f12] (f32) [] {};
        \node[state,right of=f32] (f42) [] {};
        \draw[double]
        (f12) -- (f21)
        (f12) -- (f42)
        (f22) -- (f42)
        (f32) -- (f41)
        (f12) -- (f32);
      \end{pgfonlayer}

      \begin{pgfonlayer}{main}
        \draw[double]
        (f11) -- (f41)
        (f21) -- (f41);
      \end{pgfonlayer}

    \end{tikzpicture}}
\end{minipage}
    \medskip
    \caption{
      Left: the DE-NFG $\graphN$ in Example~\ref{example:finite_graph_cover}. Right: samples of 2-covers
      of $\graphN$.
    }\label{fig:Exam_Graph_cover_DE_NFG}

  \end{center}
\end{figure}

\begin{example}
  \label{example:finite_graph_cover}

  Fig.~\ref{sec:GraCov:fig:9} (left) depicts an S-NFG $\graphN$ consisting of $4$ vertices and $5$ edges. Fig.~\ref{sec:GraCov:fig:9} (right)
  presents possible $2$-covers of $\graphN$.
  For every $ M \in \sZpp$, each $M$-cover has $M\cdot 4$ function nodes and
  $M\cdot 5$ edges.

  For comparison,
  Fig.~\ref{fig:Exam_Graph_cover_DE_NFG} (left) shows a DE-NFG $\graphN$
  consisting of $4$ vertices and $5$ double edges.
  Fig.~\ref{fig:Exam_Graph_cover_DE_NFG} (right) presents possible $2$-covers of $\graphN$.
  For every $ M \in \sZpp$, each $M$-cover
  has $M\cdot 4$ function nodes and
  $M\cdot 5$ double edges.
  \eexample
\end{example}

\begin{definition}\label{sec:GraCov:def:2}

  Let $\graphN$ be an S-NFG or a DE-NFG. For every $M \in \sZpp$, we define
  the degree-$M$ Bethe partition function of $\graphN$ to be
  \begin{align*}
    \ZBM(\graphN)
    \defeq \sqrt[M]{
             \Bigl\langle
             Z\bigl( \hgraphN \bigr)
             \Bigr\rangle_{ \! \! \hgraphN \in \hat{\set{N}}_{M}}
           } \ ,
  \end{align*}
  where
  $\bigl\langle Z\bigl( \hgraphN \bigr) \bigr\rangle_{\! \hgraphN \in \hat{\set{N}}_{M}}$
  represents the arithmetic mean of
  $Z\bigl( \hgraphN \bigr)$ over all $\hgraphN \in \hat{\set{N}}_{M}$, \ie,
  \begin{align*}
    \Bigl\langle
    Z\bigl( \hgraphN \bigr)
    \Bigr\rangle_{\! \hgraphN \in \hat{\set{N}}_{M}}
    \defeq \frac{1}{|\hat{\set{N}}_{M}|}
    \cdot \sum_{ \hgraphN \in \hat{\set{N}}_{M} } Z\bigl( \hgraphN \bigr),
  \end{align*}
  where
  $\hat{\set{N}}_{M}$ is the set of all $M$-covers of $\graphN$ (see
  Definition~\ref{sec:GraCov:def:1}).
  \edefinition
\end{definition}

\begin{theorem}\!\!\cite{Vontobel2013}
  \label{sec:GraCov:thm:1}
  For any S-NFG $\graphN$, it holds that
  \begin{align*}
    \limsup_{M\to \infty}
    \ZBM(\graphN)
     & = \ZB(\graphN). \tag*{\ensuremath{\square}}
    \nonumber
  \end{align*}
\end{theorem}

Given that DE-NFGs are in many ways a natural generalization of
S-NFGs, one wonders if there is a characterization of the Bethe partition
function of a DE-NFG in terms of its graph covers,
analogous to Theorem~\ref{sec:GraCov:thm:1}.

\begin{conjecture}
  \label{sec:GraCov:conj:1}

  For any DE-NFG $\graphN$, it holds that
  \begin{align}
    \limsup_{M\to \infty}
    \ZBM(\graphN)
     & = \ZBSPA^{*}(\graphN). \qquad \label{eqn: graph cover theorem}
  \end{align}
  \hfill$\square$
\end{conjecture}
The left-hand side of~\eqref{eqn: graph cover theorem}
features the limit superior $ \limsup_{M\to \infty} \ZBM(\graphN) $. For each DE-NFG $ \sfN $, we have $ Z(\graphN) \in \sRpp $, as assumed in Assumption~\ref{sec:DENFG:asum:2}. Because each $M$-cover is again a DE-NFG, we further have $ \ZBM(\graphN) \in \sRp $ and thus $ \limsup_{M\to \infty} \ZBM(\graphN) \in \sRp $.
The right-hand side of~\eqref{eqn: graph cover theorem}
features $\ZBSPA^{*}(\graphN)$, not $\ZB(\graphN)$, because only the
former is defined for DE-NFGs.

The proof of Theorem~\ref{sec:GraCov:thm:1} relies on the method of types
  and the primal Bethe formulation (Definition~\ref{sec:SNFG:def:1}), neither
  of which extends to a DE-NFG, whose partition function adds up complex
  values. (See also the discussion in
  Example~\ref{example:simple:complex:sum:1}.) Proving
  Conjecture~\ref{sec:GraCov:conj:1} thus requires new techniques.
Section~\ref{sec:LCT} introduces a variant of the LCT, a key technical
innovation of the present paper, which reparameterizes the partition function
of a DE-NFG into a form that is suitable for analysis. With the help of this
LCT, we prove in Theorem~\ref{sec:CheckCon:thm:1} that the limit~\eqref{eqn:
  graph cover theorem} holds for DE-NFGs satisfying an easily checkable
condition. We conjecture that the limit~\eqref{eqn: graph cover theorem} holds
for more general DE-NFGs. In the rest of this section, we provide promising
numerical results supporting Conjecture~\ref{sec:GraCov:conj:1}.

\begin{figure}
  \centering
  \subfloat[\label{sec:GraCov:fig:1}]{
    \begin{minipage}[t]{0.225\textwidth}
      \centering
      \begin{tikzpicture}[node distance=1.5cm, on grid,auto]
        \input{figures/head_files_figs.tex}
        \begin{pgfonlayer}{main}
          \node[state] (f1) at (0,0) [label=above: $f_1$] {};
          \node[state,right of=f1] (f2) [label=above: $f_2$] {};
          \node[state,below of=f1] (f3) [label=below: $f_3$] {};
          \node[state,right of=f3] (f4) [label=below: $f_4$] {};
        \end{pgfonlayer}
        \begin{pgfonlayer}{background}
          \draw[double]
          (f1) -- node[above] {$\tilde{x}_{1}$} (f2)
          (f1) -- node[left]  {$\tilde{x}_{2}$} (f3) 
          (f1) -- node[above right=.03cm of f1,xshift=.4cm] { $\tilde{x}_{3}$} (f4)
          (f2) -- node[right] {$\tilde{x}_{4}$} (f4)
          (f3) -- node[below] {$\tilde{x}_{5}$} (f4);
        \end{pgfonlayer}
      \end{tikzpicture}
    \end{minipage}
  }
  \subfloat[\label{sec:GraCov:fig:5}]{
    \begin{minipage}[t]{0.225\textwidth}
      \centering
      \begin{tikzpicture}[node distance=1.5cm, on grid,auto]
        \input{figures/head_files_figs.tex}
        \begin{pgfonlayer}{main}
          \node[state] (f1) at (1,1) [label=above: $f_1$] {};
          \node[state,right of=f1] (f2) [label=above:  $f_2$] {};
          \node[state,below of=f1] (f3) [label=below:  $f_3$] {};
          \node[state,right of=f3] (f4) [label=below:  $f_4$] {};
        \end{pgfonlayer}
        \begin{pgfonlayer}{background}
          \draw[double]
          (f1) -- node[above] { $\tilde{x}_{1}$} (f2)
          (f1) -- node[left]  { $\tilde{x}_{2}$} (f3) 
          (f1) -- node[above left= 0.3cm, xshift=.4cm] {$\tilde{x}_{3}$} (f4)
          (f2) -- node[right] {$\tilde{x}_{4}$} (f4)
          (f3) -- node[below right= 0.4cm, xshift=-.8cm] {$\tilde{x}_{5}$} (f2)
          (f3) -- node[below] {$\tilde{x}_{6}$} (f4);
        \end{pgfonlayer}
        \begin{pgfonlayer}{behind}
          \draw[double]
          (f3) -- (f2);
        \end{pgfonlayer}
      \end{tikzpicture}
    \end{minipage}\label{sec:GraCov:fig:2}
  }\\

  \subfloat[\label{sec:GraCov:fig:3}]{
    \begin{minipage}[t]{0.225\textwidth}
      \centering
      \scalebox{0.7}{\includegraphics[scale=0.4]{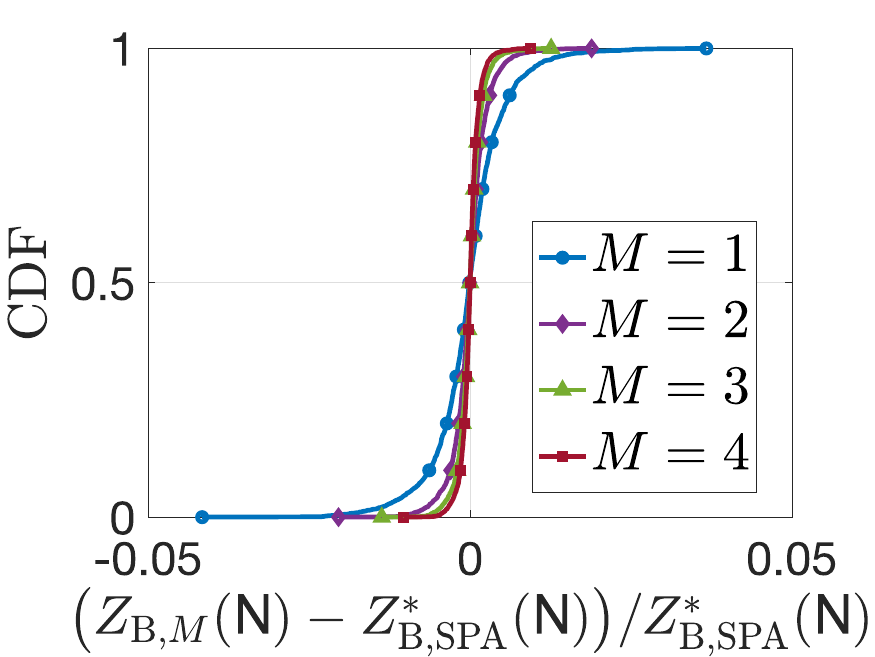}}
      \vspace{0.1 cm}
    \end{minipage}%
  }%
  \subfloat[\label{sec:GraCov:fig:7}]{
    \begin{minipage}[t]{0.225\textwidth}
      \centering
      \scalebox{0.7}{\includegraphics[scale=0.4]{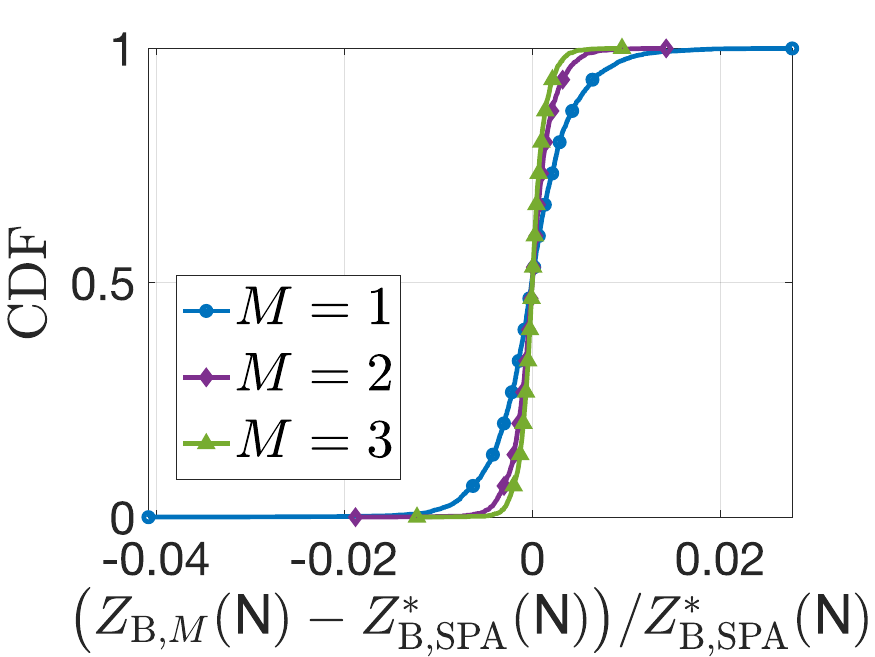}}
      \vspace{0.1 cm}
    \end{minipage}%
  }%

  \subfloat[\label{sec:GraCov:fig:4}]{
    \begin{minipage}[t]{0.225\textwidth}
      \centering
      \scalebox{0.7}{\includegraphics[scale=0.4]{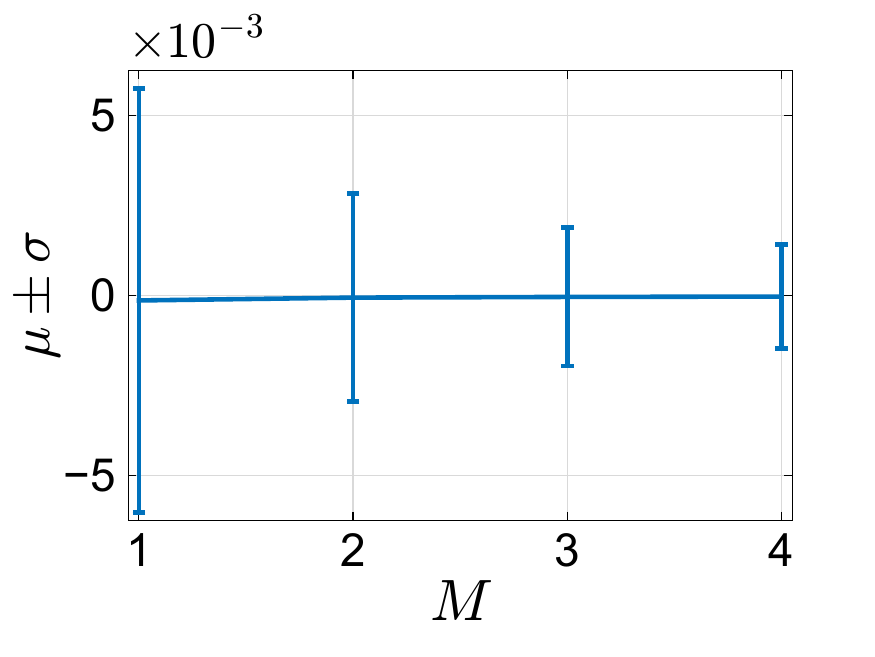}}
      \vspace{0.1 cm}
    \end{minipage}%
  }
  \subfloat[\label{sec:GraCov:fig:8}]{
    \begin{minipage}[t]{0.225\textwidth}
      \centering
      \scalebox{0.7}{\includegraphics[scale=0.4]{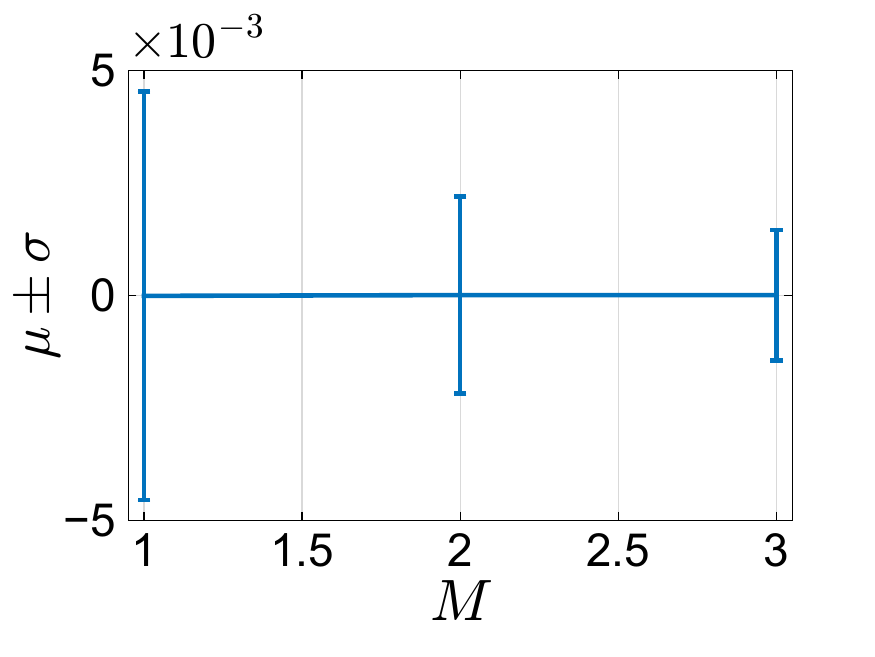}}
      \vspace{0.1 cm}
    \end{minipage}%
  }

  \caption{The DE-NFGs and the numerical results in Examples~\ref{sec:GraCov:exp:1} and~\ref{sec:GraCov:exp:2}. }
\end{figure}

\begin{example}
  \label{sec:GraCov:exp:1}

  Consider the DE-NFG $\graphN$ in Fig.~\ref{sec:GraCov:fig:1}, where
  $\set{X}_e = \{ 0, 1 \}$ and $ \tset{X}_e  = \{ (0,0),(0,1),(1,0), (1,1) \} $
  for all $e \in \setEfull$. Here, the graph $\graphN$ is a
  randomly generated DE-NFG, where the local functions are independently
  generated for each instance of $\graphN$. Specifically, for each instance, each local function $f$ is generated independently as follows: we draw $|\set{X}_{\setpf}|$ eigenvalues independently and uniformly from $[0,10]$ and collect them in a diagonal matrix $\matr{\Lambda}_f \in \sR^{|\set{X}_{\setpf}| \times |\set{X}_{\setpf}|}$; we draw a random unitary matrix $\matr{U}_f$ as the unitary factor in the QR decomposition of a $|\set{X}_{\setpf}| \times |\set{X}_{\setpf}|$ complex matrix whose real and imaginary parts have independent entries drawn uniformly from $[0,1]$, with the decomposition normalized so that the triangular factor has a positive diagonal; and we set the Choi-matrix representation of $f$ to $\matr{C}_f \defeq \matr{U}_f \cdot \matr{\Lambda}_f \cdot \matr{U}_f^{\Herm}$ (cf.~\eqref{sec:DENFG:eqn:6}). The matrix $\matr{C}_f$ is then Hermitian and PSD, so each instance is a strict-sense DE-NFG.

  For each instance of $\graphN$, we run the SPA until it reaches
  a fixed point and compute $\ZBSPA^{*}(\graphN)$. (We assume that
  $\ZBSPA^{*}(\graphN) = \ZBSPA(\graphN,\vmu)$, where $\vmu$ is the
  obtained SPA fixed-point message vector.) For
  each instance of $\graphN$, we compute $\ZBM(\graphN)$ for $ M \in [4]$. The following simulation results are for $3000$ randomly
  generated instances of $\graphN$.

  Specifically, Fig.~\ref{sec:GraCov:fig:3} shows the cumulative distribution
  function (CDF) of
  $\bigl( \ZBM(\graphN) \! - \! \ZBSPA^{*}(\graphN) \bigr) /
    \ZBSPA^{*}(\graphN)$ for $ M \in \{1,2,3,4\} $. Here, we have $ Z_{\mathrm{B},1}(\graphN) = Z(\graphN) $ because the only 1-cover of $\graphN$ is $\graphN$ itself.
  Fig.~\ref{sec:GraCov:fig:4} presents the mean $\mu$
  and standard deviation $\sigma$ of the empirical distribution of the quantity
  $\bigl( \ZBM(\graphN) \! - \! \ZBSPA^{*}(\graphN) \bigr) /
    \ZBSPA^{*}(\graphN)$.

  The simulation results indicate that, as $M$ increases, the degree-$M$ Bethe
  partition function $\ZBM(\graphN)$ approaches the SPA-based Bethe partition function $\ZBSPA^{*}(\graphN)$, in agreement with Conjecture~\ref{sec:GraCov:conj:1}.
  \eexample
\end{example}

\begin{example}
  \label{sec:GraCov:exp:2}

  The setup in this example is essentially the same as the setup in Example~\ref{sec:GraCov:exp:1}. However,
  instead of the DE-NFG in Fig.~\ref{sec:GraCov:fig:1}, we consider the DE-NFG in
  Fig.~\ref{sec:GraCov:fig:5}, and instead of $M\in [4]$, we consider $M \in [3]$. The simulation results for $3000$ randomly
  generated instances of $\graphN$ are presented in
  Figs.~\ref{sec:GraCov:fig:7} and~\ref{sec:GraCov:fig:8}.
  \eexample
\end{example}

\section{Loop Calculus Transform (LCT)}\label{sec:LCT}

This section develops the LCT for both S-NFGs and DE-NFGs. The goal of this transform is to create an equivalent NFG wherein the \finaltext{$\vmu$-based Bethe partition function at \finaltext{a} chosen SPA fixed point \finaltext{$\vmu$}} is isolated as the contribution from a single, easily identifiable configuration. This isolation simplifies the analysis of the degree-$M$ Bethe partition function.

The LCT is a specific holographic transform~\cite{AlBashabsheh2011} built on the loop calculus framework introduced by Chertkov and Chernyak~\cite{Chertkov2006, Chernyak2007}. Applying the LCT to an NFG separates its partition function into the \finaltext{$\vmu$-based Bethe partition function} and a finite series of correction terms, which correspond to weighted sums over generalized loops in the transformed graph. Prior LCT formulations have key limitations (see Table~\ref{tab:lct_comparison})\finaltext{. The} original framework is not fully explicit for non-binary alphabets\finaltext{. Mori's explicit nonbinary construction~\cite{Mori2015} uses positive beliefs; Section~IV-C \finaltext{of~\cite{Mori2015}} also discusses an extension to complex BP solutions. Cao's} extension to DE-NFGs and complex-valued messages~\cite{Cao2021} specifies only high-level constraints, not an explicit construction.

Our work overcomes these limitations by providing a unified and explicit LCT formulation for both S-NFGs and DE-NFGs. \finaltext{Our explicit LCT for nonbinary alphabets handles complex-valued SPA fixed-point messages with zero-valued components. Such \finaltext{zero-valued} components are common in DE-NFGs but were not handled by earlier explicit nonbinary constructions.} Our formulation is therefore a direct generalization of the work in~\cite{Chertkov2006, Chernyak2007,Mori2015,Cao2021}.
\begin{table*}[t]
  \centering
  \caption{Comparison of LCT formulations. \protect\finaltext{The Mori column describes the formulation for positive beliefs in~\cite[Theorem~12]{Mori2015}.}\label{tab:lct_comparison}}
  \resizebox{\textwidth}{!}{%
    \begin{tabular}{|l|c|c|c|c|}
      \hline
      {Feature} & Chertkov and Chernyak~\cite{Chertkov2006, Chernyak2007} & {Mori}~\cite{Mori2015}
                & {Cao}~\cite{Cao2021}
                & {Huang \& Vontobel (This Work)}                                                                                   \\ \hline
                                                                                                                                       {The SPA message values} & Nonnegative & Strictly Positive & Complex & Complex \\ \hline
                                                                                                                                                                                                                         {Assumes $\mu(x_e)>0$ for all $ x_e \in \setxe$ and $e \in \setEfull$?} & No & Yes & No & No \\ \hline
                                                                                                                                                                                                                                                                                                                         {Assumes $Z_e(\vmu)>0$ for all $e \in \setEfull$?}
                & Yes (implied)                                           & Yes (implied)          & Yes (implied) & Yes (explicit) \\ \hline
                                                                                                                                       {Provides explicit expressions for non-binary alphabets?} & No & Yes & No & Yes \\ \hline
    \end{tabular}
  }
\end{table*}

\subsection{The LCT for S-NFGs}
\label{sec:LCT:SNFG:1}

\begin{figure}
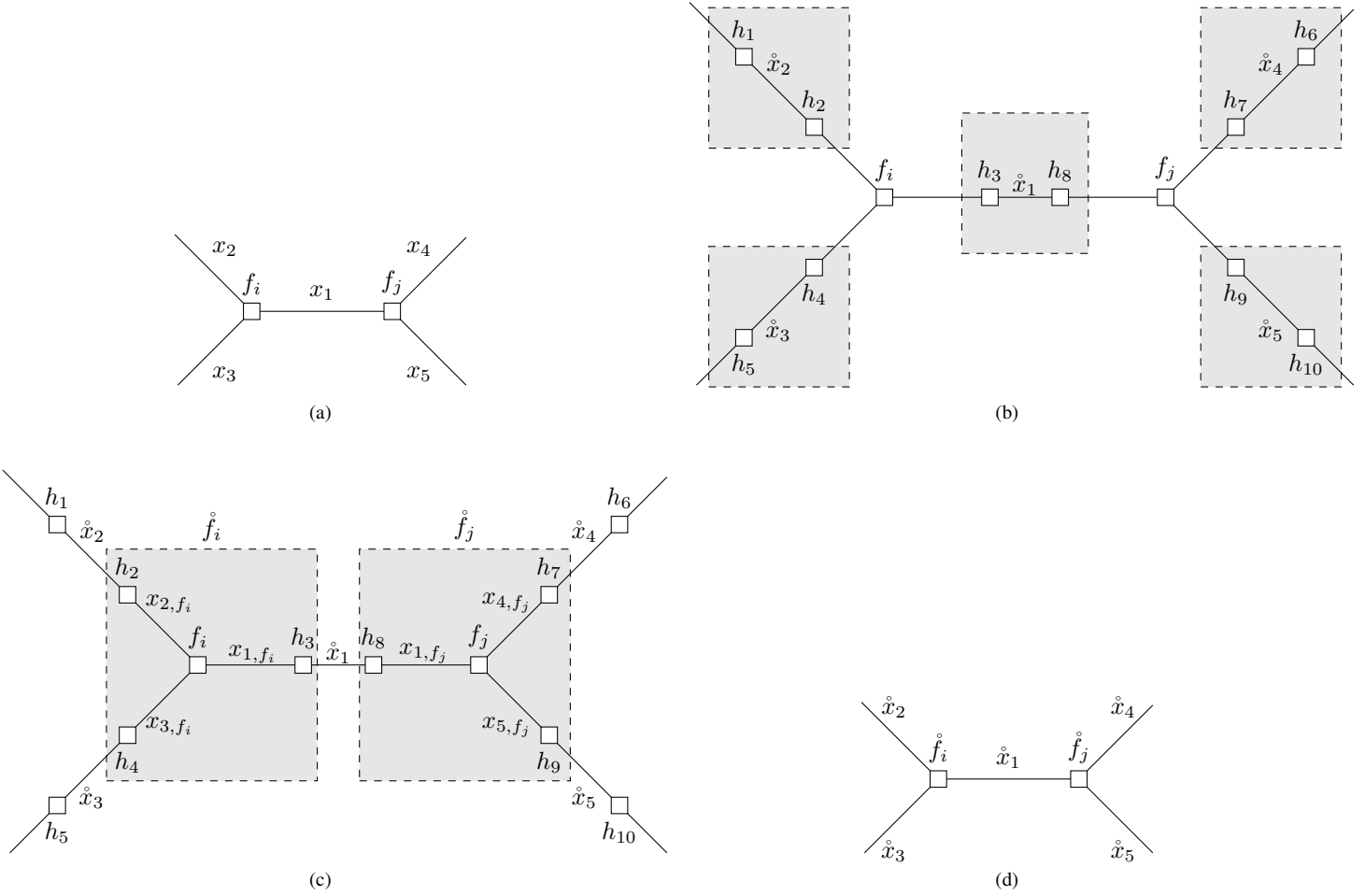

  \centering
  \subfloat[\label{sec:LCT:fig:3}]{
    \begin{minipage}[t]{0.48\textwidth}
      \centering
      \scalebox{0.9}{%
        \begin{tikzpicture}[on grid,auto]
          \input{figures/lct_examples/scale_small.tex}
          \input{figures/head_files_figs.tex}
          \input{figures/lct_examples/snfg/background_nodes.tex}
          \input{figures/lct_examples/snfg/background_lines.tex}
          \begin{pgfonlayer}{main}
            \node (x1) at (0,-0.2*\sdis)
            [label=above: $x_{1}$] {};
            \node (x3) at (-0.5*\ldis-2.1*\sdis,1.3*\sdis)
            [label=above: $x_{2}$] {};
            \node (x4) at (-0.5*\ldis-2.1*\sdis,-1.3*\sdis)
            [label=below: $x_{3}$] {};

            \node (x3) at (0.5*\ldis+2.1*\sdis,1.3*\sdis)
            [label=above: $x_{4}$] {};
            \node (x4) at (0.5*\ldis+2.1*\sdis,-1.3*\sdis)
            [label=below: $x_{5}$] {};
          \end{pgfonlayer}
        \end{tikzpicture}}
    \end{minipage}
  }\\
  \subfloat[\label{sec:LCT:fig:4}]{
    \begin{minipage}[t]{0.48\textwidth}
      \centering
      \resizebox{\linewidth}{!}{%
        \scalebox{0.7}{%
          \begin{tikzpicture}[on grid,auto]
            \input{figures/lct_examples/scale_large_snfg.tex}
            \input{figures/head_files_figs.tex}
            \input{figures/lct_examples/snfg/background_nodes.tex}
            \input{figures/lct_examples/snfg/background_lines.tex}
            \input{figures/lct_examples/otb_funs.tex}
            \input{figures/lct_examples/otb_dashed_boxes.tex}
            \input{figures/lct_examples/snfg/variables_after_lct.tex}
          \end{tikzpicture}}}
    \end{minipage}
  }\\
  \subfloat[\label{sec:LCT:fig:5}]{
    \begin{minipage}[t]{0.48\textwidth}
      \centering
      \resizebox{\linewidth}{!}{%
        \scalebox{0.7}{%
          \begin{tikzpicture}[on grid,auto]
            \input{figures/lct_examples/scale_large_snfg.tex}
            \input{figures/head_files_figs.tex}
            \input{figures/lct_examples/snfg/background_nodes.tex}
            \input{figures/lct_examples/snfg/background_lines.tex}
            \input{figures/lct_examples/otb_funs.tex}
            \input{figures/lct_examples/ctb_dashed_boxes.tex}
            \input{figures/lct_examples/snfg/variables_before_lct.tex}
            \input{figures/lct_examples/snfg/variables_after_lct.tex}
          \end{tikzpicture}}}
    \end{minipage}
  }\\
  \subfloat[\label{sec:LCT:fig:6}]{
    \begin{minipage}[t]{0.48\textwidth}
      \centering
      \scalebox{0.9}{%
        \begin{tikzpicture}[on grid,auto]
          \input{figures/lct_examples/scale_small.tex}
          \input{figures/head_files_figs.tex}
          \input{figures/lct_examples/snfg/background_nodes_after_ctb.tex}
          \input{figures/lct_examples/snfg/background_lines.tex}
          \input{figures/lct_examples/snfg/variables_after_lct.tex}
        \end{tikzpicture}}
    \end{minipage}
  }
  \caption{Illustration of the LCT for a part of an example
    S-NFG.}\label{sec:LCT:fig:1}
\end{figure}

Before giving the general definition of the LCT for an S-NFG, we consider an
example S-NFG and assume that a part of it looks as shown in
Fig.~\ref{sec:LCT:fig:3}. Let
$\vmu $ be an SPA
fixed-point message vector for this example S-NFG. The LCT is a sequence of
modifications to the original S-NFG that leave the partition function
unchanged. This is done by applying the opening-the-box and closing-the-box
operations~\cite{Loeliger2004} as follows.
\begin{enumerate}

  \item The NFG in Fig.~\ref{sec:LCT:fig:4} is obtained from the S-NFG
        in Fig.~\ref{sec:LCT:fig:3} by applying suitable opening-the-box
        operations~\cite{Loeliger2004}. Importantly, the new function nodes $h_i$
        are based on $\vmu$ and are defined in such a way that the partition function remains unchanged.

        For example, for the edge $1$ connecting function nodes $f_{i}$ and $f_{j}$, as shown in Fig.~\ref{sec:LCT:fig:3}, we
        introduce function nodes $h_3$ and $h_8$ with local functions
        $h_3: \set{X}_{1} \times \LCTset{X}_{1} \to \sC$ and
        $h_8: \set{X}_{1} \times \LCTset{X}_{1} \to \sC$, where
        $\LCTset{X}_{1} \defeq \set{X}_{1} $. The partition function of the NFG in Fig.~\ref{sec:LCT:fig:4} remains unchanged because $h_3$ and $h_8$ are
        chosen so that
        $\sum_{\LCT{x}_{1} \in \LCTset{X}_{1}}
          h_3(x_{1,f_{i}}, \LCT{x}_{1}) \cdot h_8(x_{1,f_{j}}, \LCT{x}_{1}) =
          [x_{1,f_{i}} \! = \! x_{1,f_{j}}]$ holds for all
        $ x_{1,f_{i}}, x_{1,f_{j}} \in \set{X}_{1} $.

  \item The NFG in Fig.~\ref{sec:LCT:fig:5} is essentially the same as
        the NFG in Fig.~\ref{sec:LCT:fig:4}.

  \item The NFG in Fig.~\ref{sec:LCT:fig:6} is obtained from the NFG
        in Fig.~\ref{sec:LCT:fig:5} by applying suitable closing-the-box
        operations~\cite{Loeliger2004}. For example, consider the left dashed box in Fig.~\ref{sec:LCT:fig:5}. Its exterior function $ \LCT{f}_{i} $ is the sum, over the internal variables $ x_{1,f_{i}},x_{2,f_{i}}, x_{3,f_{i}} $, of the product of the internal local functions $ f_{i},h_{2},h_{3},h_{4} $, \ie,
        \begin{align*}
          \LCT{f}_{i}(\LCT{x}_{1}, \LCT{x}_{2}, \LCT{x}_{3})
           & \defeq \!\!\! \!\!\! \sum_{x_{1,f_{i}},x_{2,f_{i}},x_{3,f_{i}}} \!\!\!
          \!\!\!\!\!\!\!\!\!
          f_{i}(x_{1,f_{i}}, x_{2,f_{i}}, x_{3,f_{i}})
          \cdot h_3(x_{1,f_{i}}, \LCT{x}_{1})
          \\
           & \qquad\qquad\qquad
          \cdot
          h_2(x_{2,f_{i}}, \LCT{x}_{2})
          \cdot
          h_4(x_{3,f_{i}}, \LCT{x}_{3}),
        \end{align*}
        where $ \LCT{x}_{1} \in \LCTset{X}_{1},\, \LCT{x}_{2}\in \LCTset{X}_{2},\, \LCT{x}_{3}\in \LCTset{X}_{3},$
        and where $\LCTset{X}_{e} \defeq \set{X}_{e}$ for any edge $ e \in \setEfull $ in the original S-NFG.
        Replacing a dashed box by a single function node that represents the associated exterior function is known as the closing-the-box operation~\cite{Loeliger2004}.
        The partition function of the NFG in Fig.~\ref{sec:LCT:fig:6} remains unchanged due to the way that $\LCT{f}$ is defined.

\end{enumerate}

The goal of the LCT is to construct matrices
$\matr{M}_{\efi},\matr{M}_{\efj} \in \sR^{|\setxe| \times |\setxe|}$ for each
$e = (f_{i},f_{j}) \in \setEfull$ that simultaneously satisfy two criteria:
(1) the equality $\matr{M}_{\efi} \cdot \matr{M}_{\efj}^{\tran} = \matr{I}$ to
preserve the partition function, and (2) a special structure that makes the
SPA fixed-point messages in the transformed NFG trivial (\ie, each
  message equals the indicator function
  $\LCT{x}_e \mapsto [\LCT{x}_e \!=\! 0]$). We now give the general
definition and main properties of the LCT for S-NFGs.

\begin{definition}
  \label{sec:LCT:def:1}

  Consider an S-NFG $\graphN$. Let
  $\vmu $ be an SPA
  fixed-point message vector for $\graphN$ such that $ Z_e(\vmu) >0 $ for all $ e \in \setEfull $. For each $ e \in \setEfull $, write $Z_e$ for $Z_e(\vmu)$, where
  (see~\eqref{sec:SNFG:eqn:6})
  \begin{align*}
    Z_e
    = \sum_{x_e} \mu_{\efi}(\xe) \cdot \mu_{\efj}(\xe).
  \end{align*}
  We apply the following steps.
  \begin{enumerate}

    \item (generalization of the first step above) For every edge $e \in
            \setEfull$, we do the following.

          \begin{itemize}

            \item \finaltext{\finaltext{Choose} a distinguished element of
                    $\set{X}_e$ and denote it by $0$.}
                  (Recall that we assume $ 0 \in \setxe $ in Definition~\ref{def: def of snfg}.)

            \item We define the alphabet
                  $\LCTsetxe \defeq \set{X}_e$.\footnote{More generally, $\LCTsetxe$ may be any alphabet satisfying $|\LCTsetxe|=|\set{X}_e|$; we identify it with $\set{X}_e$ for notational simplicity.}

            \item The edge $e$ (with associated variable $x_e$) connecting $f_{i}$ to $f_{j}$
                  is replaced by the following alternating sequence of edges and function
                  nodes: an edge with associated variable $x_{e,f_{i}} \in \set{X}_e$, a
                  function node $M_{\efi}$, an edge with associated variable
                  $\LCT{x}_e \in \LCTsetxe$, a function node $M_{\efj}$, and an edge
                  with associated variable $x_{e,f_{j}} \in \set{X}_e$.
                  \begin{itemize}

                    \item The function node $M_{\efi}$ is associated with the local function
                          $M_{\efi}: \set{X}_e \times \LCTsetxe \to \sR$.
                          Let $ \matr{M}_{\efi} \defeq \bigl( M_{\efi}(\xe, \LCT{x}_e) \bigr)_{\!
                              (\xe, \LCT{x}_e) \in \set{X}_e \times \LCTsetxe} $ be the corresponding matrix
                          with row indices $ \xe $ and column indices $ \LCT{x}_e $.

                    \item The function node $M_{\efj}$ is associated with the local function
                          $M_{\efj}: \set{X}_e \times \LCTsetxe \to \sR$.
                          Let $ \matr{M}_{\efj} \defeq \bigl( M_{\efj}(\xe, \LCT{x}_e) \bigr)_{\!
                              (\xe, \LCT{x}_e) \in \set{X}_e \times \LCTsetxe} $ be the corresponding matrix
                          with row indices $ \xe $ and column indices $ \LCT{x}_e $.

                  \end{itemize}
                  To ensure that the partition function of the resulting NFG equals the
                  partition function of the original S-NFG, the local functions $M_{\efi}$
                  and $M_{\efj}$ satisfy
                  \begin{alignat}{2}
                     & \sum_{\LCT{x}_e}
                    M_{\efi}(x_{\efi}, \LCT{x}_e)
                    \cdot
                    M_{\efj}(x_{\efj}, \LCT{x}_e)
                     & = [x_{\efi} \! = \!  x_{\efj}]
                    \label{sec:LCT:exp:1}
                  \end{alignat}
                  for all $x_{e,f_i}, x_{e,f_{j}} \in \set{X}_e.$ Here,
                    we use the shorthand $ \sum_{\LCT{x}_e} $ instead of
                    $ \sum_{\LCT{x}_e \in \LCTsetxe} $.  In matrix
                    form,~\eqref{sec:LCT:exp:1} reads
                    $ \matr{M}_{\efi} \cdot \matr{M}_{\efj}^{\tran} = \matr{I}
                    $; \ie, the matrix $ \matr{M}_{\efj}^{\tran} $ is the inverse of
                    $ \matr{M}_{\efi} $, so the two local functions are mutual
                    inverses up to transposition. This also implies that
                    $\matr{M}_{\efj}^{\tran} \cdot \matr{M}_{\efi} =
                      \matr{I}$, \ie, for all
                    $\LCT{x}_{e,f_{i}}, \LCT{x}_{e,f_{j}} \in \LCTsetxe$, it
                    holds that
                  \begin{align}
                    \sum_{x_e}
                    M_{\efi}(\xe, \LCT{x}_{e,f_{i}})
                    \cdot
                    M_{\efj}(\xe, \LCT{x}_{e,f_{j}})
                     & = \bigl[ \LCT{x}_{e,f_{i}} \! = \!  \LCT{x}_{e,f_{j}} \bigr].
                    \label{sec:LCT:exp:2}
                  \end{align}

            \item Let $\zeta_{\efi}$ and $\zeta_{\efj}$ be \finaltext{arbitrary positive real-valued constants}. At the heart of the LCT are the following definitions for
                  $M_{\efi}$ and $M_{\efj}$, which relate to the SPA fixed-point message vector $\vmu$:
                  \begin{alignat}{2}
                    M_{\efi}(\xe,\LCT{x}_e)
                     & \defeq \zeta_{\efi}
                    \cdot
                    \mu_{\efi}(\xe),
                    \quad x_e \in \set{X}_e, \ \LCT{x}_e = 0,
                    \label{sec:LCT:exp:3}  \\
                    M_{\efj}(\xe,\LCT{x}_e)
                     & \defeq \zeta_{\efj}
                    \cdot
                    \mu_{\efj}(\xe),
                    \quad x_e \in \set{X}_e, \ \LCT{x}_e = 0.
                    \label{sec:LCT:exp:4}
                  \end{alignat}
                  Evaluating~\eqref{sec:LCT:exp:2} for $\LCT{x}_{\efi} = 0$ and
                  $\LCT{x}_{\efj} = 0$ yields the constraint
                  \begin{align}
                    \zeta_{\efi}
                    \cdot
                    \zeta_{\efj}
                     & = Z_e^{-1}.
                    \label{sec:LCT:exp:5}
                  \end{align}

            \item A possible choice for the remaining function values of $M_{\efi}$ and
                  $M_{\efj}$ is
                  \begin{align}
                    \begin{aligned}
                       & M_{\efi}(\xe,\LCT{x}_e)
                      \defeq \zeta_{\efi}
                      \cdot
                      \chi_{\efi}
                      \\&\cdot
                      \begin{cases}
                        - \mu_{\efj}(\LCT{x}_e)
                         & \!\!\!\!\!\!\!\! x_e = 0, \
                        \LCT{x}_e \in \LCTsetxe\setminus\{0\}                             \\[0.2cm]
                        \begin{array}{l}
                          \delta_{\efi}
                          \cdot
                          [x_e \!=\! \LCT{x}_e]
                          \\
                          +
                          \epsilon_{\efi}
                          \cdot
                          \mu_{\efi}(\xe)
                          \cdot
                          \mu_{\efj}(\LCT{x}_e)
                        \end{array}
                         & \!\!\!\!\!\!\!\! x_e,\LCT{x}_e \in \set{X}_e \setminus \{ 0 \}
                      \end{cases}
                      \\
                       & M_{\efj}(\xe,\LCT{x}_e)
                      \defeq \zeta_{\efj}
                      \cdot
                      \chi_{\efj}
                      \\&\cdot\begin{cases}
                                - \mu_{\efi}(\LCT{x}_e)
                                 & \!\!\!\!\!\!\!\! x_e = 0, \
                                \LCT{x}_e \in \LCTsetxe\setminus\{0\}                             \\[0.2cm]
                                \begin{array}{l}
                                  \delta_{\efj}
                                  \cdot
                                  [x_e \!=\! \LCT{x}_e]
                                  \\+
                                  \epsilon_{\efj}
                                  \cdot
                                  \mu_{\efj}(\xe)
                                  \cdot
                                  \mu_{\efi}(\LCT{x}_e)
                                \end{array}
                                 & \!\!\!\!\!\!\!\! x_e,\LCT{x}_e \in \set{X}_e \setminus \{ 0 \}
                              \end{cases}
                    \end{aligned}
                    \label{eq:lct:nonzero-columns}
                  \end{align}
                  where $\chi_{\efi}$ and $\chi_{\efj}$ are arbitrary real-valued constants satisfying
                  \begin{align}
                    \chi_{\efi}
                    \cdot
                    \chi_{\efj}
                     & = 1.
                    \label{sec:LCT:exp:6}
                  \end{align}
                  Recall that the belief evaluated by the SPA fixed-point
                    message vector satisfies (see Definition~\ref{def:belief at
                      SPA fixed point for S-NFG})
                    \begin{align}
                      \beli_e(0)
                       & = (Z_e)^{-1} \cdot \mu_{\efi}(0) \cdot \mu_{\efj}(0).
                      \label{sec:LCT:exp:new:2}
                    \end{align}
                    If $ \beli_e(0) \neq 1 $, then let $\delta_{\efi}$,
                    $\delta_{\efj}$ be arbitrary real-valued constants
                    satisfying
                    \begin{align}
                      \delta_{\efi}
                      \cdot
                      \delta_{\efj}
                       & = Z_e,
                      \label{sec:LCT:exp:7}
                    \end{align}
                    and let $\epsilon_{\efi}$ and $\epsilon_{\efj}$ be
                    real-valued constants satisfying
                    \begin{align}
                      \delta_{\efi}
                      +
                      Z_e
                      \cdot
                      \bigl( 1 \! - \! \beli_e(0) \bigr)
                      \cdot
                      \epsilon_{\efi}
                       & = \mu_{\efj}(0),
                      \label{sec:LCT:exp:8} \\
                      \delta_{\efj}
                      +
                      Z_e
                      \cdot
                      \bigl( 1 \! - \! \beli_e(0) \bigr)
                      \cdot
                      \epsilon_{\efj}
                       & = \mu_{\efi}(0).
                      \label{sec:LCT:exp:9}
                    \end{align}
                    If $ \beli_e(0) = 1 $, then define
                    \begin{align}
                      \delta_{\efi}
                       & \defeq
                      \mu_{\efj}(0),
                      \label{sec:LCT:exp:new:1:part:1} \\
                      \delta_{\efj}
                       & \defeq
                      \mu_{\efi}(0),
                      \label{sec:LCT:exp:new:1:part:2}
                    \end{align}
                    and let $\epsilon_{\efi}$ and $\epsilon_{\efj}$ be arbitrary
                    real-valued constants satisfying
                    \begin{align}
                      1
                      +
                      \delta_{\efi}
                      \cdot
                      \epsilon_{\efj}
                      +
                      \delta_{\efj}
                      \cdot
                      \epsilon_{\efi} & = 0.
                      \label{extra constraint on epsilon for beli = [x = 0]}
                    \end{align}
                    Note that $\delta_{\efi}$ and $\delta_{\efj}$
                    in~\eqref{sec:LCT:exp:new:1:part:1}--\eqref{sec:LCT:exp:new:1:part:2}
                    satisfy~\eqref{sec:LCT:exp:7}--\eqref{sec:LCT:exp:9}.\footnote{The
                        verification of
                        \eqref{sec:LCT:exp:8}--\eqref{sec:LCT:exp:9} is
                        straightforward. Moreover, Eq.~\eqref{sec:LCT:exp:7} is
                        shown as follows. Namely, we obtain
                        $\delta_{\efi} \cdot \delta_{\efj} = \mu_{\efi}(0) \cdot
                          \mu_{\efj}(0) = Z_e$, where the first equality follows
                        from~\eqref{sec:LCT:exp:new:1:part:1}--\eqref{sec:LCT:exp:new:1:part:2}
                        and where the second equality follows
                        from~\eqref{sec:LCT:exp:new:2}.} The proof showing that
                    the above gives a valid choice for $M_{\efi}$ and $M_{\efj}$
                    is given in Appendix~\ref{apx:LCT SNFGs}.

          \end{itemize}

    \item (generalization of the second step above) Nothing is done in this step.

    \item (generalization of the third step above) For every function node
          $f \in \setF$, we do the following.
          \begin{itemize}

            \item Let
                  $\LCTvxf \defeq ( \LCT{x}_e )_{e \in \setpf} \in \LCTsetx_{\setpf}$ where
                  $ \LCTsetx_{\setpf} \defeq \prod_{e \in \setpf} \LCTsetxe $.

            \item For any $
                    \LCTvxf \in \LCTsetx_{\setpf}$, we define
                  \begin{align}
                    \LCT{f}(\LCTvxf)
                     & \defeq \sum_{\vx_{\setpf}}
                    f(\vx_{\setpf})
                    \cdot
                    \prod_{e \in \setpf}
                    M_{\ef}(x_{e}, \LCT{x}_e).
                    \label{sec:LCT:exp:10}
                  \end{align}

            \item Finally, the LCT NFG $\LCT{\graphN}$ is
                  defined to be the NFG with vertex set $\setF$, edge set $\setE$, and alphabet
                  $ \LCTsetx \defeq \prod_{e} \LCTsetxe $,
                  where for every vertex $f \in \setF$, the associated local function is $\LCT{f}$,
                  and where for every edge $e \in \setE$, the associated variable
                  is $\LCT{x}_e$ and the associated alphabet is $ \LCTset{X}_{e} $.
                  With this, the global function of $\LCT{\graphN}$ is
                  $\LCT{g}(\LCTv{x}) \defeq \prod_f \LCT{f}(\LCTvxf)$.\edefinition

          \end{itemize}

  \end{enumerate}
\end{definition}

\begin{remark}
  \label{sec:LCT:rem:1}
  The following remarks provide insights into the limitations and characteristics of the proposed LCT in different scenarios.
  \begin{enumerate}
    \item There are cases where it is \textbf{not} possible to obtain
          $ M_{\efi} $ and $ M_{\efj} $ for $ e =(f_{i}, f_{j}) \in \setEfull $
          based on the SPA fixed-point message vector. Consider a single-cycle S-NFG
          with local functions $\left(\begin{smallmatrix} 1 & 1 \\ 0 & 1
            \end{smallmatrix}\right)$
          and
          $\left(\begin{smallmatrix}
              1 & 0 \\ 0 & 1
            \end{smallmatrix}\right)$. The SPA fixed-point messages along each edge are
          $\left(1,\, 0\right)^{\!\tran}$
          and
          $\left(0,\, 1\right)^{\!\tran}$,
          which implies
          $Z_e = 0 $ and the LCT does not work due to~\eqref{sec:LCT:exp:3}--\eqref{sec:LCT:exp:5}.
          For more details, see also Item~\ref{sec:SNFG:remk:1:item:2} in Remark~\ref{sec:SNFG:remk:1}.

    \item\label{remk: LCT on snfg results an NFG with negative fun nodes}  The LCT NFG $\LCT{\graphN}$ of $\graphN$ is in general \textbf{not} an S-NFG. This
          is because some of the local functions of $\LCT{\graphN}$ might take on
          negative real values. With that, also the global function of $\LCT{\graphN}$
          might take on negative real values.

    \item Consider an edge $ e =(f_{i}, f_{j}) \in \setEfull $.
          For the binary alphabets $\set{X}_e = \{0,1\}$ and $\LCTsetxe = \{0,1\}$, the proposed LCT  is the same
          as the one presented in~\cite{Chertkov2006}.
          In this case, the matrices
          $ \matr{M}_{\efi} $ and $ \matr{M}_{\efj} $
          are
          \begin{align*}
            \matr{M}_{\efi}
             & =
            \begin{pmatrix}
              M_{\efi}(0,0) & M_{\efi}(0,1) \\
              M_{\efi}(1,0) & M_{\efi}(1,1)
            \end{pmatrix} \\
             & =
            \zeta_{\efi}
            \cdot
            \begin{pmatrix}
              \mu_{\efi}(0) & - \chi_{\efi} \cdot \mu_{\efj}(1) \\
              \mu_{\efi}(1) & \chi_{\efi} \cdot \mu_{\efj}(0)
            \end{pmatrix},
            \\
            \matr{M}_{\efj}
             & =
            \begin{pmatrix}
              M_{\efj}(0,0) & M_{\efj}(0,1) \\
              M_{\efj}(1,0) & M_{\efj}(1,1)
            \end{pmatrix} \\
             & =
            \zeta_{\efj}
            \cdot
            \begin{pmatrix}
              \mu_{\efj}(0) & - \chi_{\efj} \cdot \mu_{\efi}(1) \\
              \mu_{\efj}(1) & \chi_{\efj} \cdot \mu_{\efi}(0)
            \end{pmatrix}.
          \end{align*}
          Note that for this simple alphabet $\set{X}_e$ with
            $|\set{X}_e| = 2$, the choice of $\delta_{\efi}$, $\delta_{\efj}$,
            $\epsilon_{\efi}$, and $\epsilon_{\efj}$ does not affect the
            resulting matrices $\matr{M}_{\efi}$ and $\matr{M}_{\efj}$.
            However, for alphabets $\set{X}_e$ with $|\set{X}_e| > 2$, our
            LCT, although inspired by previous works such
            as~\cite{Chertkov2006, Chernyak2007, Mori2015}, is in general
            different. Notably, \finaltext{our approach allows zero-valued components in the SPA
            fixed-point messages.} \eremark
  \end{enumerate}

\end{remark}

\begin{proposition}
  \label{sec:LCT:prop:1}

  Consider an S-NFG $\graphN$. Let
  $\vmu $ be an SPA
  fixed-point message vector for $\graphN$ such that $ Z_e(\vmu) >0 $ for all $ e \in \setEfull $ and $ \ZBSPA(\graphN,\vmu) > 0 $. Then the LCT NFG $\LCT{\graphN}$ is well defined, as specified in
  Definition~\ref{sec:LCT:def:1}.
  The main properties of $\LCT{\graphN}$ are as follows.
  \begin{enumerate}

    \item\label{sec:LCT:prop:1:item:1} The transformation preserves the partition function, \ie, we have $
            Z(\graphN)
            = Z\bigl( \LCT{\graphN} \bigr).
          $

    \item\label{sec:LCT:prop:1:item:2} The \finaltext{$\vmu$-based} Bethe partition function of the original S-NFG $\graphN$ equals the global function value of the all-zero configuration of $\LCT{\graphN}$, \ie, we have $
            \ZBSPA(\graphN,\vmu)
            = \LCT{g}(\vect{0}).
          $

    \item\label{sec:LCT:prop:1:item:3} For any $f \in \setF$ and any $\LCTvxf \in \LCTsetxf$ such that $ \wh( \LCTvxf ) = 1 $, we have
          $
            \LCT{f}(\LCTvxf)
            = 0,
          $
          where $ \wh( \cdot ) $ denotes the Hamming weight of its argument.

    \item\label{sec:LCT:prop:1:item:4} We call a configuration
            $\LCTv{x}$ a generalized loop if the subgraph
            $\bigl( \setF,\setE'(\LCTv{x}) \bigr)$, where
            $\setE'(\LCTv{x}) \defeq \{ e \in \setE \ | \ \LCT{x}_{e} \neq 0
              \}$, does not contain any leaves, \ie, vertices of degree
            one. (The all-zero configuration $\LCTv{x} = \vect{0}$ is also
            considered to be a generalized loop.) Then it holds that every
            valid configuration of $\LCT{\graphN}$ is a generalized loop, and
            that, equivalently, if $\LCTv{x}$ is not a generalized loop, then
            $\LCT{g}(\LCTv{x}) = 0$.  In particular, if $\graphN$ is
            cycle-free, then so is $\LCT{\graphN}$, and $\LCTv{x} = \vect{0}$
            is the only generalized loop.

    \item\label{sec:LCT:prop:1:item:5} The partition function satisfies
          \begin{align*}
            Z(\LCT{\graphN})
             & = \ZBSPA(\graphN,\vmu)
            \cdot
            \left(
            1
            +
            \sum_{\LCTv{x}:\, \setE'(\LCTv{x}) \neq \emptyset }
            \frac{\LCT{g}(\LCTv{x})}
            {\LCT{g}(\vect{0})}
            \right),
          \end{align*}
          where the summation is over all nonzero configurations $\LCTv{x} \in \prod_{e} \LCTsetxe$, \ie, all $\LCTv{x}$ with $\setE'(\LCTv{x}) \neq \emptyset$; by Property~\ref{sec:LCT:prop:1:item:4}, only generalized loops contribute nonzero terms. The terms in the set
          $
            \bigl\{ \LCT{g}(\LCTv{x})/\LCT{g}(\vect{0})
            \bigr\}_{\! \LCTv{x}:\, \setE'(\LCTv{x}) \neq \emptyset}
          $
          are viewed as ``correction terms'' for the \finaltext{$\vmu$-based Bethe approximation}. Note
          that if $\graphN$ is cycle-free, then there is no correction term,
          implying the well-known result that
          $Z(\graphN) = \ZBSPA(\graphN,\vmu)$.

    \item\label{sec:LCT:prop:1:item:6} The SPA fixed-point message
            vector $\vmu $ for $\graphN$ induces the SPA fixed-point message
            vector $\LCTv{\mu} \defeq ( \LCTv{\mu}_{\ef} )_{f \in \setF,\,e \in
                \setpf}$ for $\LCT{\graphN}$, which satisfies
          \begin{align*}
            \LCT{\mu}_{e,f}(\LCT{x}_{e})
             & = \bigl[ \LCT{x}_{e} \! = \! 0 \bigr],
            \qquad
            \LCT{x}_{e} \in \LCTset{X}_{e}, \,
            e \in \setpf, \, f \in \setF.
          \end{align*}

    \item\label{sec:LCT:prop:1:item:7} While the above properties are shared by all variants of the LCT, the
          LCT in Definition~\ref{sec:LCT:def:1} has the following additional
          properties (that are not necessarily shared by the other LCT variants).
          \begin{enumerate}

            \item For every $e = (f_{i},f_{j}) \in \setE$, the functions $M_{\efi}$ and
                  $M_{\efj}$ are formally symmetric in the sense that swapping $f_{i}$ and
                  $f_{j}$ in the definition of $M_{\efi}$ leads to the definition of
                  $M_{\efj}$, and vice-versa.

            \item\label{sec:LCT:prop:1:item:7:b} Assume that $M_{\efi}$ and $M_{\efj}$ are defined based on the
                  choices $\zeta_{\efi} = \zeta_{\efj} \defeq Z_e^{-1/2}$,
                  $\chi_{\efi} = \chi_{\efj} \defeq 1$,
                  $\delta_{\efi} = \delta_{\efj} \defeq Z_e^{1/2}$. Let
                  $e = (f_{i},f_{j}) \in \setE$ be an edge such that
                  $\mu_{\efi}(\xe) = \mu_{\efj}(\xe) \in \sR$ for all $x_e \in \set{X}_e$. Then
                  the matrices $\matr{M}_{\efi}$
                  and $\matr{M}_{\efj}$ are equal
                  and orthogonal.

          \end{enumerate}

  \end{enumerate}
\end{proposition}

\begin{proof}
  See Appendix~\ref{apx:property of SNFGs}.
\end{proof}

We make the following remarks on Proposition~\ref{sec:LCT:prop:1}.
\begin{itemize}

  \item Property~\ref{sec:LCT:prop:1:item:2} states that the LCT is a reparameterization of the global function of $\graphN$ based on an SPA fixed point $ \vmu $ such that
        the transformed global function $\LCT{g}$ evaluated at $ \LCTv{x} = \vect{0} $ equals the $\vmu$-based Bethe partition function $\ZBSPA(\graphN,\vmu)$.

  \item Property~\ref{sec:LCT:prop:1:item:6} highlights that  $\LCT{\graphN}$ has an SPA fixed-point message vector $ \LCT{\vmu} $ with a simple structure,
        which is the crucial ingredient in the proof of the graph-cover theorem (Theorem~\ref{sec:CheckCon:thm:1}).

\end{itemize}

Property~\ref{sec:LCT:prop:1:item:2} in Proposition~\ref{sec:LCT:prop:1}
  is similar to a discrete-Fourier-transform identity. For a Fourier transform
  pair $f \leftrightarrow F$ over a finite abelian group, one has
  $\sum_{n} f(n) = F(0)$: summing the function equals the value of its
  transform at the all-zero argument. The LCT plays an analogous role. It is a
  holographic transform~\cite{AlBashabsheh2011} that leaves the partition
  function unchanged (Property~\ref{sec:LCT:prop:1:item:1} in
  Proposition~\ref{sec:LCT:prop:1}) while concentrating the \finaltext{$\vmu$-based Bethe}
  partition function at the all-zero configuration, \ie, we have $\LCT{g}(\vect{0}) = \ZBSPA(\graphN,\vmu)$, and regrouping other contributions into the loop-correction terms
  (Property~\ref{sec:LCT:prop:1:item:5} in
  Proposition~\ref{sec:LCT:prop:1}). Nothing is lost under this regrouping:
  the contributions of all configurations still sum to $Z(\graphN)$. When
  $\graphN$ is cycle-free, all correction terms vanish and
  $Z(\graphN) = \LCT{g}(\vect{0})$, so the analogy with $\sum_{n} f(n) = F(0)$
  becomes exact.

\subsection{The LCT for DE-NFGs}
\label{sec:LCT:DENFG:1}

The LCT for DE-NFGs closely parallels the LCT for S-NFGs. We therefore highlight only the differences.

\begin{definition}
  \label{def:DENFG:LCT:1}

  The LCT for a DE-NFG is the LCT of Definition~\ref{sec:LCT:def:1} with the following replacements, applied for each edge $e \in \setEfull$.
  \begin{itemize}

    \item The alphabet $\set{X}_e$ is replaced by the alphabet $\tset{X}_e$, where
          $\tset{X}_e \defeq \set{X}_e \times \set{X}_e$.

    \item The special element $0 \in \set{X}_e$ is replaced by the special element
          $\tzero = (0,0) \in \tset{X}_e$.
          (Recall that $ \tzero \in \tset{X}_e $ is well-defined in Definition~\ref{sec:DENFG:def:4}.)

    \item The variable $x_e \in \set{X}_e$ is replaced by the variable
          $\tx_e = (x_e,x'_e) \in \tset{X}_e$.

    \item The definition of the alphabet
          $\LCTset{X}_e \defeq \set{X}_e$ is
          replaced by the definition of the alphabet
          $\LCTtset{X}_e \defeq \tset{X}_e$.

    \item The element $0 \in \LCTset{X}_e$ is replaced by the element
          $\tzero\in \LCTtset{X}_e$.

    \item The variable $\LCT{x}_e \in \LCTset{X}_e$ is replaced by the variable
          $\LCTt{x}_e = (\LCT{x}_e, \LCT{x}'_e) \in \LCTtset{X}_{e}$.

  \end{itemize}
  For each $e=(f_i,f_j)\in\setEfull$, Lemmas~\ref{sec:DENFG:lem:1} and~\ref{lem: nonnegative of ZBSPA for DE-NFG} allow the constants $\zeta_{e,f}$, $\chi_{e,f}$, $\delta_{e,f}$, and $\epsilon_{e,f}$, for $f\in\{f_i,f_j\}$, to be chosen real.

  For any $ \set{I} \subseteq \setEfull $, we define
  \begin{align*}
    \LCTtset{X}_{ \set{I} }
    \defeq \prod_{e \in \set{I}} \LCTtset{X}_{ e }
    = \LCTset{X}_{ \set{I} } \times \LCTset{X}_{ \set{I} },
  \end{align*}
  where
  $
    \LCTset{X}_{ \set{I} } \defeq
    \prod_{e \in \set{I}} \LCTsetxe.
  $
  The associated variable is given by the collection
  $ \LCTtv{x}_{\set{I}} = ( \LCTv{x}_{\set{I}}, \LCTv{x}_{\set{I}}' )
    \in \LCTtset{X}_{ \set{I} } $.

  For each $ f \in \setF $ and
  $ \LCTtv{x}_{\setpf} \in \LCTtset{X}_{\setpf} $, define
  \begin{align}
    \LCT{f}(\LCTtv{x}_{\setpf})
     & \defeq \sum_{\tvx_{\setpf}}
    f(\tvx_{\setpf})
    \cdot
    \prod_{e \in \setpf}
    M_{\ef}(\tx_{e}, \LCTt{x}_e). \qquad
    \label{sec:LCT:eqn:1}
  \end{align}
  \\[-0.6 cm]{\edefinition}
\end{definition}

For any $ \set{I} \subseteq \setEfull $, we use the shorthand
  $ \sum_{ \LCTtv{x}_{\set{I}} } $ instead of
  $ \sum_{ \LCTtv{x}_{ \set{I} } \in \LCTtset{X}_{\set{I}} } $.

\begin{proposition}
  \label{prop:DENFG:LCT:1}

  Consider a DE-NFG $\graphN$. Let
  $\vmu $ be an SPA
  fixed-point message vector for $\graphN$ such that $ Z_e(\vmu) >0 $ for all $ e \in \setEfull $.
  (Recall that in Lemma~\ref{lem: nonnegative of ZBSPA for DE-NFG}, we prove that $Z_e(\vmu) \in \sRp$ for all $ e \in \setEfull $.)
  Then the LCT NFG $\LCT{\graphN}$ (see Definition~\ref{def:DENFG:LCT:1}) is well defined.
  The main properties of $\LCT{\graphN}$ are listed as follows.
  \begin{enumerate}

    \item\label{prop:DENFG:LCT:1:item:1} Natural extension of Property~\ref{sec:LCT:prop:1:item:1} in Proposition~\ref{sec:LCT:prop:1}:
          $
            Z(\graphN)
            = Z\bigl( \LCT{\graphN} \bigr).
          $
    \item\label{prop:DENFG:LCT:1:item:4} Natural extension of Property~\ref{sec:LCT:prop:1:item:2} in Proposition~\ref{sec:LCT:prop:1}:
          $
            \ZBSPA(\graphN,\vmu)
            = \LCT{g}(\tv{0}),
          $
          where $ \tv{0} \defeq ( \tzero,\ldots,\tzero ) $.

    \item\label{sec:LCT:prop:1:item:8} Natural extension of Property~\ref{sec:LCT:prop:1:item:3} in Proposition~\ref{sec:LCT:prop:1}: for any $f \in \setF$ and any $\LCTtv{x}_{\setpf} \in \LCTtset{X}_{\setpf}$ such that
            $ \wh( \LCTtv{x}_{\setpf} ) = 1 $, where $ \wh( \LCTtv{x}_{\setpf} ) \defeq
              \sum_{ e \in \setpf } \bigl[ \LCTt{x}_{e} \!\neq\! \tzero \bigr] $, we have
          \begin{align*}
            \LCT{f}(\LCTtv{x}_{\setpf})
             & = 0.
          \end{align*}

    \item Natural extension of Property~\ref{sec:LCT:prop:1:item:4} in Proposition~\ref{sec:LCT:prop:1}.

    \item Natural extension of Property~\ref{sec:LCT:prop:1:item:5} in Proposition~\ref{sec:LCT:prop:1}.

    \item Natural extension of Property~\ref{sec:LCT:prop:1:item:6} in Proposition~\ref{sec:LCT:prop:1}.
          Nevertheless, because of its importance
          for later parts of this paper, the statement is included here. Namely, the
          SPA fixed-point message vector
          $\vmu$ for
          $\graphN$ induces an SPA fixed-point message vector $ \LCTv{\mu}
            = ( \LCTv{\mu}_{e,f} )_{f \in \setF,\,e \in \setpf} $
          for
          $\LCT{\graphN}$ with
          \begin{align*}
            \LCT{\mu}_{e,f}(\LCTt{x}_{e})
             & = \bigl[ \LCTt{x}_{e} \! = \! \tzero \bigr],
            \qquad
            \LCTt{x}_{e} \in \LCTtset{X}_{e}, \,
            e \in \setpf, \, f \in \setF.
          \end{align*}

    \item\label{prop:DENFG:LCT:1:item:7} Natural extension of Property~\ref{sec:LCT:prop:1:item:7} in Proposition~\ref{sec:LCT:prop:1}.
          \begin{enumerate}
            \item For every $e = (f_{i},f_{j}) \in \setE$, the functions $M_{\efi}$ and
                  $M_{\efj}$ are formally symmetric in the sense that swapping $f_{i}$ and
                  $f_{j}$ in the definition of $M_{\efi}$ leads to the definition of
                  $M_{\efj}$, and vice-versa.

            \item Assume that $M_{\efi}$ and $M_{\efj}$ are defined based on the
                  choices $\zeta_{\efi} = \zeta_{\efj} \defeq Z_e^{-1/2}$,
                  $\chi_{\efi} = \chi_{\efj} \defeq 1$,
                  $\delta_{\efi} = \delta_{\efj} \defeq Z_e^{1/2}$.
                  Let
                  $e = (f_{i},f_{j}) \in \setE$ be an edge such that
                  $\mu_{\efi}(\txe) = \overline{\mu_{\efj}(\txe)}$ for all $\txe \in \tset{X}_e$.
                  In this case, both $ \matr{M}_{\efi} $ and $ \matr{M}_{\efj} $ are unitary matrices and satisfy
                  $ \matr{M}_{\efi} = \overline{\matr{M}_{\efj}} $.
          \end{enumerate}

    \item\label{prop:DENFG:LCT:1:item:8} For every
            $ e = (f_{i}, f_{j}) \in \setEfull $, it holds that
            $ \matr{ C }_{ M_{\efi} },\matr{ C }_{ M_{\efj} } \in \setHerm{
                \setxe \times \LCTsetxe } $.

    \item\label{prop:DENFG:LCT:1:item:9} For every $f \in \setF$, it
            holds that
            $\matr{C}_{\LCT{f}} \in \setHerm{ \LCTset{X}_{\setpf} } $.

    \item\label{LCT on strict sense DENFG results in weak sense DENFG}
          The resulting NFG $\LCT{\graphN}$ is a weak-sense DE-NFG.
          (Recall that strict-sense DE-NFG and
          weak-sense DE-NFG
          are defined in Definition~\ref{sec:DENFG:def:4}.
          In Assumption~\ref{sec:DENFG:asum:2}, we impose the condition that the original DE-NFG
          $ \sfN $ is a strict-sense DE-NFG.)

  \end{enumerate}

\end{proposition}

\begin{proof}
  See Appendix~\ref{apx:LCT DENFGs}.
\end{proof}

Proposition~\ref{prop:DENFG:LCT:1} shows that most properties of the LCT for DE-NFGs extend those established for S-NFGs. In particular, Property~\ref{LCT on strict sense DENFG results in weak sense DENFG} parallels the corresponding S-NFG property discussed in Item~\ref{remk: LCT on snfg results an NFG with negative fun nodes} of Remark~\ref{sec:LCT:rem:1}.

\section{The Average \texorpdfstring{$M$}{}-Cover}\label{sec:symmetric-subspace transform}
Let $\graphN$ be an S-NFG or a DE-NFG. In
Definition~\ref{sec:GraCov:def:2}, the degree-$M$ Bethe
partition function of $\graphN$ is defined to be
\begin{align*}
  \ZBM(\graphN)
  = \sqrt[M]{
      \Bigl\langle
      Z\bigl( \hgraphN \bigr)
      \Bigr\rangle_{ \! \hgraphN \in \hat{\set{N}}_{M}}
    }.
\end{align*}
Our
particular interest lies in the limit superior of this quantity as
$M \to \infty$, as stated in Conjecture~\ref{sec:GraCov:conj:1}. In this section, we reformulate
$\bigl( \ZBM(\graphN) \bigr)^{\!M}$ as a partition function of an NFG that represents the average $M$-cover of the considered S-NFG or DE-NFG.

The developments in this section were also motivated in part by the results presented in~\cite{Vontobel:16:1}.

\begin{figure}[t]
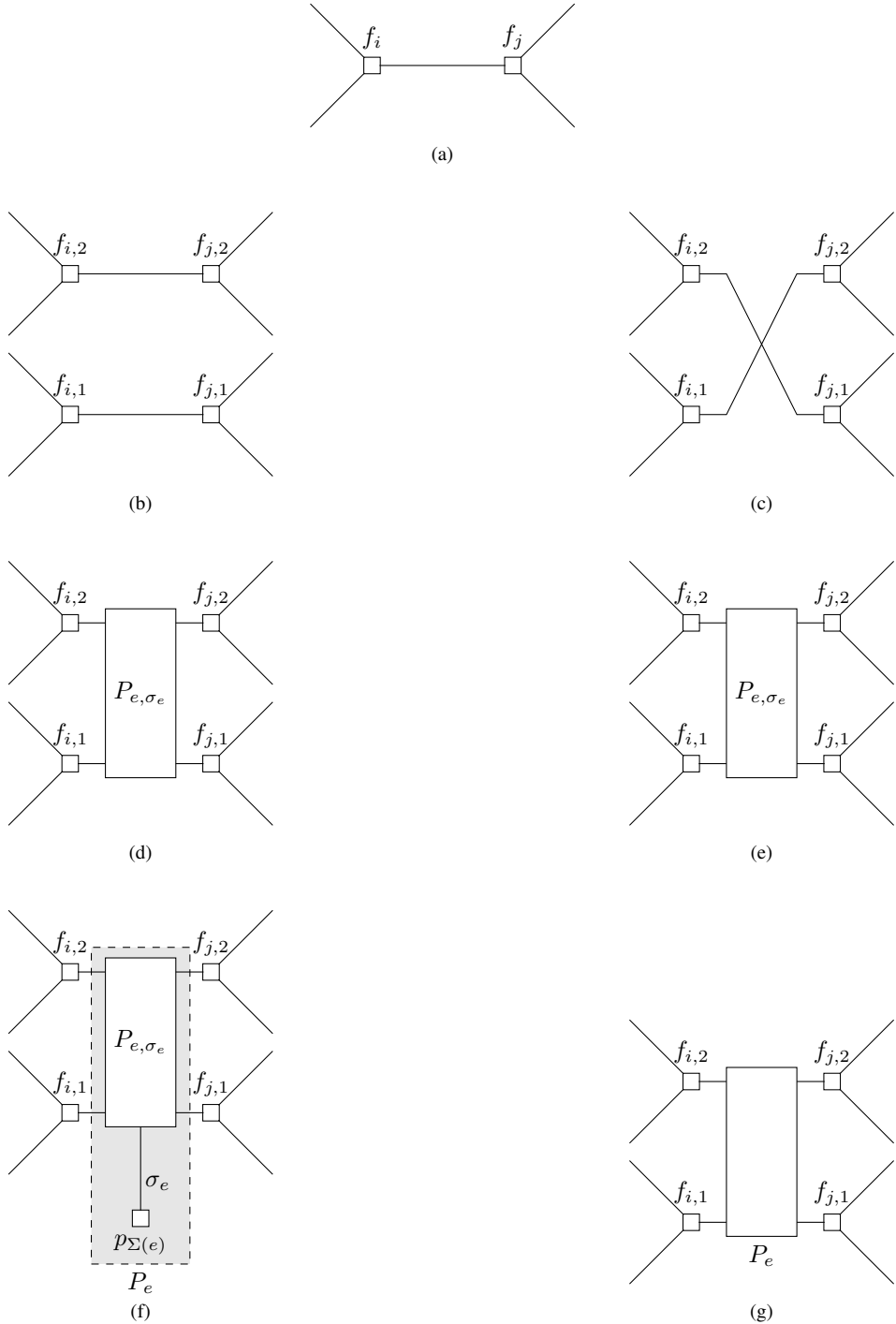

  \captionsetup{skip=0.1cm}
  \captionsetup{aboveskip=0pt}
  \subfloat[\label{sec:SST:fig:3}]{
    \begin{minipage}{0.45\textwidth}
      \centering
      \scalebox{\figscale}{%
        \begin{tikzpicture}
          \input{figures/head_files_figs.tex}
          \input{figures/sst_examples/snfg/length.tex}
          \input{figures/sst_examples/snfg/pic_1.tex}
        \end{tikzpicture}}
    \end{minipage}
  }\\
  \subfloat[\label{sec:SST:fig:4}]{
    \begin{minipage}[t]{0.225\textwidth}
      \centering
      \scalebox{\figscale}{%
        \begin{tikzpicture}
          \input{figures/head_files_figs.tex}
          \input{figures/sst_examples/snfg/length.tex}
          \input{figures/sst_examples/snfg/background_nodes_lines.tex}
          \input{figures/sst_examples/snfg/lines_not_permuted.tex}
        \end{tikzpicture}}
    \end{minipage}
  }
  \subfloat[\label{sec:SST:fig:5}]{
    \begin{minipage}[t]{0.225\textwidth}
      \centering
      \scalebox{\figscale}{%
        \begin{tikzpicture}
          \input{figures/head_files_figs.tex}
          \input{figures/sst_examples/snfg/length.tex}
          \input{figures/sst_examples/snfg/background_nodes_lines.tex}
          \begin{pgfonlayer}{background}
            \foreach \x in {0,1}{
                \coordinate (f7\x) at (-1*0.5*\ldis,\x*2*\ldis) ;
                \coordinate (f8\x) at (0.5*\ldis,\x*2*\ldis) ;
              }
            \draw[]
            (f10) -- (f70) -- (f81) -- (f21)
            (f11) -- (f71) -- (f80) -- (f20);
          \end{pgfonlayer}
        \end{tikzpicture}}
    \end{minipage}
  }\\
  \subfloat[\label{sec:SST:fig:6}]{
    \begin{minipage}[t]{0.225\textwidth}
      \centering
      \scalebox{\figscale}{%
        \begin{tikzpicture}
          \input{figures/head_files_figs.tex}
          \input{figures/sst_examples/snfg/length.tex}
          \input{figures/sst_examples/snfg/background_nodes_lines.tex}
          \input{figures/sst_examples/snfg/lines_not_permuted.tex}
          \input{figures/sst_examples/snfg/node_pe_sige.tex}
        \end{tikzpicture}}
    \end{minipage}
  }
  \subfloat[\label{sec:SST:fig:7}]{
    \begin{minipage}[t]{0.225\textwidth}
      \centering
      \scalebox{\figscale}{%
        \begin{tikzpicture}
          \input{figures/head_files_figs.tex}
          \input{figures/sst_examples/snfg/length.tex}
          \input{figures/sst_examples/snfg/background_nodes_lines.tex}
          \input{figures/sst_examples/snfg/lines_not_permuted.tex}
          \input{figures/sst_examples/snfg/node_pe_sige.tex}
        \end{tikzpicture}}
    \end{minipage}
  }\\
  \subfloat[\label{sec:SST:fig:8}]{
    \begin{minipage}[t]{0.225\textwidth}
      \centering
      \scalebox{\figscale}{%
        \begin{tikzpicture}
          \input{figures/head_files_figs.tex}
          \input{figures/sst_examples/snfg/length.tex}
          \input{figures/sst_examples/snfg/background_nodes_lines.tex}
          \input{figures/sst_examples/snfg/lines_not_permuted.tex}
          \input{figures/sst_examples/snfg/node_pe_sige.tex}
          \begin{pgfonlayer}{background}
            \node[state_dash6] (db1) at (0,0.1*\ldis) [label=below:$P_{e}$] {};
            \node[] (var1) at (-0.18*\ldis,-1*\ldis) [label=right: $\sigma_{e}$] {};
            \node[state] (Psige) at (0,-1.5*\ldis) [label=below: $p_{\Sigma(e)}$] {};
            \node[] (Psige1) at (0,-2.7*\ldis) [] {};
          \end{pgfonlayer}
          \input{figures/sst_examples/snfg/otb_back_nodes_lines.tex}
        \end{tikzpicture}}
    \end{minipage}
  }
  \subfloat[\label{sec:SST:fig:9}]{
    \begin{minipage}[t]{0.225\textwidth}
      \centering
      \scalebox{\figscale}{%
        \begin{tikzpicture}
          \input{figures/head_files_figs.tex}
          \input{figures/sst_examples/snfg/length.tex}
          \input{figures/sst_examples/snfg/background_nodes_lines.tex}
          \input{figures/sst_examples/snfg/lines_not_permuted.tex}
          \begin{pgfonlayer}{above}
            \node[state_large] (db1) at (0,1*\ldis) [label=below:$P_{e}$] {};
          \end{pgfonlayer}
        \end{tikzpicture}}
    \end{minipage}
  }
  \caption{Partial S-NFGs illustrating the construction of the average degree $M$-cover. (See main text for details.)\label{sec:SST:fig:1}}
\end{figure}

The average $M$-cover construction developed in this section applies to
  general S-NFGs. For clarity, we will illustrate the construction for the
  special case $M = 2$ and $\set{X}_e = \{ 0, 1 \}$, $e \in \setEfull$, using
  the example S-NFG fragment shown in Fig.~\ref{sec:SST:fig:3}.

\subsection{Specifying an \texorpdfstring{$M$}{}-Cover of an S-NFG}

The following construction process defines an $M$-cover of the original S-NFG $\graphN$.

\begin{definition}
  \label{def:graph:cover:construction:1}

  The construction process for an $M$-cover of $\graphN$ is outlined as follows (see also
  Definition~\ref{sec:GraCov:def:1}).
  \begin{enumerate}

    \item For each edge $e \in \setEfull$, we specify a permutation
          $\sigma_e \in \set{S}_M$. We collect all these permutations into a vector $\vsigma = (\sigma_e)_{e \in \setEfull} \in \set{S}_M^{|\setEfull|}$.

    \item\label{def:graph:cover:construction:1:item:1} For each function node $f \in \set{F}$, we introduce $M$ copies of $f$. Each copy
          of function node $f$ is associated with a collection of sockets and variables. Specifically,
          if $e \in \setpf$, then the $m$-th copy of $f$ has a socket with
          associated variable denoted by $x_{e,f,m}$.

    \item For every edge $e = (f_i, f_{j}) \in \setEfull$, we specify how the
          sockets corresponding to $(x_{\efi,m})_{m \in [M]}$ are connected to the
          sockets corresponding to $(x_{\efj,m})_{m \in [M]}$.
          The connections are determined based on the chosen permutation $\sigma_e \in \set{S}_M$. Specifically, for each $m \in [M]$, the socket corresponding to $x_{\efi,m}$ is connected to the socket corresponding to $x_{\efj,\sigma_{e}(m)}$.

    \item The resulting $M$-cover is denoted by $\hgraphN_{M,\vsigma}$.
          \edefinition
  \end{enumerate}
  Possible connections are illustrated in Figs.~\ref{sec:SST:fig:4} and~\ref{sec:SST:fig:5}, considering that $|\set{S}_M| = 2! = 2$ for $M = 2$.
\end{definition}

For notational simplicity, let $\vSigma_{M} \defeq \vSigma \defeq \bigl( \Sigmae \bigr)_{e}$ be the random vector\footnote{We use $\Sigmae$ instead of $\Sigma_e$ for these random
  variables to distinguish them from the symbol used to denote a sum
  over all edges.} \finaltext{whose components are independent random permutations,
each uniformly distributed over $\set{S}_M$.} The probability distributions of $ \Sigmae $ and $ \vSigma $ are given by, respectively,
\begin{align}
  p_{\Sigmae}(\sigma_e) = \frac{1}{|\set{S}_M|} = \frac{1}{M!}, \qquad e \in \setEfull,\,
  \sigma_e \in \set{S}_M, \label{sec:sst:eqn:31:pre:1} \\
  p_{\vSigma}(\vsigma) = \prod_{e}p_{\Sigmae}(\sigma_e)
  = \frac{1}{(M!)^{|\setEfull|}}, \qquad
  \vsigma \in \set{S}_M^{|\setEfull|}
  .\label{sec:sst:eqn:31}
\end{align}
\finaltext{\finaltext{With this,} the random graph $\hgraphN_{M,\vSigma}$ is a uniformly
sampled $M$-cover of $\graphN$.} We rewrite the expression
$\bigl( \ZBM(\graphN) \bigr)^{\! M}$ as
\begin{align}
  \bigl( \ZBM(\graphN) \bigr)^{\! M}
   & = \Bigl\langle
  Z\bigl( \hgraphN \bigr)
  \Bigr\rangle_{ \! \hgraphN \in \hat{\set{N}}_{M}}
  = \sum_{\vsigma}
  p_{\vSigma}(\vsigma)
  \cdot
  Z\bigl( \hgraphN_{M,\vsigma} \bigr), \label{sec:sst:eqn:15}
\end{align}
where $ \sum_{\vsigma} $ denotes $ \sum_{\vsigma \in \set{S}_M^{|\setEfull|}} $.

\subsection{Replacing an \texorpdfstring{$M$}{}-Cover S-NFG
  with another S-NFG}

We now replace each $M$-cover $\hgraphN_{M,\vsigma}$ with an S-NFG
  denoted by $\hgraphPsig$ that satisfies
\begin{align}
  Z\bigl( \hgraphPsig \bigr)
  = Z\bigl( \hgraphN_{M,\vsigma} \bigr). \label{sec:sst:eqn:33}
\end{align}
Before providing the general definition of $\hgraphPsig$, we briefly
discuss this replacement in terms of the illustrated example. Instead
of working with the S-NFG in Fig.~\ref{sec:SST:fig:4}, we
work with the S-NFG in Fig.~\ref{sec:SST:fig:6}. Similarly,
instead of working with the S-NFG in
Fig.~\ref{sec:SST:fig:5}, we work with the S-NFG in
Fig.~\ref{sec:SST:fig:7}.

\begin{definition}
  \label{sec:sst:def:1}

  For a given $\vsigma \in \set{S}_M^{|\setEfull|}$, the S-NFG
  $\hgraphPsig$ is defined as follows.
  \begin{enumerate}

    \item For every $f \in \set{F}$, we introduce $M$ copies of $f$.
          Each copy of $f$ is associated with a collection of sockets and variables. Specifically, if $e \in \setpf$, the $m$-th copy of $f$ has a socket with an associated variable $x_{\ef,m}$. (This step is the same as Step~\ref{def:graph:cover:construction:1:item:1} in Definition~\ref{def:graph:cover:construction:1}.)

    \item For every edge $e = (f_{i}, f_{j}) \in \setEfull$, we introduce a function node
          $P_{e,\sigma_e}$. This node is connected to the sockets associated with the collection of
          variables $(x_{\efi,m})_{m \in [M]}$ and to the sockets associated with the collection of variables
          $(x_{\efj,m})_{m \in [M]}$. The local function $P_{e,\sigma_e}: \set{X}_e^{M} \times \set{X}_e^{M}  \to  \{0,1\} $ is
          defined to be
          \begin{align*}
            P_{e,\sigma_e}\bigl( \vx_{\efi,[M]}, \vx_{\efj,[M]} \bigr)
             & \defeq
            \prod_{m \in [M]}
            \bigl[
              x_{\efi,m} \! = \! x_{\efj,\sigma_e(m)}
              \bigr],
          \end{align*}
          where $\vx_{\efi,[M]}, \vx_{\efj,[M]} \in \set{X}_e^M$ and
          \begin{align*}
            \vx_{\ef,[M]}
             & \defeq
               (x_{\ef,m})_{m \in [M]} \in  \set{X}_e^M, \qquad
            f \in \{f_{i}, f_{j}\}.
          \end{align*}
          The function $P_{e,\sigma_e}$ is an indicator function because
            it takes on only the values zero and one.\edefinition

  \end{enumerate}
\end{definition}

In the case of Fig.~\ref{sec:SST:fig:4}, the permutation
$\sigma_e$ satisfies $\sigma_e(1) = 1$ and $\sigma_e(2) = 2$. Therefore, the corresponding local function $P_{e,\sigma_e}$ is given by
\begin{align*}
  P_{e,\sigma_e}\bigl( \vx_{\efi,[M]}, \vx_{\efj,[M]} \bigr)
   & =
  [x_{\efi,1} \! = \! x_{\efj,1}]
  \cdot
  [x_{\efi,2} \! = \! x_{\efj,2}].
\end{align*}
This function has a matrix representation whose rows are indexed by
  $(x_{\efi,1}, x_{\efi,2}) $ and whose columns are indexed by
  $(x_{\efj,1}, x_{\efj,2})$. Ordering the possible vectors for each pair as
  $ (0,0), \, (0,1), \, (1,0), \, (1,1) $, we obtain
\begin{align*}
  \matr{P}_{e,\sigma_e}
   & = \begin{pmatrix}
         1 & 0 & 0 & 0 \\
         0 & 1 & 0 & 0 \\
         0 & 0 & 1 & 0 \\
         0 & 0 & 0 & 1
       \end{pmatrix}.
\end{align*}
The corresponding S-NFG is shown in Fig.~\ref{sec:SST:fig:6}.

On the other hand, in the case of Fig.~\ref{sec:SST:fig:5}, the
permutation $\sigma_e$ satisfies $\sigma_e(1) = 2$ and $\sigma_e(2) = 1$.  Thus, the corresponding local function $P_{e,\sigma_e}$ is given by
\begin{align*}
  P_{e,\sigma_e}\bigl( \vx_{\efi,[M]}, \vx_{\efj,[M]} \bigr)
   & =
  [x_{\efi,1} \! = \! x_{\efj,2}]
  \cdot
  [x_{\efi,2} \! = \! x_{\efj,1}].
\end{align*}
Representing this function in matrix form with the row and column indexing as above, we obtain
\begin{align*}
  \matr{P}_{e,\sigma_e}
   & = \begin{pmatrix}
         1 & 0 & 0 & 0 \\
         0 & 0 & 1 & 0 \\
         0 & 1 & 0 & 0 \\
         0 & 0 & 0 & 1
       \end{pmatrix}.
\end{align*}
The corresponding S-NFG is shown in Fig.~\ref{sec:SST:fig:7}.

\subsection{An NFG Representing the Average \texorpdfstring{$M$}{}-Cover
  of an S-NFG}
\label{sec:sst:NFG:average:cover:1}

The next step is to construct an NFG $\hgraphNavg$ whose partition function
equals
$\sum_{\vsigma} p_{\vSigma}(\vsigma) \cdot Z\bigl( \hgraphN_{M,\vsigma}
  \bigr)$. We do this as follows. (See Fig.~\ref{sec:SST:fig:8} for an
illustration.)

\begin{definition}
  \label{sec:sst:def:2}

  The construction of $\hgraphNavg$ is essentially the same as the
  construction of $\hgraphPsig$ for any $\vsigma$ in Definition~\ref{sec:sst:def:1}, but with the
  following modifications and additions.
  Namely, for every $e \in \setEfull$, the
  permutation $\sigma_e$ is not fixed, but a variable. With this, the function
  $P_{e,\sigma_e}\bigl( \vx_{\efi,[M]}, \vx_{\efj,[M]} \bigr)$ is now considered to be a function
  not only of $\vx_{\efi,[M]}$ and $\vx_{\efj,[M]}$, but also of
  $\sigma_e$. Graphically, this is done as follows.
  \begin{itemize}

    \item For every $e \in \setEfull$, we draw a function node $p_{\Sigmae}$.

    \item For every $e \in \setEfull$, we draw an edge that connects
          $P_{e,\sigma_e}$ and $p_{\Sigmae}$ and that represents the variable
          $\sigma_e$. Recall that $ p_{\Sigmae}( \sigma_{e} ) = (M!)^{-1} $ for
          all $ e \in \setEfull $ and $ \sigma_{e} \in \set{S}_{M} $, as defined
          in~\eqref{sec:sst:eqn:31:pre:1}.\edefinition

  \end{itemize}
\end{definition}

The construction here will be used in the proof of
  Theorem~\ref{sec:CheckCon:thm:1}, which averages the partition functions of
  the $M$-covers of the LCT DE-NFG, whose local functions are complex
  valued. The averaging therefore acts on the transformed DE-NFG rather than
  on the original graph, and it must handle complex-valued local functions
  rather than real-valued local functions. Note that a related averaging
  operation over finite covers appears in~\cite{Csikvari2022}. 

\begin{lemma}
  \label{sec:sst:lem:1}

  The partition function of $\hgraphNavg$ satisfies
  \begin{align*}
    Z\bigl( \hgraphNavg \bigr)
     & = \sum_{\vsigma}
    p_{\vSigma}(\vsigma)
    \cdot Z\bigl( \hgraphPsig \bigr)
    = \bigl( \ZBM(\graphN) \bigr)^{\! M}.
  \end{align*}
\end{lemma}

\begin{proof}
  The first equality  follows directly from the definition of $\hgraphNavg$ in Definition~\ref{sec:sst:def:2}, and the last equality follows from the expressions in~\eqref{sec:sst:eqn:15} and~\eqref{sec:sst:eqn:33}.
\end{proof}

Fix an arbitrary edge $e = (f_{i}, f_{j}) \in \setEfull$.

\begin{definition}\label{sec:sst:def:4}
  We define
  \begin{align*}
    \!P_e\bigl( \vx_{\efi,[M]}, \vx_{\efj,[M]} \bigr)
    \!\! & \defeq\!\!\!\!\!
    \sum_{\sigma_e \in \set{S}_M}
    \!\!\!\!p_{\Sigmae}(\sigma_e)
    \!\cdot\!
    P_{e,\sigma_e}\bigl( \vx_{\efi,[M]}, \vx_{\efj,[M]} \bigr)
    \\&= \frac{1}{M!}
    \cdot
    \sum_{\sigma_e \in \set{S}_M} P_{e,\sigma_e}\bigl( \vx_{\efi,[M]}, \vx_{\efj,[M]} \bigr),
  \end{align*}
  where the last equality follows from $p_{\Sigmae}(\sigma_e) = 1/M!$ for all
  $\sigma_e \in \set{S}_M$. Also, we define
  \begin{align*}
    \vx_{[M]}       & \defeq ( x_{e,f,m} )_{f \in \setF,\,e \in \setpf,\,m \in [M]}
    \in \setx^{2M},
    \nonumber                                                                       \\
    \vx_{\setpff,m} & \defeq ( x_{e,f,m} )_{e \in \setpf} \in \set{X}_{\setpf},
    \qquad \qquad \quad f \in \setF, \ m \in [M],
  \end{align*}
  where the set $ \setx^{2M} $ comes from the fact that the collection $ \vx_{[M]} $
  can be rewritten as
  $ \vx_{[M]} = ( x_{e,f_{i},m},x_{e,f_{j},m} )_{e =(f_{i},f_{j}) \in \setEfull,m \in [M]} $.
  When unambiguous, we use the shorthand $ \sum_{\vx_{[M]}} $ instead of
    $ \sum_{\vx_{[M]} \in \setx^{2M}} $.
  \edefinition
\end{definition}

In the case of Fig.~\ref{sec:SST:fig:9}, the function $P_e$ corresponds to the exterior function of the dashed box in Fig.~\ref{sec:SST:fig:8}. This function is obtained through the closing-the-box operation~\cite{Loeliger2004}. For the specific illustrated example in Fig.~\ref{sec:SST:fig:8}, the
matrix representation of $ P_e $ is given by
\begin{align}
  \matr{P}_e
   & \defeq
  \sum_{\sigma_{e} \in \set{S}_2}
  \frac{1}{2!}
  \cdot
  \matr{P}_{e,\sigma_e}
  = \begin{pmatrix}
      1 & 0   & 0   & 0 \\
      0 & 1/2 & 1/2 & 0 \\
      0 & 1/2 & 1/2 & 0 \\
      0 & 0   & 0   & 1
    \end{pmatrix}
  \label{sec:sst:eqn:1},
\end{align}
where the rows and columns of $\matr{P}_e$ are indexed and ordered as those of $\matr{P}_{e,\sigma_e}$.

\begin{lemma}\label{sec:sst:lem:6}
  The partition function $ Z\bigl( \hgraphNavg \bigr) $ equals
  \begin{align*}
    Z\bigl( \hgraphNavg \bigr) & \!\! =\!\! \sum_{\vx_{[M]}}
    \Biggl( \prod_{m,f} f(\vx_{\setpff,m}) \!\! \Biggr)    \!\cdot \! \prod_{e} \! P_e\bigl( \vx_{\efi,[M]}, \vx_{\efj,[M]} \bigr).
  \end{align*}
\end{lemma}
\begin{proof}
  See Appendix~\ref{apx:alternative expression of ZBM by Pe}.
\end{proof}

{
Now we can express the $M$-th power of the degree-$M$ Bethe partition function as the partition function of the average $M$-cover.
\begin{theorem}
  \label{thm: expression of Z for degree M Bethe partition function}
  The degree-$M$ Bethe partition function satisfies
  \begin{align*}
    \bigl( \ZBM(\graphN) \bigr)^{\! M}
     & \!\!\!\! = \!\!
    \sum_{\vx_{[M]}}
    \!\Biggl( \prod_{m ,f} f(\vx_{\setpff,m}) \!\! \Biggr)
    \!\!\cdot \!\! \prod_{e} \! P_e\bigl( \vx_{\efi,[M]}, \vx_{\efj,[M]} \bigr).
  \end{align*}
\end{theorem}
\begin{proof}
  This follows from Lemmas~\ref{sec:sst:lem:1} and~\ref{sec:sst:lem:6}.
\end{proof}
}

We use types and type classes\footnote{See Cover and
    Thomas~\cite{Cover:Thomas:06:1} for a detailed exposition; our notation
    differs.} to characterize the function $P_e$.

{
\begin{definition}\label{sec:sst:def:6}
  We introduce the following objects.\footnote{For brevity, we use shorthand $x$ instead of $\xe$.}
  
  \begin{itemize}

    \item The type
          $\vt_e(\vv_e) \defeq \bigl( t_{e,x}(\vv_e) \bigr)_{\! x \in \set{X}_e} \in \pmf{\setxe}$ of a
          vector $\vv_e \defeq (v_{e,m})_{m \in [M]} \in \set{X}_e^M$ is defined to be
          \begin{align*}
            t_{e,x}(\vv_e)
             & \defeq
            \frac{1}{M}
            \cdot
            \bigl| \hskip0.5mm
            \left\{
            m \in [M]
            \ \middle| \
            v_{e,m} = x
            \right\} \hskip0.5mm
            \bigr|,
            \qquad x \in \set{X}_e.
          \end{align*}

    \item Let $\set{B}_{\set{X}_e^M}$ be the set of possible types of vectors of
          length $M$ over $\set{X}_e$, \ie,
          \begin{align*}
            \set{B}_{\set{X}_e^M}
             & \defeq
            \bigl\{
            \vt_e \in \pmf{\setxe}
            \bigm|
            \exists\, \vv_e \in \set{X}_e^M
            \text{ s.t. } \vt(\vv_e) = \vt_e
            \bigr\}.
          \end{align*}
          Let $ \set{B}_{\set{X}^M} $ be the Cartesian product $ \prod_{e} \set{B}_{\set{X}_e^M} $.

    \item Let $\vt_e \in \set{B}_{\set{X}_e^M}$. Then the type class of $\vt_e$ is defined to be the set
          \begin{align*}
            \set{T}_{e,\vt_e}
             & \defeq
            \bigl\{
            \vv_e \in \set{X}_e^M
            \bigm|
            \vt_e(\vv_e) = \vt_e
            \bigr\}. \tag*{\ensuremath{\triangle}}
          \end{align*}

  \end{itemize}
\end{definition}
}

{
\begin{lemma}
  \label{sec:sst:lem:2}

  For every $\vt_e \in \set{B}_{\set{X}_e^M}$, it holds that
  
  \begin{align*}
    \bigl| \set{B}_{\set{X}_e^M} \bigr|
     & = \binom{|\set{X}_e| + M - 1}{M},                                               \qquad
    |\set{T}_{e,\vt_e}|
    = \frac{M!}{ \prod\limits_{x \in \set{X}_e} \bigl( (M \cdot t_{e,x}) ! \bigr) }.
  \end{align*}
\end{lemma}
}

\begin{proof}
  These expressions can be derived from standard combinatorial results.
\end{proof}

For the illustrated example, the number of possible types is
$\bigl| \set{B}_{\set{X}_e^M} \bigr| = \binom{2 + 2 - 1}{2} = 3$ with
$\set{B}_{\set{X}_e^M} = \bigl\{ (1,0), \, (1/2,1/2), \, (0,1) \bigr\} $. The
corresponding type classes have the following sizes:
$|\set{T}_{e,(1,0)}| = 1$, $|\set{T}_{e,(1/2,1/2)}| = 2$, and
$|\set{T}_{e,(0,1)}| = 1$.

{
\begin{lemma}\label{sec:sst:lem:4}
  The function $P_e$ for $ e = (f_{i}, f_{j}) $ satisfies
  
  \begin{align*}
    P_e\bigl( \vx_{\efi,[M]}, \vx_{\efj,[M]} \bigr)
     & = \begin{cases}
           |\set{T}_{e,\vt_e}|^{-1}
            &
           \begin{aligned}[t]
             \vt_e & = \vt_e\bigl(\vx_{\efi,[M]}\bigr)   \\
                   & = \vt_e\bigl( \vx_{\efj,[M]} \bigr)
           \end{aligned} \\
           0
            & \text{otherwise}
         \end{cases}.
  \end{align*}
\end{lemma}
}

\begin{proof}
  This result follows from the definition of the function $P_{e,\sigma_e}$ in
  Definition~\ref{sec:sst:def:1}, along with the properties of the uniform distribution $p_{\Sigmae}$ and the symmetry of $ P_e $, as shown in Definition~\ref{sec:sst:def:4}.
\end{proof}

For the illustrated example, the expression in the above lemma is
corroborated by the matrix in~\eqref{sec:sst:eqn:1}. Here, the matrix is reproduced with horizontal and vertical lines to highlight the three type classes of the row and column indices, respectively:
\begin{align*}
  \matr{P}_e
   & = \left(
  \begin{array}{c|cc|c}
    1 & 0   & 0   & 0 \\
    \hline
    0 & 1/2 & 1/2 & 0 \\
    0 & 1/2 & 1/2 & 0 \\
    \hline
    0 & 0   & 0   & 1
  \end{array}
  \right).
\end{align*}

The matrix $\matr{P}_e$ associated with $P_e$ is known as the
symmetric-subspace projection operator (see, \eg,~\cite{Harrow2013}).

\subsection{The Average \texorpdfstring{$M$}{}-Cover of a DE-NFG}
\label{sec:sst:DENFG:1}

The average $M$-cover construction for S-NFGs extends directly to DE-NFGs.
  In an $M$-cover of a DE-NFG, the two strands of each double edge are permuted together; Definition~\ref{sec:GraCov:def:1} and Fig.~\ref{fig:Exam_Graph_cover_DE_NFG} give the definition and an example, respectively.
  The only change from the S-NFG construction is that, for
every edge $e \in \setEfull$, the alphabet $\set{X}_e$ is replaced with
the alphabet $\tset{X}_e = \set{X}_e \times \set{X}_e$.
Define the paired configuration
\begin{align*}
  \tvx_{[M]} \defeq
             (\tx_{e,f,m})_{f\in\setF,\,e\in\setpf,\,m\in[M]}
  \in \tset{X}^{2M}.
\end{align*}

Applying the proof strategy of Theorem~\ref{thm: expression of Z for degree M Bethe partition function} to an arbitrary DE-NFG $\graphN$ yields
\begin{align}
  \bigl( \ZBM(\graphN) \bigr)^{\! M}
   & \!\!\! =\!\!
  \sum_{\tvx_{[M]}}\!
  \Biggl( \prod_{m ,f} f(\tvx_{\setpff,m}) \!\! \Biggr) \!\!\cdot \! \prod_{e} \!P_e\bigl( \tvx_{\efi,[M]}, \tvx_{\efj,[M]} \bigr),
  \label{eqn: Z for degree M average cover of DENFG}
\end{align}
where
\begin{align*}
  P_e\bigl( \tvx_{\efi,[M]}, \tvx_{\efj,[M]} \bigr)
  = \begin{cases}
      |\set{T}_{e,\vt_e}|^{-1}
       &
      \begin{aligned}[t]
        \vt_e & = \vt_e\bigl(\tvx_{\efi,[M]}\bigr)   \\
              & = \vt_e\bigl( \tvx_{\efj,[M]} \bigr)
      \end{aligned} \\
      0
       & \text{otherwise}
    \end{cases}.
\end{align*}
Here $\vt_e$, $\vt_e(\cdot)$, and $\set{T}_{e,\vt_e}$ are defined as in
Definition~\ref{sec:sst:def:6}, with $\set{X}_e$ replaced by
$\tset{X}_e$; in particular, we have $\set{T}_{e,\vt_e}\subseteq\tset{X}_e^M$.

The average $M$-cover construction for S-NFGs and DE-NFGs motivated us
to introduce the symmetric-subspace transform for S-NFGs and DE-NFGs
in~\cite[Chapter 5]{Huang2024}. For any S-NFG or DE-NFG, denoted by
$ \graphN $, the symmetric-subspace transform (SST) rewrites
$ \bigl( \ZBM( \graphN ) \bigr)^{\! M} $ as an integral, offering a novel
perspective on the $M$-cover of $ \graphN $. Although the SST-style
calculations are widely employed in quantum
physics~\cite{Wood2015,Harrow2013}, the use of the SST in the factor-graph
  literature is, to our knowledge, new.

\section{The Graph-Cover Theorem for Special DE-NFGs} \label{sec:CheckCon}

This section proves Conjecture~\ref{sec:GraCov:conj:1} for a class of
  DE-NFGs that satisfy an easily checkable condition. As mentioned in the
introduction, the proof of the graph-cover theorem for S-NFGs relies heavily
on the method of types, which does not extend directly to DE-NFGs because
  the partition function of a DE-NFG adds up complex-valued terms. (See also the discussion
  in Example~\ref{example:simple:complex:sum:1}.) Our approach circumvents
this obstacle by leveraging the LCT developed in Section~\ref{sec:LCT}, which
reparameterizes the partition function of the average $M$-cover into a form
amenable to analysis via its finite graph covers.

Consider a DE-NFG $\graphN$. The setup of our approach is as follows.

Let $\vmu$ be the SPA fixed-point message vector that achieves the SPA-based Bethe partition function, \ie,
\begin{align}
  \ZBSPA(\graphN,\vmu) = \ZBSPA^{*}(\graphN) \in \sR_{>0}, \qquad
  Z_{e}(\vmu) >0, \qquad e \in \setEfull. \label{eqn:positive_ZBSPA}
\end{align}
(See also Definition~\ref{sec:DENFG:def:3} and Assumption~\ref{sec:DENFG:remk:1}.)
Based on $\vmu$, we apply the LCT to $\graphN$ and obtain the LCT NFG $ \LCT{\graphN} $.
The NFG $ \LCT{\graphN} $ has the following properties, as proven in Proposition~\ref{prop:DENFG:LCT:1}.
\begin{enumerate}

  \item The partition functions of $ \graphN $ and
        $ \LCT{\graphN} $ are equal, \ie, we have $ Z\bigl( \LCT{\graphN} \bigr) = Z(\graphN) $.

  \item For any $ M \in \sZpp $, the degree-$M$ Bethe partition functions of $ \graphN $ and
        $ \LCT{\graphN} $ are equal, \ie,
        \begin{align*}
          \!\!\!\!\!\!\!\!\ZBM(\graphN)
           & \! \overset{(a)}{=}\! \!\! \sqrt[M]{
                                          \bigl\langle
                                          Z\bigl( \hgraphN \bigr)
                                          \bigr\rangle_{ \hgraphN \in \hat{\set{N}}_{M}}
                                        }
          \!\overset{(b)}{=} \!\! \sqrt[M]{
                                    \Bigl\langle
                                    Z\bigl( \hLCTgraphN \bigr)
                                    \Bigr\rangle_{ \hLCTgraphN \in \hLCT{\set{N}}_{M} }
                                  }
          \! \overset{(c)}{=} \! \ZBM\bigl( \LCT{\graphN} \bigr).
        \end{align*}
        Step $(a)$ follows from the definition of the scalar $ \ZBM $ in
        Definition~\ref{sec:GraCov:def:2}.  Step $(b)$ follows from
        Proposition~\ref{prop:DENFG:LCT:1}. Step $(c)$ again follows from
        Definition~\ref{sec:GraCov:def:2}.

  \item The SPA fixed-point message vector $\vmu$ induces an SPA
        fixed-point message vector
        $\LCTv{\mu} \defeq \bigl( \LCTv{\mu}_{\ef} \bigr)_{ \! e \in \setpf,\,
            f \in \setF}$ for $\LCT{\graphN}$ with a simple structure
        \begin{align*}
          \LCT{\mu}_{e,f}(\LCTt{x}_{e})
           & = \bigl[ \LCTt{x}_{e} \! = \! \tzero \bigr],
          \qquad
          \LCTt{x}_{e} \in \LCTtset{X}_{e}, \,
          e \in \setpf, \, f \in \setF.
        \end{align*}
        Proposition~\ref{prop:DENFG:LCT:1} implies that
        \begin{align}
          \ZBSPA^{*}(\graphN)
          = \ZBSPA(\graphN,\vmu)
          = \ZBSPA(\LCT{\graphN}, \LCTv{\mu})
          =
          \prod_{f} \LCT{f}(\tv{0}).
          \label{sec:CheckCon:eqn:31}
        \end{align}
\end{enumerate}

For any $ M \in \sZpp $, we consider an arbitrary $M$-cover of $ \LCT{\graphN} $, which is denoted by $ \hLCTgraphN $. This graph cover is specified by a collection of permutations
$ \vsigma = (\sigma_e)_{e \in \setEfull}
  \in \set{S}^{|\setEfull|}_{M} $.
The following vectors
\begin{align*}
  \hat{\LCTv{\mu}}_{(e,m),(f_{i},m)}
   & \defeq \bigl( \hat{\LCT{\mu}}_{(e,m),(f_{i},m)}(\LCTt{x}_{e})
  \bigr)_{\! \LCTt{x}_{e} \in \LCTtset{X}_{e}},                                \\
  \hat{\LCTv{\mu}}_{(e,m),(f_{j},\sigma_{e}(m))}
   & \defeq \bigl( \hat{\LCT{\mu}}_{(e,m),(f_{j},\sigma_{e}(m))}(\LCTt{x}_{e})
  \bigr)_{\! \LCTt{x}_{e} \in \LCTtset{X}_{e}},
\end{align*}
where $m \in [M]$ and $e = (f_{i},f_{j}) \in \setEfull,$ and
whose elements are specified by
\begin{align*}
  \hat{\LCT{\mu}}_{(e,m),(f_{i},m)}(\LCTt{x}_{e})
   & =
  \hat{\LCT{\mu}}_{(e,m),(f_{j},\sigma_{e}(m))}(\LCTt{x}_{e}) \! = \bigl[ \LCTt{x}_{e} \! = \! \tzero \bigr],
  \quad
  \LCTt{x}_{e} \in \LCTtset{X}_{e},
\end{align*}
form an SPA fixed-point message vector for $ \hLCTgraphN $.
For each pair $(f,m') \in \{ (f_{i},m), (f_{j}, \sigma_{e}(m)) \}$, each integer $ m \in [M] $, and each edge $ e=(f_{i},f_{j}) \in \setEfull $, the vector $ \hat{\LCTv{\mu}}_{(e,m),(f,m')} $ corresponds to the SPA fixed-point message from the edge $ (e,m) $ to the \finaltext{function node $ (f,m') $} in $ \hLCTgraphN $.
Indeed, each lifted node of the $M$-cover $\hLCTgraphN$ has the same incident-edge structure and local function as its counterpart in $\LCT{\graphN}$. Therefore, repeating each component of the SPA fixed-point message vector $\LCTv{\mu}$ on its $M$ lifted copies yields the vector $\hat{\LCTv{\mu}}$ defined above.\footnote{This local equivalence originally motivated graph-cover analysis of message-passing iterative decoders~\cite{Koetter2003,Koetter2007}.}
By Proposition~\ref{prop:DENFG:LCT:1}, the all-zero configuration contributes $ \ZBSPA(\hLCTgraphN, \hat{\LCTv{\mu}}) = \bigl( \ZBSPA^{*}(\graphN) \bigr)^{M}$ to $Z\bigl( \hLCTgraphN \bigr)$, whereas every other configuration \finaltext{with a nonzero contribution} corresponds to a generalized loop in $\hLCTgraphN$.
  Constructing a graph cover before or after applying the LCT gives the same partition function $Z\bigl( \hLCTgraphN \bigr)$.

To prove the graph-cover theorem for $\sfN$ (equivalently, Conjecture~\ref{sec:GraCov:conj:1} for $\sfN$), we analyze the average $M$-cover $ \hat{\LCT{\graphN}}_{M,\mathrm{avg}} $, whose partition function is
$ Z(\hat{\LCT{\graphN}}_{M,\mathrm{avg}}) =  \bigl\langle
  Z\bigl( \hLCTgraphN \bigr)
  \bigr\rangle_{ \hLCTgraphN \in \hLCT{\set{N}}_{M} } $, where $\hLCT{\set{N}}_{M}$ is the set of all $M$-covers of $ \LCT{\graphN} $. We then evaluate
$ \ZBM\bigl( \LCT{\graphN} \bigr) = \sqrt[M]{ Z(\hat{\LCT{\graphN}}_{M,\mathrm{avg}}) } $ as $ M \to \infty $.
We next give a checkable sufficient condition for the limit $ \lim\limits_{M \to \infty} \ZBM\bigl( \LCT{\graphN} \bigr) = \ZBSPA^{*}(\graphN)$, and hence the graph-cover theorem for $\sfN$, to hold.
\begin{theorem}\label{sec:CheckCon:thm:1}
  Consider a DE-NFG $ \graphN $, along with the associated LCT NFG $ \LCT{\graphN} $. If
  the checkable inequality
  \begin{align}
    \ZBSPA^{*}(\graphN)
     & >
    \frac{2}{3}
    \cdot
    \prod_{f}
    \left(
    \sum_{\LCTtv{x}_{\setpf}}
    \bigl| \LCT{f}( \LCTtv{x}_{\setpf} ) \bigr|
    \right)
    \label{sec:CheckCon:eqn:70}
  \end{align}
  holds,
  then we have
  \begin{align*}
    \lim_{M \to \infty}
    \ZBM(\graphN)
    = \lim_{M \to \infty}
    \ZBM(\LCT{\graphN})
    = \ZBSPA^{*}(\graphN).
  \end{align*}
\end{theorem}
\begin{proof}
  See Appendix~\ref{apx:check_graph_thm}.
\end{proof}

Theorem~\ref{sec:CheckCon:thm:1} provides a combinatorial characterization of the SPA-based Bethe partition function for a certain class of DE-NFGs. This result elevates the SPA-based Bethe approximation for this class of DE-NFGs from a heuristic to a quantity with a rigorous combinatorial interpretation, mirroring the established understanding for S-NFGs in the classical case. Beyond the theorem, the proof shows how the generalized LCT reorganizes the average-cover partition function around its all-zero contribution, thereby making the correction terms tractable. The transform is also of independent interest: it is the first explicit LCT for non-binary alphabets that admits \finaltext{complex-valued SPA fixed-point messages with possibly zero-valued components}. The condition in~\eqref{sec:CheckCon:eqn:70}, while technical, is the critical element that makes this proof possible, and its role warrants a detailed discussion.

\subsection{Discussion of the Condition in \texorpdfstring{Theorem~\ref{sec:CheckCon:thm:1}}{Theorem 46}}

We make three remarks on the condition~\eqref{sec:CheckCon:eqn:70}.

  \emph{An equivalent, per-node form.} By Proposition~\ref{prop:DENFG:LCT:1},
  the LCT places the SPA-based Bethe partition function at the all-zero
  configuration, where $\LCT{g}(\tv{0}) = \ZBSPA^{*}(\graphN)$, while every other
  configuration contributes a complex-valued loop-correction term. Using
  $\ZBSPA^{*}(\graphN) = \prod_{f} \LCT{f}(\tv{0})$
  from~\eqref{sec:CheckCon:eqn:31}, with $\LCT{f}(\tv{0}) \in \sR_{>0}$,
  condition~\eqref{sec:CheckCon:eqn:70} is equivalent to
  \begin{equation}
    \prod_{f} \, ( 1 + L_f ) < \frac{3}{2}, \quad \text{with} \quad
    L_f
    \defeq
    \sum_{\LCTtv{x}_{\setpf} \neq \tv{0}} \,
    \Biggl|
    \frac{\LCT{f}(\LCTtv{x}_{\setpf})}
    {\LCT{f}(\tv{0})}
    \Biggr|
    \ge 0,
    \ f \in \set{F}.
      \label{eq:per-node-condition}
  \end{equation}
  We call $L_f$ the \emph{per-node relative loop weight}. The condition
  bounds their combined contribution.

  \emph{Role in the proof.} This dominance bound \finaltext{keeps the ratio of the average $M$-cover partition function to the leading term}
  $\bigl(\ZBSPA^{*}(\graphN)\bigr)^M$ between positive constants independent
  of $M$. The ratio's $M$-th root therefore converges to one, yielding the limit in
  Theorem~\ref{sec:CheckCon:thm:1} despite the numerical sign
  problem~\cite{LohJr1990} (see Appendix~\ref{apx:check_graph_thm}).

  \emph{Scope.} \finaltext{The condition holds when the per-node relative loop weights
  $L_f$ are small, as in the near-identity single-cycle example of
  Section~\ref{sec:extra_example_thm} for sufficiently small
  $|\epsilon|$.\footnote{\finaltext{After unit-trace normalization, the local Choi
    matrix in that example is $\matr{I}/4+\epsilon\matr{J}/4$.
    Small $|\epsilon|$ therefore describes weak local correlations
    around the maximally mixed product state $\matr{I}/4$.}}} The condition is sufficient
  but not necessary: the numerical results in Examples~\ref{sec:GraCov:exp:1}
  and~\ref{sec:GraCov:exp:2} indicate that the graph-cover theorem should hold
  more broadly, and proving Conjecture~\ref{sec:GraCov:conj:1} in general
  remains a key open problem, discussed in the next section.

\subsection{Checking the Condition for DE-NFGs Consisting of a Single
    Cycle}
  \label{sec:extra_example_thm}

  By the equivalent form established in~\eqref{eq:per-node-condition}, verifying
  condition~\eqref{sec:CheckCon:eqn:70} amounts to checking
  $\prod_{f}(1+L_f) < 3/2$. Let $\graphN$ be a DE-NFG consisting of a
  single cycle. For such a DE-NFG, all function nodes have degree two, and
  with this the SPA fixed point, the SPA-based Bethe partition function
  $\ZBSPA^{*}(\graphN)$, and the LCT local functions $\LCT{f}$ are all
  determined by the collection of Choi matrices $\{ \matr{C}_f
    \}_{f}$. Condition~\eqref{sec:CheckCon:eqn:70} therefore reduces to an
  explicit inequality in $\{ \matr{C}_f \}_{f}$, which we now make concrete.

  Specifically, consider a DE-NFG $\graphN$ that consists of a single cycle
  with three function nodes $f_1$, $f_2$, $f_3$ and three edges
  $e_1 =(f_1,f_2)$, $e_2 = (f_2,f_3)$, $e_3 = (f_3,f_1)$, where each edge
  $e \in \{e_1,e_2,e_3\}$ has the binary alphabet $\set{X}_e = \{ 0, 1
    \}$. Throughout this example, edge subscripts are interpreted modulo $3$; in particular, we have $e_0=e_3$. Suppose that every local function has the same Choi
  matrix $\matr{C}_{f_k}=\matr{C}\defeq\matr{I}+\epsilon\cdot\matr{J}$, drawn from
  the near-identity family $\matr{I}+\epsilon\cdot\matr{H}_{f_k}$ with
  $\matr{H}_{f_k}$ Hermitian and $\epsilon\in\sR$ satisfying $|\epsilon|<1$. Here $\matr{I}$ is the
  $4 \times 4$ identity matrix and $\matr{J}$ is the anti-diagonal (exchange)
  matrix
  \begin{align*}
    \matr{J} =
    \begin{pmatrix}
      0 & 0 & 0 & 1 \\
      0 & 0 & 1 & 0 \\
      0 & 1 & 0 & 0 \\
      1 & 0 & 0 & 0
    \end{pmatrix}.
  \end{align*}
  Following the Choi-matrix convention of~\eqref{sec:DENFG:eqn:6}, the
    rows and columns of $\matr{C}_{f_k}$ are indexed by
    $(x_{e_{k-1}},x_{e_k})$ and $(x'_{e_{k-1}},x'_{e_k})$, respectively, in
    the order $(0,0),(0,1),(1,0),(1,1)$. The eigenvalues of $\matr{C}_{f_k}$
    are $1\pm\epsilon$ (each with multiplicity two), so $\matr{C}_{f_k}$ is
    positive definite, and hence a valid strict-sense Choi matrix, for
    $|\epsilon|<1$. Evaluating the LCT of Definition~\ref{def:DENFG:LCT:1} at
    the SPA fixed point gives $\LCT{f_k}(\tv{0})\in\sR_{>0}$ and
    $L_{f_k}=2|\epsilon|$ for every $k\in[3]$.
  Condition~\eqref{sec:CheckCon:eqn:70} thus becomes
  $(1 + 2|\epsilon|)^{3} < 3/2$, \ie, we have $ |\epsilon| < \epsilon^{*} $, where
  $\epsilon^{*}\defeq (1/2)\cdot \bigl( (3/2)^{1/3} - 1 \bigr) \approx
    0.0724.$ Theorem~\ref{sec:CheckCon:thm:1} therefore applies to this DE-NFG
  $\sfN$ whenever $|\epsilon| < \epsilon^{*}$. The details of these
  calculations are given in Appendix~\ref{apx:single-cycle-loop-weight}.

\section{Conclusion and Open Problems}
\label{sec: conclusion and open problems}

This paper establishes a graph-cover-based characterization of the Bethe
partition function for a certain class of DE-NFGs. By proving that the limit
of the degree-$M$ Bethe partition function equals the SPA-based Bethe
partition function for DE-NFGs satisfying an easily checkable condition, we
establish a combinatorial foundation for the Bethe approximation in this
setting.  The proof relies on a generalized LCT that is explicit for
  non-binary alphabets and \finaltext{accommodates zero-valued components of SPA fixed-point
  messages} that can arise in DE-NFGs (Table~\ref{tab:lct_comparison}).

Despite this progress, several fundamental questions remain open.
\begin{itemize}
  \item \textbf{The general graph-cover conjecture.} The primary open problem is
        to prove or disprove Conjecture~\ref{sec:GraCov:conj:1}, which posits that
        the graph-cover theorem holds for all DE-NFGs, not just those satisfying the
        condition in Theorem~\ref{sec:CheckCon:thm:1}. The numerical evidence
        in Section~\ref{sec:GraCov} supports the conjecture.

  \item \textbf{Overcoming the numerical sign problem in the limit.} The
          central barrier is the numerical sign problem: in the limit $M \to \infty$
          of the average $M$-cover partition function, the sum over generalized loop
          configurations involves significant destructive interference.

  \item \textbf{The symmetric-subspace transform.} This transform,
          introduced in~\cite[Chapter 5]{Huang2024}, reformulates the average
          $M$-cover partition function as an integral and may provide suitable tools
          to analyze the complex-valued sum in the general case.

  \item \textbf{Analysis of correction terms.} The generalized LCT gives a
          loop-series expansion~\cite{Chertkov2006,Chernyak2007} of the partition
          function for DE-NFGs. A second open problem is to determine how the
          magnitude and cancellation of the correction terms in the expansion depend
          on the graph-cover degree $M$.

  \item \textbf{Weakening the condition via correlation decay.} A
          complementary route to proving Conjecture~\ref{sec:GraCov:conj:1} beyond
          the dominance condition~\eqref{sec:CheckCon:eqn:70} is to control the loop
          corrections through correlation-decay criteria of Dobrushin
          type~\cite{Dobrushin1968}, or through polymer and cluster-expansion
          methods~\cite{Kotecky1986,Friedli2017}; the challenge is to adapt such
          tools to the complex-valued LCT correction terms of a DE-NFG, \finaltext{while controlling their magnitudes and cancellations.}

  \item \textbf{The primal formulation of the Bethe free energy \finaltext{function}.}
        Extending the primal Bethe free energy \finaltext{function}
        (Definition~\ref{sec:SNFG:def:1}) from S-NFGs to DE-NFGs remains challenging
        because the complex logarithm is multi-valued.
\end{itemize}

\begin{appendices}

  {\section{Properties of \texorpdfstring{$Z(\graphN)$}{} for DE-NFGs}
  \label{apx:property of ZN}}

  We first consider the case where $\graphN$ is a strict-sense DE-NFG,
  afterwards the case where $\graphN$ is a weak-sense DE-NFG.

  Let $\graphN$ be a strict-sense DE-NFG. Fix some $f \in \setF$. Because
  $\matr{C}_f$ is a PSD matrix, the eigenvalue decomposition of $ \matr{C}_f  $ in~\eqref{sec:DENFG:eqn:9} implies that $ \lambda_{f}(\ell_f) \in \sR_{\geq 0} $ for all $\ell_f \in \set{L}_f$ and $ f \in \setF $.
  By the decomposition of $ f $ in~\eqref{sec:DENFG:eqn:5}, we obtain
  \begin{align*}
    Z(\graphN)
     & = \sum_{\tvx}
    \prod_{f \in \setF}
    f(\tvx_{\setpf})                                  \\
     & = \sum_{\tvx}
    \prod_{f \in \setF}
    \left(
    \sum_{\ell_f \in \set{L}_f}
    \lambda_{f}(\ell_f) \cdot
    u_{f}(\vx_{\setpf},\ell_f)
    \cdot
    \overline{ u_{f}(\vx_{\setpf}',\ell_f) }
    \right)                                           \\
     & = \!\!\!\! \sum_{\ell_{f_1} \in \set{L}_{f_1}}
    \!\!\!\!\cdots \!\!\!\!
    \sum_{\ell_{f_{|\setF|} \in \set{L}_{|\setF|}}}
    \!\!\!
    \Biggl( \prod_{f \in \setF} \lambda_{f}(\ell_f) \Biggr)
    \!\!\cdot\!\!
    \Biggl(
    \sum_{ \vx_{\setEfull} \in \setx_{\setEfull} }
    \prod_{f \in \setF} u_{f}(\vx_{\setpf},\ell_f)
    \Biggr)                                           \\
     & \qquad \cdot
    \Biggl(
    \sum_{ \vx_{\setEfull}' \in \setx_{\setEfull} }
    \prod_{f \in \setF}
    \overline{u_{f}(\vx_{\setpf}',\ell_f)}
    \Biggr)
    \\
     & = \!\!\!\! \sum_{\ell_{f_1} \in \set{L}_{f_1}}
    \!\!\!\!\cdots\!\!\!\!
    \sum_{\ell_{f_{|\setF|} \in \set{L}_{|\setF|}}}
    \!\!\!\Biggl(
    \underbrace{
      \prod_{f \in \setF}
      \lambda_{f}(\ell_f)
    }_{\geq 0} \Biggr)
    \!\!\cdot\!\!
    \underbrace{
      \Biggl|
      \sum_{ \vx_{\setEfull} \in \setx_{\setEfull} }
      \prod_{f \in \setF} u_{f}(\vx_{\setpf},\ell_f)
      \Biggr|^{2}
    }_{\geq 0}
    \nonumber                                         \\
     & \geq 0.
  \end{align*}

  Let $\graphN$ be a weak-sense DE-NFG. Fix some $f \in \setF$. Because
  $\matr{C}_f$ is a Hermitian matrix, the eigenvalue decomposition of $ \matr{C}_f  $ in~\eqref{sec:DENFG:eqn:9} implies that $ \lambda_{f}(\ell_f) \in \sR $ for all $ \ell_f \in \set{L}_f $ and $ f \in \set{F} $.
  Similar calculations as above lead to
  \begin{align*}
    Z(\graphN)
     & = \!\!\!\! \sum_{\ell_{f_1} \in \set{L}_{f_1}}
    \!\!\!\!\cdots\!\!\!\!
    \sum_{\ell_{f_{|\setF|} \in \set{L}_{|\setF|}}}
    \!\!\!\!
    \Biggl(
    \underbrace{
      \prod_{f \in \setF}
      \lambda_{f}(\ell_f)
    }_{\in \sR}
    \Biggr) \!\! \cdot \!\!
    \underbrace{
      \Biggl|
      \sum_{ \vx_{\setEfull} \in \setx_{\setEfull} }
      \prod_{f \in \setF} u_{f}(\vx_{\setpf},\ell_f)
      \Biggr|^{2}
    }_{\geq 0}
    \\&\in \sR.
  \end{align*}

  {
{\section{Proof of Lemma~\ref{sec:DENFG:lem:1}}
\label{apx: proof of lemma17}}
\begin{proof}
  At $t=0$, Definition~\ref{sec:DENFG:def:2} gives
    $\matr{C}_{\mu_{\ef}^{(0)}}\in\setPD{\set{X}_e}
      \subseteq\setPSD{\set{X}_e}$, proving the message assertion in
    Item~\ref{sec:DENFG:lem:1:item:1} of this lemma. Assume that all message
    Choi matrices are PSD at iteration $t-1$. For $e=(f_i,f_j)$, order the
    $e$-coordinate first. Let $\bm{1}_e$ be the all-one vector indexed by
    $\set{X}_e$, and set
    $\matr{J}_e\defeq\bm{1}_e\cdot\bm{1}_e^{\Herm}\succeq0$.
    Let $\bm{1}_{-e}$ denote the all-one vector indexed by
    $\set{X}_{\setpfi\setminus e}$. Then
  \begin{align*}
    \matr{A}_{e,f_j}^{(t)}
     & \defeq \matr{C}_{f_i}\circ
         \left(\matr{J}_e\otimes
         \bigotimes_{e'\in\setpfi\setminus e}
         \matr{C}_{\mu_{e',f_i}^{(t-1)}}\right)
         \overset{(a)}{\succeq}0, \\
    \widehat{\matr{C}}_{\mu_{e,f_j}^{(t)}}
     & \defeq
         (\matr{I}_e\otimes\bm{1}_{-e})^{\Herm}
         \cdot\matr{A}_{e,f_j}^{(t)}\cdot
         (\matr{I}_e\otimes\bm{1}_{-e})
         \overset{(b)}{\succeq}0.
  \end{align*}
  Here, the symbol $\circ$ denotes the entrywise product. Step $(a)$
    follows from closure under tensor products and Schur's product theorem.
    Step $(b)$ follows because $K^{\Herm}\cdot A\cdot K$ is PSD whenever
    $A$ is PSD, for every matrix $K$ of compatible dimensions. For every
    nonzero unnormalized message, the normalization factor chosen in
    Definition~\ref{sec:DENFG:def:2} is positive, so division preserves positive
    semidefiniteness. An identically zero message triggers a fresh
    PD-normalized initialization. Thus, every message Choi matrix remains PSD.
    Induction proves the message assertion in
    Item~\ref{sec:DENFG:lem:1:item:1}. Similarly,
  \begin{align*}
    \widehat{\matr{C}}_{\beli_e^{(t)}}
     & \defeq
         \matr{C}_{\mu_{e,f_i}^{(t)}}\circ
         \matr{C}_{\mu_{e,f_j}^{(t)}}\succeq0, \qquad
    \widehat{\matr{C}}_{\beli_f^{(t)}}
    \defeq
      \matr{C}_f\circ
      \bigotimes_{e\in\setpf}
      \matr{C}_{\mu_{e,f}^{(t)}}\succeq0,
  \end{align*}
  Division by a positive belief normalization factor preserves
    positive semidefiniteness. This proves the belief assertion in
    Item~\ref{sec:DENFG:lem:1:item:1} whenever the corresponding belief is defined.
    Definition~\ref{sec:DENFG:def:2}
    restricts fixed points to PSD-normalized messages, proving
    Item~\ref{sec:DENFG:lem:1: PSD of SPA fixed point message}. At a fixed
    point, the message-update equation makes
    $\sum_{\tvx_{\setpf\setminus e}}\beli_f(\tvx_{\setpf})$ proportional to
    $\mu_{e,f_i}(\txe) \cdot \mu_{e,f_j}(\txe)$, and hence to $\beli_e(\txe)$.
    Under the conditions in Item~\ref{sec:DENFG:lem:1:item:2} of this lemma,
    both sides sum to one and are therefore equal.
\end{proof}
}

{
{\section{Proof of Proposition~\ref{prop: sufficient condition for pd messages}}
  \label{apx: sufficient condition for pd messages}}
}
  The proof generalizes the main idea of~\cite[Proposition 4]{Yedidia2005}. Throughout this appendix, the vector $ \vmu $ is an SPA fixed-point message vector, and the hypothesis of Proposition~\ref{prop: sufficient condition for pd messages} is in force: $\matr{C}_f$ is positive definite for every $f\in\setF$.

  \begin{lemma}
    \label{lem: property of mu}
    The SPA fixed-point message vector $ \vmu $
    has the following properties.
    \begin{enumerate}
      \item For each $ f \in \setF $ and $ e \in \setpf $,
            $ \matr{C}_{\mu_{e,f}} \in \setPSD{\setxe} $.

      \item For each $ f \in \setF $ and $ e \in \setpf $, the matrix $\mathbf{C}_{\mu_{e,f}}$ has at least one positive eigenvalue.

      \item For each edge $e=(f_i,f_j)$, both unnormalized
            messages have positive total entry sums. Their normalization factors
            therefore equal these sums:
            {
            \begin{align*}
              \kappa_{e,f_i}
               &=\sum_{\tvx_{\setpfj}} f_j(\tvx_{\setpfj})
                 \cdot\prod_{e'\in\setpfj\setminus e}\mu_{\epfj}(\tx_{e'}),\\
              \kappa_{e,f_j}
               &=\sum_{\tvx_{\setpfi}} f_i(\tvx_{\setpfi})
                 \cdot\prod_{e'\in\setpfi\setminus e}\mu_{\epfi}(\tx_{e'}).
            \end{align*}}

    \end{enumerate}
  \end{lemma}
  \begin{proof}
    We prove the three properties in turn.
    \begin{enumerate}
      \item This follows directly from Lemma~\ref{sec:DENFG:lem:1}.

      \item By Definition~\ref{sec:DENFG:def:2}, every message
            Choi matrix is nonzero and PSD. Hence, it has at least one positive
            eigenvalue.

      \item We prove the claim for $\kappa_{e,f_j}$; the proof for
            $\kappa_{e,f_i}$ is identical. Let
            \begin{align*}
              \matr{Q}_{e,f_i}
              \defeq \bigotimes_{e'\in\setpfi}\matr{A}_{e'},
              \qquad
              \matr{A}_{e'} \defeq
              \begin{cases}
                \bm{1}\cdot\bm{1}^{\tran}    & e'=e     \\
                \matr{C}_{\mu_{e',f_i}} & e'\neq e
              \end{cases},
            \end{align*}
            The tensor factors are ordered according to
              $\setpfi$. By Items~1 and~2, the matrix $\matr{Q}_{e,f_i}$ is nonzero and
              PSD. Let $s_{e,f_j}$ be the total entry sum of the unnormalized
              message. Its definition gives
              $s_{e,f_j}=\Tr\!\left(\matr{C}_{f_i}\cdot
                \matr{Q}_{e,f_i}^{\tran}\right)>0$,
              because $\matr{C}_{f_i}$ is positive definite and
              $\matr{Q}_{e,f_i}^{\tran}$ is nonzero and PSD.
              Definition~\ref{sec:DENFG:def:2} therefore selects
              $\kappa_{e,f_j}=s_{e,f_j}$.\qedhere
    \end{enumerate}
  \end{proof}

  \begin{definition}
    \label{def: def of eig of mu ef}
    \finaltext{For each $f\in\setF$ and $e\in\setpf$, define the following
      objects \finaltext{based on} the message vector $\vmu_{\etof}$ of the
      SPA~fixed~point~$\vmu$.}
    \begin{enumerate}
      \item The mappings
            $\matr{u}_{ \mu_{\etof} }: \set{L}_{e,f} \to \sC^{|\setxe|}$ and
            $\lambda_{\mu_{\etof}}: \set{L}_{e,f} \to \sR_{\geq 0}$ are defined
              so that
            \begin{align}
              \matr{C}_{\mu_{\etof}}
               & =  \!\!\! \sum_{ \ell_{e,f} \in \set{L}_{e,f} }
              \!\!\!\!\!\!
              \matr{u}_{ \mu_{\etof} }( :, \ell_{e,f} )
              \cdot \lambda_{ \mu_{\etof} }(\ell_{e,f} ) \cdot \bigl( \matr{u}_{ \mu_{\etof} }(:, \ell_{e,f} ) \bigr)^{\!\Herm},
              \label{eqn: decomposition of the Choi matrix for SPA message}
            \end{align}
            where $ \set{L}_{e,f} $ is a finite set with $ |\set{L}_{e,f}| = |\setxe| $.
            For each $ \ell_{e,f} \in \set{L}_{e,f} $, the column vector
            \begin{align*}
              \matr{u}_{ \mu_{\etof} }( :, \ell_{e,f} )
              \defeq \bigl( u_{ \mu_{\etof} }( \xe, \ell_{e,f} )
              \bigr)_{ \! \xe \in \setxe } \in \sC^{|\setxe|}
            \end{align*}
            is the right-eigenvector associated with the eigenvalue
            $ \lambda_{ \mu_{\etof} }( \ell_{e,f} ) \in \sR_{\geq 0} $. Note
            the 2-norm of $\matr{u}_{ \mu_{\etof} }( :, \ell_{e,f} )$ is one.

      \item The set $ \set{L}_{e,f,\mathrm{p}} $ is defined to be
            \begin{align}
              \set{L}_{e,f,\mathrm{p}}
              \defeq \bigl\{ \ell_{e,f} \ \bigl| \
              \ell_{e,f} \in \set{L}_{e,f},\,
              \lambda_{ \mu_{\etof} }( \ell_{e,f} )
              \in \sR_{>0} \bigr\}.  \label{eqn: property of ell e p}
            \end{align}
            A generic element of $ \set{L}_{e,f,\mathrm{p}} $ is denoted by
            $ \ell_{e,f,\mathrm{p}} $.\edefinition
    \end{enumerate}
  \end{definition}

  \begin{lemma}
    \label{lem: positive of message}
    If for all $ f \in \setF $, the Choi matrix
    $ \matr{C}_{f} $ is a positive definite matrix,
    then for each $ f \in \setF $ and $ e \in \setpf $, the Choi-matrix representation
    $\matr{C}_{\mu_{\etof}}$ is also a positive definite matrix.
  \end{lemma}
  \begin{proof}
    Without loss of generality, we adopt the following indexing.
    \begin{itemize}
      \item The edge $ 1 \in \setEfull $ connects function nodes $ f_{1} $ and $ f_{2} $.

      \item The set $ \setpf_{1} $ is given by
            $ \setpf_{1} = \{1,2,\ldots,|\setpf_{1}|\} $, which implies
            $ \vx_{\setpf_{1}} = ( x_{1}, x_{2},\ldots,x_{|\setpf_{1}|} ) $.

    \end{itemize}
    We use the shorthand notation
    $ \sum_{ \ell_{2},\ldots,\ell_{|\setpf_{1}|} } $ instead of
    $ \sum_{ \ell_{2,f_1} \in \set{L}_{2,f_1}, \ldots,\ell_{|\setpf_{1}|,f_1}
        \in \set{L}_{|\setpf_1|,f_1} } $. Every product and tensor product
      below runs over $e \in \setpf_{1}\setminus\{1\}$. We prove the positive
      definiteness of $ \matr{C}_{\mu_{\onetoftwo}} $ by contradiction.

    Suppose that for the matrix $ \matr{C}_{\mu_{\onetoftwo}} $,
    there exists an index $ \ell_{1,f_{2},0} \in \set{L}_{1,f_2} $
    such that $ \lambda_{ \mu_{\onetoftwo} }( \ell_{1,f_{2},0} ) = 0 $.
    Then we obtain a contradiction. Namely,
    \begin{align}
      0
       & =
      \kappa_{\onetoftwo}\lambda_{ \mu_{\onetoftwo} }( \ell_{1,f_{2},0} )
      \nonumber                                \\
       & \overset{(a)}{=}
      \kappa_{\onetoftwo}
      \!\cdot\!
      \bigl( \matr{u}_{ \mu_{\onetoftwo} }( :, \ell_{1,f_{2},0} ) \bigr)^{\!\Herm}
      \!\cdot\! \matr{C}_{\mu_{\onetoftwo}}
      \!\cdot\!
      \matr{u}_{ \mu_{\onetoftwo} }( :, \ell_{1,f_{2},0} )
      \nonumber                                \\
       & =
      \sum_{\tx_{1}}
      \overline{u_{ \mu_{\onetoftwo} }( x_{1}, \ell_{1,f_{2},0} )}
      \cdot u_{ \mu_{\onetoftwo} }( x_{1}', \ell_{1,f_{2},0} )
      \nonumber                                                                 \cdot \kappa_{1,f_{2}} \cdot \mu_{1,f_{2}}(\tx_{1})
      \nonumber                                \\
       & \overset{(b)}{=}
      \sum_{\tvx_{\setpf_{1}}}
      \overline{u_{ \mu_{\onetoftwo} }( x_{1}, \ell_{1,f_{2},0} )}
      \cdot
      u_{ \mu_{\onetoftwo} }( x_{1}', \ell_{1,f_{2},0} )
      \cdot
      f_{1}\bigl( \tvx_{\setpf_{1}} \bigr)
      \nonumber                                \\
       & \qquad
      \cdot \prod_{e \in \setpf_{1} \setminus \{1\}}
      \mu_{e,f_{1}}(\tx_{e})
      \nonumber                                \\
       & \overset{(c)}{=}
      \!\!\!\!\!\!
      \sum_{
        \ell_{2},\ldots,\ell_{|\setpf_{1}|}
      }\!\!\!
      \Biggl(
      \prod_{e \in \setpf_{1} \setminus \{1\}}
      \underbrace{ \lambda_{\mu_{\etofone}}( \ell_{e,f_{1}} ) }_{\overset{(d)}{\geq 0} }
      \Biggr)
      \nonumber                                \\
       & \quad\cdot
      \underbrace{
        \begin{aligned}
           & \Bigl(
          \bigl( \matr{u}_{ \mu_{\onetoftwo} }
          ( :, \ell_{1,f_{2},0} ) \bigr)^{\!\Herm}
          \otimes
          \mathop{\bigotimes}_{\substack{e \in \setpf_{1}\\ e \neq 1}}
          \bigl( \matr{u}_{\mu_{\etofone}}
          ( :, \ell_{e,f_1} ) \bigr)^{\tran}
          \Bigr)
          \cdot \matr{C}_{f_{1}}
          \\[-0.25ex]
           & \quad \cdot \Bigl(
          \matr{u}_{ \mu_{\onetoftwo} }
          ( :, \ell_{1,f_{2},0} )
          \otimes
          \mathop{\bigotimes}_{\substack{e \in \setpf_{1}\\ e \neq 1}}
          \overline{ \matr{u}_{\mu_{\etofone}}
            ( :, \ell_{e,f_1} ) }
          \Bigr)
        \end{aligned}
      }_{\overset{(e)}{>0}}
      \nonumber                                \\
       & \overset{(f)}{\geq}
      \prod_{e \in \setpf_{1} \setminus \{1\}}
      \underbrace{ \lambda_{\mu_{\etofone}}( \ell_{e,f_{1},\mathrm{p}} ) }_{\overset{(g)}{>0}}
      \nonumber                                \\
       & \quad\cdot
      \underbrace{
        \begin{aligned}
           & \Bigl(
          \bigl( \matr{u}_{ \mu_{\onetoftwo} }
          ( :, \ell_{1,f_{2},0} ) \bigr)^{\!\Herm}
          \otimes
          \mathop{\bigotimes}_{\substack{e \in \setpf_{1}\\ e \neq 1}}
          \bigl( \matr{u}_{\mu_{\etofone}}
          ( :, \ell_{e,f_{1},\mathrm{p}} ) \bigr)^{\tran}
          \Bigr)
          \cdot \matr{C}_{f_{1}}
          \\[-0.25ex]
           & \quad \cdot \Bigl(
          \matr{u}_{ \mu_{\onetoftwo} }
          ( :, \ell_{1,f_{2},0} )
          \otimes
          \mathop{\bigotimes}_{\substack{e \in \setpf_{1}\\ e \neq 1}}
          \overline{ \matr{u}_{\mu_{\etofone}}
            ( :, \ell_{e,f_{1},\mathrm{p}} ) }
          \Bigr)
        \end{aligned}
      }_{>0}
      \nonumber                                \\
       & >0. \label{eqn: positive of messages}
    \end{align}
    The above derivation uses the following observations. Step $(a)$ uses
    $ \kappa_{\onetoftwo} \in \sR_{>0} $, as proved in Lemma~\ref{lem:
      property of mu}, and the eigenvalue equation for
    $\lambda_{ \mu_{\onetoftwo} }( \ell_{1,f_{2},0} )$ from~\eqref{eqn:
      decomposition of the Choi matrix for SPA message}. Step $(b)$ follows
    from the SPA fixed-point Eq.~\eqref{sec:DENFG:eqn:15}. Step $(c)$
    follows from the Choi-matrix representation $ \matr{C}_{f_{1}} $
    in~\eqref{sec:DENFG:eqn:9} and the eigendecomposition~\eqref{eqn:
      decomposition of the Choi matrix for SPA message} of each message in
    $ ( \vmu_{e,f_{1}} )_{e \in \setpf_{1} \setminus \{1\}} $. Step $(d)$
    follows because $ \matr{C}_{\mu_{e,f_1}} \in \setPSD{\setxe} $ by
    Lemma~\ref{lem: property of mu}; hence
    $ \lambda_{\mu_{\etofone}}( \ell_{e,f_1} ) \geq 0 $. Step $(e)$ follows
    because $ \matr{C}_{f_1} $ is positive definite. Step $(f)$ selects
    $ \ell_{e,f_{1}, \mathrm{p}} \in \set{L}_{e,f_{1},\mathrm{p}} $ for every
    $ e \in \setpf_{1}\setminus\{1\} $. Each such set is nonempty by
    Lemma~\ref{lem: property of mu}. Step $(g)$ follows from the definition of
    $ \ell_{e,f_{1}, \mathrm{p}} $ in~\eqref{eqn: property of ell e p}.

    The proof for every other message in
    $ \vmu =  (\vmu_{e,f})_{f \in \setF,\,e \in \setpf} $ is analogous.
  \end{proof}

  We now prove the proposition.
  \begin{enumerate}
    \item For each $ f \in \setF $, we have
          \begin{align*}
            \hspace{-0.5cm}
            Z_f(\vmu)
             & =
            \sum_{ \substack{
                (\ell_{e,f})_{e \in \setpf} \in \prod_{e \in \setpf} \set{L}_{e,f}
              } }
            \Biggl(
            \prod_{ e \in \setpf }
            \lambda_{\mu_{\etof}}( \ell_{e,f} )
            \!\Biggr)
            \\&\quad
            \cdot \
            \Biggl(
            \mathop{\bigotimes}_{ e \in \setpf }
            \bigl( \matr{u}_{\mu_{\etof}}(:, \ell_{e,f} ) \bigr)^{\tran}
            \!\!\Biggr)\!\cdot \matr{C}_{f}\!\cdot
            \Biggl(
            \mathop{\bigotimes}_{ e \in \setpf }
            \overline{ \matr{u}_{\mu_{\etof}}(:, \ell_{e,f} ) }
            \Biggr) \\& >0,
          \end{align*}
          where the last inequality follows by the same argument used to prove the strict inequality in~\eqref{eqn: positive of messages}.

    \item  For each $ e = (f_{i},f_{j}) \in \setEfull $, we have
          \begin{align*}
            Z_e(\vmu)
             & = \Tr \bigl( \matr{C}_{\mu_{e,f_{i}}}
            \cdot \matr{C}_{\mu_{e,f_{j}}}^{\tran} \bigr)
            \\
             & = \Tr \bigl( \sqrt{ \matr{C}_{\mu_{e,f_{j}}}^{\tran} } \cdot
            \matr{C}_{\mu_{e,f_{i}}}
            \cdot \sqrt{ \matr{C}_{\mu_{e,f_{j}}}^{\tran} } \bigr)
            \\&> 0,
          \end{align*}
          where the last inequality follows from Lemma~\ref{lem: positive of message}.
  \end{enumerate}
  Therefore, by Definition~\ref{sec:DENFG:def:3}, we have
  \begin{align*}
    \ZBSPA(\graphN,\vmu)
    = \frac{\prod\limits_f Z_f(\vmu)}
      {\prod\limits_{e} Z_{e}(\vmu)}
    >0.
  \end{align*}

  {
{\section{Details of the LCT for S-NFGs}
  \label{apx:LCT SNFGs}}
}
  Given $ |\set{X}_e| \in \sZpp $, without loss of generality, we suppose that
  $\set{X}_e = \LCTset{X}_e = \{ 0, 1, \ldots, |\set{X}_e| \! - \! 1 \}$.  We have
  to verify that the functions $M_{\efi}$ and $M_{\efj}$
  satisfy~\eqref{sec:LCT:exp:1}, \ie,
  \begin{align*}
     & \sum_{\LCT{x}_e}
    M_{\efi}(x_{\efi}, \LCT{x}_e)
    \cdot
    M_{\efj}(x_{\efj}, \LCT{x}_e)
    = [x_{\efi} \! = \!  x_{\efj}]
  \end{align*}
  for all $x_{\efi}, x_{\efj} \in \set{X}_e.$
  We consider the following four cases.
  \begin{enumerate}

    \item For $x_{\efi} = 0$ and $x_{\efj} = 0$, we obtain the constraint
          \begin{align*}
            \zeta_{\efi}
             & \cdot
            \mu_{\efi}(0)
            \cdot
            \zeta_{\efj}
            \cdot
            \mu_{\efj}(0)
            +
            \sum_{\LCT{x}_e: \, \LCT{x}_e \neq 0}
            \zeta_{\efi}
            \cdot
            \chi_{\efi}
            \\
             & \cdot
            \bigl( - \mu_{\efj}(\LCT{x}_e) \bigr)
            \cdot
            \zeta_{\efj}
            \cdot
            \chi_{\efj} \cdot
            \bigl( - \mu_{\efi}(\LCT{x}_e) \bigr)
            = 1.
          \end{align*}
          Using~\eqref{sec:LCT:exp:5}
          and~\eqref{sec:LCT:exp:6}, and multiplying both sides by
          $Z_e$, we obtain the equivalent constraint
          \begin{align*}
            \mu_{\efi}(0)
            \cdot
            \mu_{\efj}(0)
            +
            \sum_{\LCT{x}_e: \, \LCT{x}_e \neq 0}
            \mu_{\efj}(\LCT{x}_e)
            \cdot
            \mu_{\efi}(\LCT{x}_e)
             & = Z_e.
          \end{align*}
          This constraint holds because its left-hand side is the
            definition of $Z_e$.

    \item For $x_{\efi} = 0$ and $x_{\efj} \in \set{X}_e \setminus \{ 0 \}$, we obtain
          the constraint
          \begin{align*}
             &
            \zeta_{\efi}
            \cdot
            \mu_{\efi}(0)
            \cdot
            \zeta_{\efj}
            \cdot
            \mu_{\efj}(x_{\efj})
            +
            \sum_{\LCT{x}_e: \, \LCT{x}_e \neq 0}
            \zeta_{\efi}
            \cdot
            \chi_{\efi}
            \\
             &
            \cdot
            \bigl( - \mu_{\efj}(\LCT{x}_e) \bigr)
            \cdot
            \Bigl(
            \zeta_{\efj}
            \cdot
            \chi_{\efj}
            \cdot
            \bigl(
            \delta_{\efj}
            \cdot
            [x_{\efj} \!=\! \LCT{x}_e]
            \\
             &
            \qquad\quad\qquad
            +
            \epsilon_{\efj}
            \cdot
            \mu_{\efj}(x_{\efj})
            \cdot
            \mu_{\efi}(\LCT{x}_e)
            \bigr)
            \Bigr)
            = 0.
          \end{align*}
          Using~\eqref{sec:LCT:exp:5}
          and~\eqref{sec:LCT:exp:6}, we obtain the equivalent
          constraint
          \begin{align*}
             & \mu_{\efi}(0)
            \cdot
            \mu_{\efj}(x_{\efj})
            -
            \sum_{\LCT{x}_e: \, \LCT{x}_e \neq 0}
            \mu_{\efj}(\LCT{x}_e)
            \\
             & \cdot
            \Bigl(
            \delta_{\efj}
            \cdot
            [x_{\efj} \!=\! \LCT{x}_e]
            +
            \epsilon_{\efj}
            \cdot
            \mu_{\efj}(x_{\efj})
            \cdot
            \mu_{\efi}(\LCT{x}_e)
            \Bigr)
            = 0.
          \end{align*}
          \finaltext{After simplifying the sum in the preceding expression, we obtain
          the equivalent constraint}
          \begin{align*}
             & \mu_{\efi}(0)
            \cdot
            \mu_{\efj}(x_{\efj})
            -
            \mu_{\efj}(x_{\efj})
            \cdot
            \delta_{\efj}
            \\
             & -
            \epsilon_{\efj}
            \cdot
            \mu_{\efj}(x_{\efj})
            \cdot
            \sum_{\LCT{x}_e: \, \LCT{x}_e \neq 0}
            \mu_{\efj}(\LCT{x}_e)
            \cdot
            \mu_{\efi}(\LCT{x}_e)
            = 0.
          \end{align*}
          The sum appearing in the above expression equals
          $Z_e \cdot \bigl( 1 - \beli_e(0) \bigr)$. Therefore, this constraint can be
          rewritten as
          \begin{align*}
             & \mu_{\efj}(x_{\efj})
            \!\cdot\!
            \Bigl(
            \mu_{\efi}(0)
            -
            \delta_{\efj}
            -
            \epsilon_{\efj}
            \! \cdot\!
            Z_e
            \!\cdot\!
            \bigl( 1 - \beli_e(0) \bigr)
            \Bigr)
            \!=0.
          \end{align*}
          The expression in the parentheses equals zero because
            of~\eqref{sec:LCT:exp:9}, and so this constraint is satisfied.

    \item For $x_{\efi} \in \set{X}_e \setminus \{ 0 \}$ and $x_{\efj} = 0$, we obtain
          \finaltext{expressions similar to those in the second case, with the roles of
          $f_{i}$ and $f_{j}$ exchanged.}

    \item For $x_{\efi}, x_{\efj} \in \set{X}_e \setminus \{ 0 \}$, we obtain the
          constraint (we have already used~\eqref{sec:LCT:exp:5}
          and~\eqref{sec:LCT:exp:6} to simplify the expression)
          \begin{align*}
             & Z_e
            \cdot
            [x_{\efi} \!=\! x_{\efj}]= \mu_{\efi}(x_{\efi})
            \cdot
            \mu_{\efj}(x_{\efj})
            \\&+
            \!\! \!\!\sum_{\LCT{x}_e: \, \LCT{x}_e \neq 0} \!\!
            \Bigl(
            \delta_{\efi}
            \cdot
            [x_{\efi} \!=\! \LCT{x}_e]
            +
            \epsilon_{\efi}
            \cdot
            \mu_{\efi}(x_{\efi})
            \cdot
            \mu_{\efj}(\LCT{x}_e)
            \Bigr) \\
             &
            \qquad\,\,\cdot
            \Bigl(
            \delta_{\efj}
            \cdot
            [x_{\efj} \!=\! \LCT{x}_e]
            +
            \epsilon_{\efj}
            \cdot
            \mu_{\efj}(x_{\efj})
            \cdot
            \mu_{\efi}(\LCT{x}_e)
            \Bigr).
          \end{align*}
          This constraint is equivalent to the constraint
          \begin{align*}
            \hspace{0.25cm} & \hspace{-0.25cm}
            \mu_{\efi}(x_{\efi})
            \cdot
            \mu_{\efj}(x_{\efj})
            +
            \delta_{\efi}
            \cdot
            \delta_{\efj}
            \cdot
            [x_{\efi} \!=\! x_{\efj}]          \\
                            &
            +
            \delta_{\efi}
            \cdot
            \epsilon_{\efj}
            \cdot
            \mu_{\efj}(x_{\efj})
            \cdot
            \mu_{\efi}(x_{\efi})
            \\
                            &
            +
            \delta_{\efj}
            \cdot
            \epsilon_{\efi}
            \cdot
            \mu_{\efi}(x_{\efi})
            \cdot
            \mu_{\efj}(x_{\efj})               \\
                            &
            +
            \epsilon_{\efi}
            \cdot
            \epsilon_{\efj}
            \cdot
            \mu_{\efi}(x_{\efi})
            \cdot
            \mu_{\efj}(x_{\efj})
            \\&\quad\cdot \!\!\!\!\sum_{\LCT{x}_e: \, \LCT{x}_e \neq 0}\!\!\!\!
            \mu_{\efj}(\LCT{x}_e)
            \cdot
            \mu_{\efi}(\LCT{x}_e)
            = Z_e
            \cdot
            [x_{\efi} \!=\! x_{\efj}].
          \end{align*}
          The sum appearing in the above expression equals
          $Z_e \cdot \bigl( 1 - \beli_e(0) \bigr)$. Therefore, after rewriting this
          constraint, we obtain
          \begin{align*}
             &
            [x_{\efi} \!=\! x_{\efj}]
            \!\cdot\!
            \bigl(
            \delta_{\efi}
            \!\cdot\!
            \delta_{\efj}
            \!-\!
            Z_e
            \bigr)
            \!+\!
            \mu_{\efi}(x_{\efi})
            \!\cdot\!
            \mu_{\efj}(x_{\efj}) \\
             & \!\cdot\!\Bigl(\!
            1
            \!+\!
            \delta_{\efi}
            \!\cdot\!
            \epsilon_{\efj}
            \!+\!
            \delta_{\efj}
            \!\cdot\!
            \epsilon_{\efi}
            \!+\!
            \epsilon_{\efi}
            \!\cdot\!
            \epsilon_{\efj}
            \!\!\cdot\!\!
            Z_e
            \!\cdot\!
            \bigl( 1 \!-\! \beli_e(0) \bigr)
            \!\Bigr)
            \!=\! 0.
          \end{align*}
          We claim that this constraint is satisfied.
          This follows from the above two parenthetical
          expressions being equal to zero. For the first parenthetical
          expression, this observation follows
          from~\eqref{sec:LCT:exp:7}. For the second parenthetical
          expression, this observation follows from the following
          considerations.

          If $ \beli_e(0) = 1$, the second parenthetical expression equals
          \begin{align*}
            1
            +
            \delta_{\efi}
            \cdot
            \epsilon_{\efj}
            +
            \delta_{\efj}
            \cdot
            \epsilon_{\efi}
          \end{align*}
          which, due to~\eqref{extra constraint on epsilon for beli =
              [x = 0]}, equals $0$.

          Now we consider $ \beli_e(0) \neq 1$.
          Because the product of the left-hand sides
          of~\eqref{sec:LCT:exp:8}
          and~\eqref{sec:LCT:exp:9} must equal the product of the
          right-hand sides of~\eqref{sec:LCT:exp:8}
          and~\eqref{sec:LCT:exp:9}, we obtain
          \begin{align*}
             &
            \delta_{\efi}
            \cdot
            \delta_{\efj}
            +
            \delta_{\efi}
            \cdot
            Z_e
            \cdot
            \bigl( 1 - \beli_e(0) \bigr)
            \cdot
            \epsilon_{\efj}
            +
            \delta_{\efj}
            \cdot
            Z_e
            \\
             & \cdot
            \bigl( 1 - \beli_e(0) \bigr)
            \cdot
            \epsilon_{\efi}
            +
            (Z_e)^2
            \cdot
            \bigl( 1 - \beli_e(0) \bigr)^2
            \cdot
            \epsilon_{\efi}
            \cdot
            \epsilon_{\efj} \\
             &
            = \mu_{\efi}(0)
            \cdot
            \mu_{\efj}(0).
          \end{align*}
          Using~\eqref{sec:LCT:exp:new:2}--\eqref{sec:LCT:exp:7}, we obtain
          \begin{align*}
             & Z_e
            +
            \delta_{\efi}
            \cdot
            Z_e
            \cdot
            \bigl( 1 - \beli_e(0) \bigr)
            \cdot
            \epsilon_{\efj}
            +
            \delta_{\efj}
            \cdot
            Z_e
            \cdot
            \bigl( 1 - \beli_e(0) \bigr)
            \\
             & \cdot
            \epsilon_{\efi}
            +
            (Z_e)^2
            \cdot
            \bigl( 1 - \beli_e(0) \bigr)^2
            \cdot
            \epsilon_{\efi}
            \cdot
            \epsilon_{\efj} \nonumber  = Z_e \cdot \beli_e(0).
          \end{align*}
          Subtracting $Z_e\cdot\beli_e(0)$ from both sides and then dividing both sides by $Z_e\cdot\bigl(1-\beli_e(0)\bigr)$ results in
          \begin{align*}
             & 1
            \!+\!
            \delta_{\efi}
            \!\cdot\!
            \epsilon_{\efj}
            \!+\!
            \delta_{\efj}
            \!\cdot\!
            \epsilon_{\efi}
            \!+\!
            \epsilon_{\efi}
            \!\cdot\!
            \epsilon_{\efj}
            \!\cdot\! Z_e
            \!\cdot\!
            \bigl( 1 \!- \!\beli_e(0) \bigr)
            \!=\! 0.
          \end{align*}
          This is the promised result.

  \end{enumerate}

  {
{\section{Proof of Proposition~\ref{sec:LCT:prop:1}}
  \label{apx:property of SNFGs}}
}

  \begin{enumerate}

    \item This follows from the fact that in every step of the LCT, the partition
          function is unchanged.

    \item This follows from
          \begin{align*}
            \LCT{g}(\vect{0})
             & = \prod_f
            \LCT{f}(\vect{0})                         \\
             & \overset{(a)}{=} \prod_f
            \sum_{\vx_{\setpf}}
            f(\vx_{\setpf})
            \cdot
            \prod_{e \in \setpf}
            M_{\etof}(x_{e}, 0)
            \\
             & \overset{(b)}{=} \prod_f
            \sum_{\vx_{\setpf}}
            f(\vx_{\setpf})
            \cdot
            \prod_{e \in \setpf}
            \bigl(
            \zeta_{\etof}
            \cdot
            \mu_{\etof}(x_{e})
            \bigr)
            \\
             & = \left(
            \prod_f
            \sum_{\vx_{\setpf}}
            f(\vx_{\setpf})
            \cdot
            \prod_{e \in \setpf}
            \mu_{\etof}(x_{e})
            \right)
            \cdot
            \!\!\!\!\!\!\!
            \prod_{e = (f_{i},f_{j}) \in \setEfull}
            \!\!\!\!\!\!\!\!
            \bigl(
            \zeta_{\efj}
            \!\cdot\!
            \zeta_{\efi}
            \bigr)                                    \\
             & \overset{(c)}{=}
            \Biggl(
            \prod_f
            Z_f( \vmu)
            \Biggr)
            \cdot
            \left(
            \prod_{e}
            \bigl( Z_e( \vmu) \bigr)^{\!-1}
            \right)                                   \\
             & \overset{(d)}{=} \ZBSPA(\graphN,\vmu).
          \end{align*}
          Step $(a)$ follows from~\eqref{sec:LCT:exp:10}.  Step $(b)$ follows
          from~\eqref{sec:LCT:exp:3} and~\eqref{sec:LCT:exp:4}.  Step $(c)$
          uses~\eqref{sec:LCT:exp:5} and the definition of $ Z_{f} $
          in~\eqref{eqn: def of zf for snfg}.  Step $(d)$ follows
          from~\eqref{sec:SNFG:eqn:5}.

    \item Let $e = (f_{i}, f_{j}) \in \setEfull $ and
          $\LCTv{x}_{\setpfi} \in \LCTset{X}_{\setpfi}$ be
          such that $\LCT{x}_e \neq 0$ and $\LCT{x}_{e'} = 0$ for all
          $e' \in \setpfi \setminus e$. For this setup, we obtain
          \begin{align*}
             & \hspace{-0.25 cm} \LCT{f_{i}}(\LCTvxfi)
            \\ & \overset{(a)}{=} \sum_{x_{e}}
            M_{\efi}(x_{e}, \LCT{x}_{e})
            \cdot \!
            \sum_{\vx_{\setpfi \setminus e}}
            f_{i}(\vx_{\setpfi})
            \cdot\!\!\!\!\!  \prod_{e' \in \setpfi \setminus e}
            \!\!\!\! M_{\epfi}(x_{e'}, \LCT{x}_{e'})   \\
             & \overset{(b)}{=} \sum_{x_{e}}
            M_{\efi}(x_{e}, \LCT{x}_e)
            \cdot \!\!\!\!
            \sum_{\vx_{\setpfi \setminus e}}
            \!\!\!f_{i}(\vx_{\setpfi})
            \cdot\!\!\!\!\!\! \prod_{e' \in \setpfi \setminus e}
            \!\!\!\!\bigl(
            \zeta_{\epfi}
            \cdot
            \mu_{\epfi}(x_{e'})
            \bigr)                                     \\
             & \overset{(c)}{=}
            \left(
            \prod_{e' \in \setpfi \setminus e}
            \zeta_{\epfi}
            \right)
            \cdot
            \sum_{x_e} M_{\efi}(x_e,\LCT{x}_e)
            \\ &\quad\cdot
            \underbrace{
              \sum_{\vx_{\setpfi\setminus e}} f_i(\vx_{\setpfi})
              \cdot\prod_{e'\in\setpfi\setminus e}\mu_{\epfi}(x_{e'})
            }_{\overset{(d)}{=}\;\kappa_{\efj}\cdot\mu_{\efj}(x_e)}
            \\
             & \overset{(e)}{=} \left(
            \!\prod_{e' \in \setpfi \setminus e}
            \!\!\!\!\zeta_{\epfi}
            \right)
            \!\cdot
            \frac{\kappa_{\efj}}{\zeta_{\efj}}
            \!\cdot\!
            \sum_{x_{e}}
            \!M_{\efi}(x_{e}, \LCT{x}_e)
            \cdot M_{\efj}(x_{e},0)                    \\
             & \overset{(f)}{=} \left(
            \prod_{e' \in \setpfi \setminus e}
            \zeta_{\epfi}
            \right)
            \cdot
            \frac{\kappa_{\efj}}{\zeta_{\efj}}
            \cdot
            [\LCT{x}_e \! = \! 0]                      \\
             & \overset{(g)}{=} 0.
          \end{align*}
          Step $(a)$ follows from Eq.~\eqref{sec:LCT:exp:10}.  Step $(b)$ uses
          $\LCT{x}_{e'} = 0$ for $e' \in \setpfi \setminus e$ and
          Eqs.~\eqref{sec:LCT:exp:3} and~\eqref{sec:LCT:exp:4}.  Step
          $(c)$ only reorders terms.
          Step $(d)$ uses the SPA fixed-point relation,
            with the actual positive message normalization factor
            $\kappa_{\efj}$ from
            Definition~\ref{def:belief at SPA fixed point for S-NFG}.
          Step $(e)$ follows from~\eqref{sec:LCT:exp:4}.  Step $(f)$ follows
          from~\eqref{sec:LCT:exp:2}, and step $(g)$ uses $\LCT{x}_e \neq 0$.

    \item Assume that $\LCTv{x}$ is not a generalized loop. Based on the
          third property and the definition of a generalized loop, we can
          conclude that there is at least one $f \in \setF$ such that
          $ \wh(\LCTvxf) = 1 $, which implies $\LCT{f}(\LCTvxf) = 0$ and
          $\LCT{g}(\LCTv{x}) = 0$.

    \item Follows immediately from the fourth property.

    \item From the end of the first step in the construction of
          $\LCT{\graphN}$, it follows that the SPA fixed-point message vector
          $\vmu$ for $\graphN$ induces an SPA fixed-point message vector
          $\LCTv{\mu} $ for $\LCT{\graphN}$ via
          \begin{align*}
            \mu_{\etof}(x_e)
             & = ( \nu_{\etof} )^{-1}
            \cdot
            \sum_{\LCT{x}_e}
            M_{\etof}(x_e, \LCT{x}_e)
            \cdot
            \LCT{\mu}_{\etof}(\LCT{x}_e),
          \end{align*}
          where $\nu_{\etof}$ is a nonzero constant. \finaltext{The message
          $\LCT{\mu}_{\etof}(\LCT{x}_e)=[\LCT{x}_e\! =\! 0]$ specified in
          Property~\ref{sec:LCT:prop:1:item:6} satisfies the displayed relation
          with $\nu_{\etof}=\zeta_{\etof}$ by~\eqref{sec:LCT:exp:3}
          and~\eqref{sec:LCT:exp:4}. The solution is unique up to rescaling
          because $\matr{M}_{\etof}$ is invertible
          by~\eqref{sec:LCT:exp:1}.}

    \item
          \begin{enumerate}

            \item This follows immediately from the expressions in
                  Definition~\ref{sec:LCT:def:1}.

            \item \finaltext{If $\beli_e(0)\neq1$, then
                  \eqref{sec:LCT:exp:8} and~\eqref{sec:LCT:exp:9} imply
                  $\epsilon_{\efi}=\epsilon_{\efj}$. If $\beli_e(0)=1$, then
                  $\mu_{\efi}(x_e)=\mu_{\efj}(x_e)=0$ for all
                  $x_e\in\set{X}_e\setminus\{0\}$.
                  The terms containing $\epsilon_{\efi}$ or $\epsilon_{\efj}$
                  in~\eqref{eq:lct:nonzero-columns} therefore vanish.
                  In either case, the symmetric parameter choices give
                  $\matr{M}_{\efi}=\matr{M}_{\efj}$, and
                  \eqref{sec:LCT:exp:1} implies that the matrices are orthogonal.}

          \end{enumerate}

  \end{enumerate}

  {
{\section{Proof of Proposition~\ref{prop:DENFG:LCT:1}}
  \label{apx:LCT DENFGs}}
}
  Properties~\ref{prop:DENFG:LCT:1:item:1}--\ref{prop:DENFG:LCT:1:item:7} follow by directly generalizing the
    corresponding S-NFG proofs of Properties~\ref{sec:LCT:prop:1:item:1}--\ref{sec:LCT:prop:1:item:7} in Proposition~\ref{sec:LCT:prop:1}.

  \begin{enumerate}

    \item[8)] For each $ f \in \setF $ and $ e \in \setpf $, we obtain the kernels
          $ M_{\ef} $ by applying the substitutions introduced in Definition~\ref{def:DENFG:LCT:1} to the LCT for an S-NFG in Definition~\ref{sec:LCT:def:1}.
          The SPA fixed-point message vector $ \vmu $ satisfies
          $ \matr{C}_{ \mu_{\ef} } \in \setPSD{\set{X}_{e}} $ by Lemma~\ref{sec:DENFG:lem:1}. Moreover, Definition~\ref{def:DENFG:LCT:1} specifies real constants $\zeta_{e,f}$, $\chi_{e,f}$, $\delta_{e,f}$, and $\epsilon_{e,f}$ for both endpoints $f\in\{f_i,f_j\}$. These facts imply
          \begin{align*}
             & M_{\ef}\bigl( (\xe, \xe') ,(\LCT{x}_{e}, \LCT{x}_{e}') \bigr)
            = \overline{ M_{\ef}\bigl( (\xe', \xe) ,(\LCT{x}_{e}', \LCT{x}_{e}) \bigr) }
          \end{align*}
          for all $
            \txe = (\xe, \xe') \in \tset{X}_{e}
          $ and $
            \LCTt{x}_{e} = ( \LCT{x}_{e}, \LCT{x}_{e}' ) \in \LCTtset{X}_{e}$.
          Therefore, the matrices $ \matr{C}_{ M_{\efi} }$ and
          $\matr{C}_{ M_{\efj} }$ belong to
          $\setHerm{\set{X}_e\times\LCTset{X}_e}$.

    \item[9)] For arbitrary $ f \in \setF $ and
          $ \LCTtv{x}_{\setpf} = (\LCTv{x}_{\setpf}', \LCTv{x}_{\setpf}) \in
            \LCTtset{X}_{\setpf} $, we have
          \begin{align*}
             & \hspace{-0.5cm}
            \LCT{f}(\LCTv{x}_{\setpf}', \LCTv{x}_{\setpf})
            \\& \overset{(a)}{=}
            \sum_{ \tvx_{\setpf} }
            f( \vx_{\setpf}, \vx_{\setpf}' )
            \cdot
            \prod_{e \in \setpf}
            M_{\ef}\bigl( (x_{e}, x_{e}'), (\LCT{x}_e', \LCT{x}_e) \bigr)
            \nonumber                                                                        \\
             & \overset{(b)}{=}
            \sum_{ \tvx_{\setpf} }
            f( \vx_{\setpf}', \vx_{\setpf} )
            \cdot
            \prod_{e \in \setpf}
            M_{\ef}\bigl( (x_{e}', x_{e}), (\LCT{x}_e', \LCT{x}_e) \bigr)
            \nonumber                                                                        \\
             & \overset{(c)}{=}
            \sum_{ \tvx_{\setpf} }
            \overline{f( \vx_{\setpf}, \vx_{\setpf}' )}
            \cdot
            \prod_{e \in \setpf}
            \overline{
              M_{\ef}\bigl( (x_{e}, x_{e}'), (\LCT{x}_e, \LCT{x}_e') \bigr)
            }
            \nonumber                                                                        \\
             & \overset{(d)}{=} \overline{ \LCT{f}(\LCTv{x}_{\setpf}, \LCTv{x}_{\setpf}') }.
          \end{align*}
          Step $(a)$ follows from the definition of $ \LCT{f} $
          in~\eqref{sec:LCT:eqn:1}.  Step $(b)$ changes the summation
          variables.  Step $(c)$ uses
          $ \matr{C}_{f} \in \setPSD{\set{X}_{\setpf}} $
          from~\eqref{sec:DENFG:eqn:6} and
          Property~\ref{prop:DENFG:LCT:1:item:8}, namely, the condition $\matr{ C }_{ M_{\efi} },\matr{ C }_{ M_{\efj} } \in
            \setHerm{\set{X}_{e}\times\LCTset{X}_{e}}$.  Step $(d)$ again
          follows from~\eqref{sec:LCT:eqn:1}.

    \item[10)] This follows from Property~\ref{prop:DENFG:LCT:1:item:9} in
          Proposition~\ref{prop:DENFG:LCT:1} and the definition of a
          weak-sense DE-NFG in Definition~\ref{sec:DENFG:def:4}.

  \end{enumerate}

  {
{\section{Proof of Lemma~\ref{sec:sst:lem:6}}
  \label{apx:alternative expression of ZBM by Pe}}
}

  The partition function of $ \hgraphNavg $, as defined in
  Definition~\ref{sec:sst:def:2}, satisfies
  {
\begin{align*}
    Z(\hgraphNavg) &\overset{(a)}{=}  \sum_{\vsigma}
    p_{\vSigma}(\vsigma) \cdot Z(\hgraphPsig)
    \nonumber           \\
     & \overset{(b)}{=}
    \sum_{\vsigma}
    p_{\vSigma}(\vsigma)
    \!\cdot\! \sum_{\vx_{[M]}}\!
    \Bigl( \prod_{m,f}\! f(\vx_{\setpff,m}) \Bigr)
    \\&\qquad\cdot \prod_{e}
    \! P_{e,\sigma_e}(\vx_{\efi,[M]}, \vx_{\efj,[M]})
    \nonumber           \\
     & \overset{(c)}{=}
    \sum_{\vx_{[M]}}\!
    \Bigl(\prod_{m,f}\! f(\vx_{\setpff,m})\Bigr)
    \!\cdot\!\prod_e\!\sum_{\sigma_e\in\set{S}_M}\!
    p_{\Sigmae}(\sigma_e)
    \\&\qquad\cdot P_{e,\sigma_e}(\vx_{\efi,[M]},\vx_{\efj,[M]})
    \nonumber \\
     & \overset{(d)}{=}
    \sum_{\vx_{[M]}}\!
    \Bigl( \prod_{m,f}\! f(\vx_{\setpff,m}) \Bigr)
    \!\cdot\! \prod_{e}\!
    P_{e}(\vx_{\efi,[M]}, \vx_{\efj,[M]}).
  \end{align*}
}
  Step $(a)$ follows from Lemma~\ref{sec:sst:lem:1}. Step $(b)$ follows from
  Definition~\ref{sec:sst:def:1}, \ie,
  the partition function of $ \hgraphPsig $ is given by
  \begin{align*}
     & \hspace{-0.5cm}
    Z(\hgraphPsig)
    \\&=  \sum_{\vx_{[M]}}
    \Biggl( \prod_{m,f} f(\vx_{\setpff,m}) \Biggr)
    \cdot \prod_{e}  P_{e,\sigma_e}(\vx_{\efi,[M]}, \vx_{\efj,[M]}).
  \end{align*}
  Step $(c)$ uses $ p_{\vSigma}(\vsigma) = \prod_{e}p_{\Sigmae}(\sigma_e) $ from~\eqref{sec:sst:eqn:31} and $ \vsigma = (\sigma_e)_{e \in \setEfull} $ from Definition~\ref{def:graph:cover:construction:1}.
  Step $(d)$ follows from Definition~\ref{sec:sst:def:4}.

  {
\vspace{0.5\baselineskip}
{\section{Proof of Theorem~\ref{sec:CheckCon:thm:1}}
  \label{apx:check_graph_thm}}
}

We prove Theorem~\ref{sec:CheckCon:thm:1} by separating the all-zero
    contribution, which equals $\bigl(\ZBSPA^*(\graphN)\bigr)^M$, from the other
    contributions in $Z(\hat{\LCT{\graphN}}_{M,\mathrm{avg}})$.
    The LCT makes the weight-one local terms vanish, allowing us to bound
    the other contributions using bounds on the product $\prod_e P_e$.
    Under the sufficient condition in Theorem~\ref{sec:CheckCon:thm:1}, the ratio
    of $Z(\hat{\LCT{\graphN}}_{M,\mathrm{avg}})$ to the all-zero contribution $\bigl(\ZBSPA^*(\graphN)\bigr)^M$
    stays between positive constants independent of $M$.
    Taking $M$-th roots gives the required limit.

  \begin{definition}\label{sec:CheckCon:def:1}
    The following definitions pertain to the LCT NFG $ \LCT{\sfN} $.
    \begin{enumerate}

      \item The set of all possible configurations for the variables on the
            edges is $ \LCTtset{X} \defeq \prod_{e} \LCTtset{X}_e$.

      \item Two collections of variables, $ \LCTvxs $ and $ \LCTvxl $, are
            defined based on an ordering of the function nodes. Assuming
            $ \set{F} = \{ f_{1},\ldots,f_{|\set{F}|} \} $ (from
            Definition~\ref{sec:DENFG:def:4}), for each edge
            $e = (f_{i},f_{j}) \in \setEfull$ with $i<j$, we define

            \begin{align*}
              \LCTvxs & \defeq
              \bigl( \LCTt{x}_{e,f_{i}}
              \bigr)_{ e = (f_{i},f_{j}) \in \setEfull },   \quad
              \LCTvxl \defeq \bigl( \LCTt{x}_{e,f_{j}}
                             \bigr)_{ e = (f_{i},f_{j}) \in \setEfull }
              \in \LCTtset{X},
            \end{align*}
            where ``$\mathrm{s}$'' stands for ``small-index'' and
            ``$\mathrm{l}$'' stands for ``large-index''.  The collections
            $ \LCTvxs $ and $\LCTvxl$ consist of distinct variables, \ie, the relation $ \LCTt{x}_{e,f_{i}} \neq \LCTt{x}_{e,f_{j}} $ is possible for any
            $ e = (f_{i},f_{j}) \in \setEfull $.
            For each $f\in\setF$, define
            $\LCTtv{x}_{\setpf,f}\defeq
              (\LCTt{x}_{e,f})_{e\in\setpf}\in\LCTtset{X}_{\setpf}$.
            Distinguish $\LCTtv{x}_{\setpf,f}$ from
            $\LCTtv{x}_{\setpf}$. The former collects the socket variables
            at $f$, whereas the latter \finaltext{collects the variables on edges incident to $f$ in $\LCT{\graphN}$, with each variable being shared by the function nodes at the two endpoints of the corresponding edge.}

      \item The functions $ \gzero $ and $ \gone $ partition the product
            of LCT local functions based on the Hamming weight of
            $ \LCTvxs \in \LCTtset{X} $ and $ \LCTvxl \in \LCTtset{X} $:
            \begin{alignat}{3}
              \gzero
              (\LCTvxs, \LCTvxl)
                                      & \defeq
              \Biggl( \prod_{f}\LCT{f}( \LCTtv{x}_{\setpf,f}) \Biggr)
              \cdot
              \bigl[ \wh(\LCTvxs) + \wh( \LCTvxl) \!=\! 0 \bigr]
              \nonumber
              \\&\overset{(a)}{=} \ZBSPA^{*}(\graphN)
              \cdot \bigl[ \LCTvxs\!=\!\LCTvxl \!=\! \tv{0} \bigr],
              \label{sec:CheckCon:eqn:60}      \\
              \gone(\LCTvxs, \LCTvxl) & \defeq
              \Biggl( \prod_{f}\LCT{f}( \LCTtv{x}_{\setpf,f}) \Biggr) \cdot
              \bigl[ \wh(\LCTvxs) + \wh( \LCTvxl) \!\geq\! 1 \bigr].
              \label{sec:CheckCon:eqn:61}
            \end{alignat}
            Eq.~\eqref{sec:CheckCon:eqn:31} gives
            $ \prod_{f}\LCT{f}( \tv{0} ) = \ZBSPA^{*}(\graphN) $, while
            $ \wh(\LCTvxs) + \wh( \LCTvxl) = 0 $ implies
            $ \LCTvxs = \LCTvxl = \tv{0} $. These facts prove step $(a)$.
            Here, the Hamming weight $\wh(\mathbf v)$ is the number of nonzero
            entries of $\mathbf v$.  Note that the support of
              the function $ \gzero $ and the support of the function
              $ \gone $ do not overlap. It follows then immediately from the
            definitions of $ \gzero $ and $ \gone $, along with
            $\ZBSPA^{*}(\graphN) \geq 0$, that
            \begin{align}
               & \prod_{f}\LCT{f}( \LCTtv{x}_{\setpf,f})=
              \gzero
              (\LCTvxs, \LCTvxl)
              + \gone(\LCTvxs, \LCTvxl). \label{eqn: property of g0 g1} \\
               & \prod_{f} \bigl| \LCT{f}( \LCTtv{x}_{\setpf,f}) \bigr| =
              \gzero
              (\LCTvxs, \LCTvxl)
              + \bigl| \gone(\LCTvxs, \LCTvxl) \bigr|. \label{eqn: property of g0 g1:2}
            \end{align}
    \end{enumerate}
    In order to analyze the average $M$-cover of $ \LCT{\sfN} $,
      introduce the following variables, indexed by $m \in [M]$, for
      configurations across the $M$ copies.
    \begin{enumerate}
      \setcounter{enumi}{3} 
      \item For every $ f \in \setF $, $ e \in \setpf $, and $ m \in [M] $,
              let $ \LCTt{x}_{\ef,m} \in \LCTtset{X}_e$. Collect these variables
              across the $M$ copies as
            $ \LCTtv{x}_{\ef,[M]} \defeq ( \LCTt{x}_{\ef,m} )_{m \in [M]} \in
              \LCTtset{X}_{e}^{M} $.

      \item Define the collections of ``small-index'' variables by
            \begin{align*}
               & \LCTvxsm \defeq ( \LCTt{x}_{\efi,m} )_{ e = (f_{i},f_{j}) \in \setEfull:\, i <j }
              \in \LCTtset{X}, \qquad m \in [M],
              \\
               & \LCTvxsM \defeq (\LCTvxsm)_{m \in [M]} \in \LCTtset{X}^{M}.
            \end{align*}
            We define $ \LCTvxlm $ and $ \LCTvxlM $ analogously for the
            ``large-index'' variables.
    \end{enumerate}

    When unambiguous, we use the shorthands $ ( \cdot )_{m} $,
    $\prod_{\LCTt{x}_{e}}$, $ \sum_{ \LCTvxs, \LCTvxl } $,
    $ \sum_{\LCTvxsM} $, and $ \sum_{\LCTvxlM} $ instead of
      $ (\cdot)_{m \in [M]} $, $\prod_{\LCTt{x}_{e} \in \LCTtset{X}_{e}}$,
    $ \sum_{ \LCTvxs, \LCTvxl \in \LCTtset{X} } $,
    $ \sum_{\LCTvxsM \in \LCTtset{X}^{M} } $, and
    $ \sum_{\LCTvxlM \in \LCTtset{X}^{M} } $, respectively.  \edefinition
  \end{definition}

  \begin{remark}
    \label{remark: expression of Pe}
    For the LCT NFG $ \LCT{\sfN} $, the analysis in
    Section~\ref{sec:sst:DENFG:1} remains valid after replacing $ \txe $ and
    $ \tset{X}_{e} $ with $ \LCTt{x}_{e} $ and $ \LCTtset{X}_{e} $,
    respectively.  In particular, for each $ e = (f_{i},f_{j}) \in \setEfull $
    in $ \LCT{\sfN} $, the function $ P_{e} $ (related to the average over
    permutations in the $M$-cover) is
    \begin{align*}
       & P_{e}\bigl( \LCTtv{x}_{\efi,[M]}, \LCTtv{x}_{\efj,[M]} \bigr)
      =\begin{cases}
         \binom{M}{ M \cdot \vt_{e} }^{\!\! -1}
           & \begin{aligned}[t]
               \text{if } \vt_{e}
                & =\vt_e(\LCTtv{x}_{\efi,[M]})  \\
                & = \vt_e(\LCTtv{x}_{\efj,[M]})
             \end{aligned} \\
         0 & \text{otherwise}
       \end{cases},
    \end{align*}
    where
    \begin{align*}
      \binom{M}{ M \cdot \vt_{e} }
      \defeq \frac{M!}{ \prod\limits_{\LCTt{x}_{e} \in \LCTtset{X}_{e}}
               \bigl( ( M \cdot t_{e,\LCTt{x}_{e}} ) ! \bigr) }.
    \end{align*}
    Here, the quantities $ \vt_{e} $, $ \vt_{e}(\cdot) $, and $ t_{e,\LCTt{x}_{e}} $ are defined in Definition~\ref{sec:sst:def:6}.
    \eremark
  \end{remark}

  \begin{lemma}
    \label{lem: expression of go and g1 and pe}
    The $M$-th power of the degree-$M$ Bethe partition function is
    \begin{align*}
       & \bigl( \ZBM(\graphN) \bigr)^{\! M}
      \\&=
      \sum_{ k = 0 }^{M}
      \binom{M}{k}
      \cdot
      \bigl( \ZBSPA^{*}( \graphN) \bigr)^{\! k}
      \cdot\!\!\!\!\!\! \!\!\! \sum_{ \LCTvxsM, \LCTvxlM }\!\!\! \!\!\!
      \bigl[ \LCTvxsm \! = \! \LCTvxlm \! = \! \tv{0},\, \forall m \in [k] \bigr]
      \\&\qquad\cdot
      \left(
      \prod_{ m \in [M] \setminus [k] }
      \gone( \LCTvxsm, \LCTvxlm )
      \right) \cdot \prod_{e}
      P_{e}\bigl( \LCTtv{x}_{\efi,[M]}, \LCTtv{x}_{\efj,[M]} \bigr).
    \end{align*}
  \end{lemma}
  \begin{proof}
    The derivation begins with the expression for
    $ \bigl( \ZBM(\graphN) \bigr)^{\! M} $ from~\eqref{eqn: Z for degree M
      average cover of DENFG} and applies a series of algebraic manipulations:
    \begin{align}
       &
      \bigl( \ZBM(\graphN) \bigr)^{\! M}
      \nonumber                                                  \\
       & \overset{(a)}{=}
      \!\!\!\!\sum_{ \LCTvxsM, \LCTvxlM }\!\!\!\!
      \Biggl(
      \prod_{m \in [M]} \prod_{f}
      \LCT{f}( \LCTtv{x}_{\setpf,f,m})
      \Biggr)
      \!\!\cdot\!\! \prod_{e}
      P_{e}(\LCTtv{x}_{\efi,[M]}, \LCTtv{x}_{\efj,[M]})
      \nonumber                                                  \\
       & \overset{(b)}{=}
      \!\!\!\!\sum_{ \LCTvxsM, \LCTvxlM }\!\!\!\!
      \Biggl(
      \prod_{m \in [M]}
      \Bigl( \gzero( \LCTvxsm, \LCTvxlm )
      + \gone( \LCTvxsm, \LCTvxlm )
      \Bigr)
      \Biggr)
      \nonumber                                                  \\&\qquad \qquad \qquad\qquad\qquad\qquad\qquad \cdot \prod_{e}
      P_{e}(\LCTtv{x}_{\efi,[M]}, \LCTtv{x}_{\efj,[M]})
      \nonumber                                                  \\
       & =
      \sum_{ k = 0 }^{M} 
      \sum_{ \set{M} \subseteq [M]: |\set{M}| = k}\, 
      \sum_{ \LCTvxsM, \LCTvxlM }
      \Biggl( \prod_{ m \in \set{M} } \gzero( \LCTvxsm, \LCTvxlm ) \Biggr)
      \nonumber                                                  \\
       & \qquad
      \cdot \Biggl( \prod_{ m \in [M] \setminus \set{M} } \gone( \LCTvxsm, \LCTvxlm ) \Biggr)
      \cdot \prod_{e}
      P_{e}(\LCTtv{x}_{\efi,[M]}, \LCTtv{x}_{\efj,[M]})
      \nonumber                                                  \\
       & \overset{(c)}{=}
      \sum_{ k = 0 }^{M}
      \sum_{ \set{M} \subseteq [M]: |\set{M}| = k}
      \sum_{ \LCTvxsM, \LCTvxlM } \!\!\!
      \Biggl( \prod_{ m \in [k] } \gzero( \LCTtv{x}_{\mathrm{s}, \sigma_{\set{M}}(m)},
      \LCTtv{x}_{\mathrm{l}, \sigma_{\set{M}}(m)} ) \!\! \Biggr)
      \nonumber                                                  \\
       &
      \cdot \!\! \Biggl( \prod_{ m \in [M] \setminus [k] } \!\!\!\!\!\!  \gone( \LCTtv{x}_{\mathrm{s}, \sigma_{\set{M}}(m)},
      \LCTtv{x}_{\mathrm{l}, \sigma_{\set{M}}(m)} ) \!\! \Biggr)
      \! \cdot
      \prod_{e}
      \!\!\!\!\!
      \underbrace{
        P_{e}(\LCTtv{x}_{\efi,[M]}, \LCTtv{x}_{\efj,[M]})
      }_{ \!\!\!\!\!\!  \!\!\!\!\!\! \overset{(d)}{=} P_{e}(\LCTtv{x}_{\efi,\sigma_{\set{M}}([M])}, \LCTtv{x}_{\efj,\sigma_{\set{M}}([M])}) }
      \nonumber                                                  \\
       & \overset{(e)}{=}
      \sum_{ k = 0 }^{M}
      \binom{M}{k}
      \sum_{ \LCTvxsM, \LCTvxlM }
      \Biggl( \prod_{ m \in [k] } \gzero( \LCTvxsm, \LCTvxlm ) \Biggr)
      \nonumber                                                  \\
       & \quad
      \cdot \Biggl( \prod_{ m \in [M] \setminus [k] } \gone( \LCTvxsm, \LCTvxlm ) \Biggr)
      \cdot \prod_{e}
      P_{e}(\LCTtv{x}_{\efi,[M]}, \LCTtv{x}_{\efj,[M]})\nonumber \\
       & \overset{(f)}{=}
      \sum_{ k = 0 }^{M}
      \binom{M}{k}
      \cdot
      \bigl( \ZBSPA^{*}( \graphN) \bigr)^{\! k}
      \! \cdot \!\!\!\!\! \!\!\!\!\! \sum_{ \LCTvxsM, \LCTvxlM } \!\!\!\!\!
      \bigl[ \LCTvxsm \! = \! \LCTvxlm \! = \! \tv{0},\, \forall m \in [k] \bigr]
      \nonumber
      \\&\qquad\cdot
      \left(
      \prod_{ m \in [M] \setminus [k] }
      \gone( \LCTvxsm, \LCTvxlm )
      \right) \!\! \cdot  \prod_{e}
      P_{e}\bigl( \LCTtv{x}_{\efi,[M]}, \LCTtv{x}_{\efj,[M]} \bigr).\nonumber
    \end{align}
    Step $(a)$ follows by
    replacing $ \tx_{e} $, $ \tset{X}_e $, and
    $ f $ with $ \LCTt{x}_{e} $, $ \LCTtset{X}_{e} $, and $ \LCT{f} $, respectively, in~\eqref{eqn: Z for degree M average cover of DENFG}
    (the equation remains valid after these substitutions).
    Step $(b)$ follows from~\eqref{eqn: property of g0 g1}.
    Step $(c)$ introduces a permutation $ \sigma_{\set{M}} \in \set{S}_{M} $ satisfying
    $ \sigma_{\set{M}}( [k] ) = \set{M} $
    and
    $ \sigma_{\set{M}}( [M] \setminus [k] ) = [M] \setminus \set{M} $.
    Step $(d)$ first uses
    \begin{align}
      \bm{t}_{e}(\LCTtv{x}_{\efi,[M]}) & = \bm{t}_{e}(\LCTtv{x}_{\efi,\sigma_{\set{M}}([M])}), \label{eqn: the type of lct vx2} \\   \bm{t}_{e}(\LCTtv{x}_{\efj,[M]}) &= \bm{t}_{e}(\LCTtv{x}_{\efj,\sigma_{\set{M}}([M])})\finaltext{.}
      \label{eqn: the type of lct vx}
    \end{align}
    The expression for $ P_{e} $ in Remark~\ref{remark: expression of Pe} then implies that
    if~\eqref{eqn: the type of lct vx2} and~\eqref{eqn: the type of lct vx} hold,
    then
    \begin{align*}
      P_{e}( \LCTtv{x}_{\efi,[M]},\LCTtv{x}_{\efj,[M]} )
      = P_{e}( \LCTtv{x}_{\efi,\sigma_{\set{M}}([M])}, \LCTtv{x}_{\efj,\sigma_{\set{M}}([M])} )\finaltext{.}
    \end{align*}
    Step $(e)$ uses relabeling invariance: the sum
    $\sum_{ \LCTvxsM, \LCTvxlM } (\ \cdots\ )$ has the same value for every choice
    of the $k$-element set $\set{M}$, and step $(d)$ shows that $P_e$ is
    invariant under the same relabeling. The $\binom{M}{k}$ choices of
    $\set{M}$ therefore produce the displayed binomial factor.  Step $(f)$
    follows from~\eqref{sec:CheckCon:eqn:60}.
  \end{proof}

  In order to prove Theorem~\ref{sec:CheckCon:thm:1}, we bound each term
    in the expansion of the $M$-th power of the degree-$M$ Bethe partition
    function in Lemma~\ref{lem: expression of go and g1 and pe}. The next
    three lemmas establish the binomial, LCT-support, and cover-weight
    estimates needed for the key results.

  \begin{lemma}\label{sec:CheckCon:lem:2}
    \noindent
    \begin{enumerate}
      \item The binomial coefficient $ \binom{M}{k} $ is increasing in
              $ k \in \sZ $ for $ 0 \leq k \leq M/2 $ and decreasing in
              $ k \in \sZ $ for $M/2 \leq k \leq M $.

      \item Consider an arbitrary integer $ M' \in \sZpp $ and an arbitrary
            sequence of $M'$ integers $ k_{1},\ldots,k_{M'} \in \sZp $.  We define
            $K_{M'} \defeq \sum_{m' \in [M']} k_{m'}$. If $ K_{M'} \leq M $,
            then we have
            \begin{align}
              \prod\limits_{m' \in [M']}\binom{M}{ k_{m'} }
              \geq
              \binom{M}{K_{M'}}. \label{sec:CheckCon:eqn:82}
            \end{align}
    \end{enumerate}

  \end{lemma}
  \begin{proof}
    \noindent
    \begin{enumerate}
      \item The ratio $\binom{M}{k+1}/\binom{M}{k}=(M-k)/(k+1)$ is at least
              one exactly when $k\leq(M-1)/2$, which proves the claim.

      \item  We have
            \begin{align}
               & \Biggl( \prod\limits_{m' \in [M']}\binom{M}{ k_{m'} } \Biggr)
              \cdot
              \frac{
                \biggl( M-K_{M'} \biggr)!
                \cdot \prod\limits_{m' \in [M']}(k_{m'}!)
              }{
                M!
              }
              \nonumber                                                        \\
               & \quad = \frac{
                           \prod\limits_{m' \in [M']}
                           \bigl( M \cdot ( M - 1 ) \cdots
                           (M +1 -k_{m'}) \bigr)
                         }{
                           M \cdot ( M - 1 ) \cdots
                           \left( M +1 - K_{M'} \right)
                         } \geq 1. \label{eqn: bound on binon coeff: 1}
            \end{align}
            Moreover, by the definition of binomial coefficients, we obtain
            \begin{align}
              \frac{
                M!
              }{
                \biggl( M-K_{M'} \biggr)!
                \cdot \prod\limits_{m' \in [M']}(k_{m'}!)
              }\geq \binom{M}{K_{M'}}. \label{eqn: bound on binon coeff: 2}
            \end{align}
            Combining the inequalities in~\eqref{eqn: bound on binon coeff: 1}
            and~\eqref{eqn: bound on binon coeff: 2} yields~\eqref{sec:CheckCon:eqn:82}.\qedhere
    \end{enumerate}

  \end{proof}
  
  \begin{lemma}
    \label{sec:CheckCon:prop:properties of N(0)}
    \noindent
    \begin{enumerate}

      \item It holds that
            \begin{align*}
               & \sum_{ \LCTvxs, \LCTvxl }
              \gone( \LCTvxs, \LCTvxl )
              = \prod_{f}
              \left(
              \sum_{\LCTtv{x}_{\setpf,f}}
              \LCT{f}( \LCTtv{x}_{\setpf,f} )
              \right)
              -
              \ZBSPA^{*}(\graphN).
            \end{align*}

      \item It holds that
            \begin{align*}
               & \sum_{ \LCTvxs, \LCTvxl }
              \bigl| \gone( \LCTvxs, \LCTvxl ) \bigr|
              = \prod_{f}
              \left(
              \sum_{\LCTtv{x}_{\setpf,f}}
              \bigl| \LCT{f}( \LCTtv{x}_{\setpf,f} ) \bigr|
              \right)
              -
              \ZBSPA^{*}(\graphN).
            \end{align*}

      \item The function $\gone$ satisfies
            \begin{align*}
              \gone( \LCTvxs, \LCTvxl ) = \gone( \LCTvxs, \LCTvxl) \cdot
              \bigl[ \wh(\LCTvxs) \! + \! \wh( \LCTvxl) \! \geq \! 2 \bigr].
            \end{align*}

          \end{enumerate}
          
  \end{lemma}

  \begin{proof}
    Summing~\eqref{eqn: property of g0 g1} over $\LCTvxs$ and $\LCTvxl$ and
    using~\eqref{sec:CheckCon:eqn:60} proves the first
    identity. Summing~\eqref{eqn: property of g0 g1:2} over $\LCTvxs$ and
    $\LCTvxl$ and using~\eqref{sec:CheckCon:eqn:60} proves
    the second identity. (Note that for the first and the second identity we
    used the fact that the variable collections $\LCTtv{x}_{\setpf,f}$ for
    different function nodes $f$ are disjoint.) For the third identity,
    suppose that $\wh(\LCTvxs)+\wh(\LCTvxl)=1$. Exactly one transformed
    half-edge variable is then nonzero, so one local configuration has Hamming
    weight one. Property~\ref{sec:LCT:prop:1:item:8} of
    Proposition~\ref{prop:DENFG:LCT:1} makes the corresponding LCT local
    function, and hence also $\gone(\LCTvxs,\LCTvxl)$, equal to zero. Thus
    $\gone$ is supported only on configurations of total Hamming weight at
    least two.
  \end{proof}

  \begin{lemma} \label{sec:CheckCon:lem: lower bound of Pe}
    Consider an integer $ M \in \sZpp $ and an integer $ k $ satisfying $ 0 \leq k \leq M-1 $.
    For any $ \LCTvxsM,\LCTvxlM \in \LCTtset{X}^{M} $, if
    \begin{alignat}{4}
      \LCTvxsm      & = \LCTvxlm = \tv{0},\qquad     & \forall & m \in [k], \label{eqn: assum on k} \\
      \wh(\LCTvxsm) & + \wh( \LCTvxlm) \geq 2,\qquad & \forall & m \in [M] \setminus [k],
      \label{eqn: assum on M - k}
    \end{alignat}
    then
    \begin{align*}
      \prod_{e}
      P_{e}\bigl( \LCTtv{x}_{\efi,[M]}, \LCTtv{x}_{\efj,[M]} \bigr)
      \leq \binom{M}{k}^{\!\! -1}.
    \end{align*}
  \end{lemma}

  \begin{proof}
    In this proof, the collections of variables $ \LCTvxsM,\LCTvxlM \in \LCTtset{X}^{M} $ are fixed.
    For brevity, we use the shorthand $ P_{e} $ instead of
    $ P_{e}\bigl( \LCTtv{x}_{\efi,[M]}, \LCTtv{x}_{\efj,[M]} \bigr)$.

    If there exists an edge
    $ e_{1} \in \setEfull $ such that
    $ P_{e_{1}} \leq \binom{ M }{ k }^{\!\! -1} $,
    then we have
    $
      \prod_{e}
      P_{e}
      =
      P_{e_{1}}
      \cdot
      \prod_{e \in \setEfull \setminus \{e_{1}\}}
      P_{e}
      \overset{(a)}{\leq}\binom{M}{k}^{\!-1},
    $
    where step $(a)$ follows from $ P_{e} \leq 1 $ for all $ e \in \setEfull $, as is immediate from Remark~\ref{remark: expression of Pe}.

    If no such edge exists, then we have
    $
      P_{e}
      > \binom{ M }{ k }^{\!\! -1}
    $ for all $ e \in \setEfull $.

    Fix an arbitrary edge $ e = (f_{i},f_{j}) \in \setEfull $.
    From the expression of $P_{e}$ in Remark~\ref{remark: expression of Pe},
    the following expressions hold:
    \begin{alignat}{3}
      \binom{M}{ M \cdot \vt_{e} }^{\!\!-1}
       & > \binom{ M }{ k }^{\!\! -1},
      \label{eqn: ineq of types te}     \\
      \vt_{e}
      = \vt_e(\LCTtv{x}_{\efi,[M]})
       & = \vt_e(\LCTtv{x}_{\efj,[M]}).
      \label{eqn: positive of pe}
    \end{alignat}
    We claim that
    \begin{alignat}{3}
      M \cdot t_{e,\tzero} \geq M/2.
      \label{eqn: lower bound of tzero}
    \end{alignat}
    In order to prove~\eqref{eqn: lower bound of tzero}, we note that expression~\eqref{eqn: assum on k} implies
    \begin{align}
      M \cdot t_{e,\tzero} \geq k. \label{eqn: ineq of M tero: 1}
    \end{align}
If $k\geq M/2$, then \eqref{eqn: ineq of M tero: 1}
    immediately proves~\eqref{eqn: lower bound of tzero}.
    Therefore, assume $k<M/2$. We also have
    {
    \begin{align}
      \binom{M}{M\cdot t_{e,\tzero}}
      &=\frac{M!}{(M\cdot t_{e,\tzero})!\,(M-M\cdot t_{e,\tzero})!}\nonumber\\
      &\overset{(a)}{\leq}\binom{M}{M\cdot\vt_e}\nonumber\\
      &\overset{(b)}{<}\binom{M}{k}\nonumber\\
      &=\binom{M}{M-k}.\label{eqn: ineq of M tero: 2}
    \end{align}
    }%
    Step $(a)$ follows from the definition of the type in
    Definition~\ref{sec:sst:def:6} and the inequality
    $(M-M\cdot t_{e,\tzero})! \geq \prod_{\LCTt z\neq\tzero}(M\cdot t_{e,\LCTt z})!$.
    Step $(b)$ follows from~\eqref{eqn: ineq of types te}.
    Now, by the first statement of Lemma~\ref{sec:CheckCon:lem:2},
    $\binom{M}{n}\geq\binom{M}{k}$ whenever $k\leq n\leq M-k$.
    Hence, \eqref{eqn: ineq of M tero: 1} and
    \eqref{eqn: ineq of M tero: 2} force $M\cdot t_{e,\tzero}>M-k$.
    Since $k<M/2$, we have $M-k>M/2$, proving
    \eqref{eqn: lower bound of tzero} in this case.%

    We now allow $ e = (f_{i},f_{j}) $ to range over $\setEfull$.
    We have
    \begin{align}
      2(M-k)
       & \overset{(a)}{\leq}
      \sum_{m \in [M]} \bigl( \wh(\LCTvxsm) + \wh( \LCTvxlm) \bigr)
      \nonumber                            \\
       & = \wh(\LCTvxsM) + \wh(\LCTvxlM)
      \nonumber                            \\
       & \overset{(b)}{=}
      2 \wh(\LCTvxsM)
      \nonumber                            \\
       & =
      2 \cdot \sum_{e}
      \sum_{ \LCTt{z}_{e} \in \LCTtset{X}_{e} \setminus \{\tzero\} }
      M \cdot t_{ e, \LCTt{z}_{e} }
      \nonumber                            \\
       & \overset{(c)}{=} 2 \cdot \sum_{e}
      ( M - M \cdot t_{ e, \tzero } ).
      \label{existence of ke}
    \end{align}
    Step $(a)$ follows from~\eqref{eqn: assum on M - k}.
    Step $(b)$ follows from~\eqref{eqn: positive of pe}, which holds for all $ e \in \setEfull $,
    and from the property that
    if $ \vt_e(\LCTtv{x}_{\efi,[M]}) = \vt_e(\LCTtv{x}_{\efj,[M]}) $,
    then $ \wh(\LCTtv{x}_{\efi,[M]}) = \wh(\LCTtv{x}_{\efj,[M]}) $.
    Step $(c)$ follows from $ \sum_{ \LCTt{z}_{e} \in \LCTtset{X}_{e} } t_{ e, \LCTt{z}_{e} } = 1 $.
    The inequality in~\eqref{existence of ke} implies the existence of a collection of integers
    $ (k_{e})_{e \in \setEfull} \in \sZp^{|\setEfull|} $ such that
    \begin{align}
      M - M \cdot t_{e,\tzero} - k_{e}     & \geq 0, \qquad e \in \setEfull,
      \nonumber                                                              \\
      \sum_{e}
      ( M - M \cdot t_{e,\tzero} - k_{e} ) & = M-k.
      \label{sec:CheckCon:eqn:91}
    \end{align}
    Therefore, it holds that
    \begin{align*}
      \prod_{e}
      P_{e}^{-1}
       & =
      \prod_{ e }
      \binom{M}{ M \cdot \vt_{e} }
      \nonumber                             \\
       &
      \overset{(a)}{\geq}
      \prod_{ e } \binom{ M }{
                    M \cdot t_{ e, \tzero }
                  }
      \nonumber                             \\
       & \overset{(b)}{\geq}
      \prod_{ e }
      \binom{ M }{
        M \cdot t_{e,\tzero} + k_{e}
      } \nonumber         \\
       & = \prod_e\binom{M}{M-M\cdot t_{e,\tzero}-k_e} \\
       & \overset{(c)}{\geq}
      \binom{M}{\sum\limits_{e}
        ( M - M \cdot t_{e,\tzero} - k_{e} )}
      \nonumber                             \\
       & \overset{(d)}{=} \binom{ M }{ k }.
    \end{align*}
    Step $(a)$ follows from
    \begin{align*}
      \binom{M}{ M \cdot \vt_{e} }
       & = \frac{ M! }{
             \bigl( M \cdot t_{ e, \tzero } \bigr)!
             \cdot
             \prod\limits_{ \LCTt{x}_{e} \in \LCTtset{X}_{e} \setminus \{\tzero\} }
             \Bigl(
             \bigl( M \cdot t_{ e, \LCTt{x}_{e} } \bigr)!
             \Bigr)
           }
      \\
       & \geq
      \frac{ M! }{
        \bigl( M \cdot t_{ e, \tzero } \bigr)!
        \cdot
        \Bigl(
        \bigl( M - M \cdot t_{ e, \tzero } \bigr)!
        \Bigr)
      }
      \\
       & =
      \binom{ M }{
        M \cdot t_{ e, \tzero }
      }.
    \end{align*}
    Step $(b)$ follows from the first statement in Lemma~\ref{sec:CheckCon:lem:2}, together with the facts that $ k_{e} \in \sZp $ and
    $ M \cdot t_{ e, \tzero } \geq M/2 $, as proved in~\eqref{eqn: lower bound of tzero}.
    Step $(c)$ follows from~\eqref{sec:CheckCon:eqn:82}, and step $(d)$ follows from~\eqref{sec:CheckCon:eqn:91}.
  \end{proof}

  We need one more definition and lemma before being able to prove
  Theorem~\ref{sec:CheckCon:thm:1}.

  \begin{definition}
    \label{def:alpha:N:constant:1}      
    Define the constant
    \begin{align*}
      \alpha_{\graphN}
       & \defeq
      \frac{
        \sum\limits_{ \LCTvxs, \LCTvxl }
        \bigl|\gone( \LCTvxs, \LCTvxl )\bigr|
      }{
        \ZBSPA^{*}(\graphN) }.
    \end{align*}
    (Recall from Assumption~\ref{sec:DENFG:remk:1} that
    $ \ZBSPA^{*}(\graphN) \in \sRpp $.)
    \edefinition
  \end{definition}

  \begin{lemma}
    \label{lemma:alpha:N:bounds}
    It holds that
    \begin{align*}
      0
        &\leq
           \alpha_{\graphN}
         < \frac{1}{2}.
    \end{align*}
  \end{lemma}

  \begin{proof}
    The first inequality is straightforward. For the second inequality, we have
    \begin{align*}
      \alpha_{\graphN}
      &\overset{(a)}{=}
        \frac{\sum_{\LCTvxs,\LCTvxl} \bigl| \gone(\LCTvxs,\LCTvxl) \bigr|}
             {\ZBSPA^{*}(\graphN)} \\
      &\overset{(b)}{=}
         \frac{\prod_f\Bigl(\sum_{\LCTtv{x}_{\setpf,f}}
                 \bigl| \LCT f(\LCTtv{x}_{\setpf,f})\bigr| \Bigr) - \ZBSPA^*(\graphN)}
              {\ZBSPA^{*}(\graphN)} \\ 
      &= \underbrace{\frac{\prod_f\Bigl(\sum_{\LCTtv{x}_{\setpf,f}}
                 \bigl| \LCT f(\LCTtv{x}_{\setpf,f}) \bigr| \Bigr)}
              {\ZBSPA^{*}(\graphN)}}_{\overset{(c)}{<} 3/2}
         -
         1 \\ 
      &< \frac{1}{2},
    \end{align*}
    where step~$(a)$ follows from the definition of $\alpha_{\graphN}$,
    step~$(b)$ follows from the second identity in
    Lemma~\ref{sec:CheckCon:prop:properties of N(0)}, and step~$(c)$ follows
    from~\eqref{sec:CheckCon:eqn:70}.
  \end{proof}

  We now prove Theorem~\ref{sec:CheckCon:thm:1}, which states that
  $
    \lim\limits_{M\to\infty}\ZBM(\graphN)
    = \ZBSPA^{*}(\graphN)
  $
  under condition~\eqref{sec:CheckCon:eqn:70}, by showing that
  $
    \liminf\limits_{M\to\infty}\ZBM(\graphN)
    \geq \ZBSPA^{*}(\graphN)$ and
  $
    \limsup\limits_{M\to\infty}\ZBM(\graphN)
    \leq \ZBSPA^{*}(\graphN).
  $
  \begin{lemma}\label{sec:CheckCon:lem:liminf Z N0 greater than ZSPA}
    If the strict inequality in~\eqref{sec:CheckCon:eqn:70} holds, then
    $
      \liminf_{M \to \infty}
      \ZBM(\graphN)
      \geq \ZBSPA^{*}(\graphN).
    $
  \end{lemma}

  \begin{proof}

    For each $M$, set
    \begin{align}
      R_M
      \defeq
      \Biggl( \frac{ \ZBM(\graphN) }{ \ZBSPA^{*}(\graphN) } \Biggr)^{\!M} \overset{(a)}{\in} \sRp.
      \label{sec:CheckCon:eqn:R-def}
    \end{align}
    Each finite graph cover of $\graphN$ is a DE-NFG, so its partition
      function is a nonnegative real number. Hence, we have $\ZBM(\graphN)\in\sRp$. Assumption~\ref{sec:DENFG:remk:1} gives
      $\ZBSPA^{*}(\graphN)\in\sRpp$. Therefore, we have $R_M\in\sRp$, which proves
      step $(a)$. Lemma~\ref{lem: expression of go and g1 and pe} then gives
      \begin{align}
        R_M
         & = 1
        +
        \sum_{k=0}^{M-1}
        \bigl( \ZBSPA^{*}(\graphN) \bigr)^{\! k-M}
        \cdot
        T_{M,k},
        \label{sec:CheckCon:eqn:R-expansion}
      \end{align}
      where, for $0 \leq k \leq M-1$, we have defined
      \begin{align*}
        T_{M,k}
         & \defeq
        \binom{M}{k}
        \cdot
        \sum_{ \LCTvxsM, \LCTvxlM }
        \bigl[ \LCTvxsm \! = \! \LCTvxlm \! = \! \tv{0},\, \forall m \in [k] \bigr]
        \\
         & \
        \cdot
        \Biggl(
        \prod_{ m \in [M] \setminus [k] }
        \gone( \LCTvxsm, \LCTvxlm )
        \Biggr)
        \! \cdot
        \prod_{e}
        P_{e}\bigl( \LCTtv{x}_{\efi,[M]}, \LCTtv{x}_{\efj,[M]} \bigr).
      \end{align*}
      In order to obtain a lower and an upper bound on $R_M$, we study
      $|T_{M,k}|$. We obtain
      \begin{align}
         & |T_{M,k}|                                                                                   \\
         & \overset{(a)}{\leq}
        \binom{M}{k}
        \cdot
        \sum_{ \LCTvxsM, \LCTvxlM }
        \bigl[ \LCTvxsm \! = \! \LCTvxlm \! = \! \tv{0},\, \forall m \in [k] \bigr]
        \nonumber
        \\
         & \quad\ \cdot
        \Biggl(
        \prod_{ m \in [M] \setminus [k] }
        \bigl| \gone( \LCTvxsm, \LCTvxlm ) \bigr|
        \Biggr)
        \cdot
        \prod_{e} P_{e}\bigl( \LCTtv{x}_{\efi,[M]}, \LCTtv{x}_{\efj,[M]} \bigr)
        \nonumber                                                                                      \\
         & \overset{(b)}{\leq} \!\!\!
        \sum_{ \LCTvxsM, \LCTvxlM } \!\!\!\!\!\!\! \bigl[ \LCTvxsm
                                                     \! = \! \LCTvxlm  \! = \! \tv{0},\, \forall m \in [k] \bigr] \cdot
        \!\!\!\!\!\!\!
        \prod_{ m \in [M] \setminus [k] } \bigl| \gone( \LCTvxsm, \LCTvxlm ) \bigr|
        \nonumber                                                                                      \\
         & \overset{(c)}{=}
        \Biggl( \prod\limits_{m \in [k]} \underbrace{\sum\limits_{ \LCTvxsm
                                             \atop \LCTvxlm } \bigl[ \LCTvxsm \! = \! \LCTvxlm \! = \! \tv{0}
                                             \bigr]}_{=1} \! \Biggr)
        \! \cdot \!\!\!\!\! \prod\limits_{m \in [M] \setminus [k] } \sum\limits_{ \LCTvxsm
                                                                          \atop \LCTvxlm } \! \bigl| \gone( \LCTvxsm, \LCTvxlm ) \bigr|
        \nonumber                                                                                      \\
         & = \Biggl( \sum\limits_{ \LCTvxs, \LCTvxl } |\gone( \LCTvxs, \LCTvxl )| \Biggr)^{\!\!\! M-k}
        \nonumber                                                                                      \\
         &
        \overset{(d)}{=}
        \bigl( \ZBSPA^{*}( \graphN) \bigr)^{\! M-k}
        \cdot
        \alpha_{\graphN}^{\!M-k}.
        \label{sec:CheckCon:eqn:T-def}
      \end{align}
    Step $(a)$ follows from the triangle inequality. Step $(b)$ follows
      termwise. Namely, if $\gone(\LCTvxsm,\LCTvxlm)=0$ for some
      $m\in[M]\setminus[k]$, the corresponding summand vanishes. Otherwise,
      the LCT support identity in
      Lemma~\ref{sec:CheckCon:prop:properties of N(0)} gives
      \begin{align*}
        \wh(\LCTvxsm)+\wh(\LCTvxlm)\geq2,
        \qquad m\in[M]\setminus[k].
      \end{align*}%
      Together with the prescribed all-zero configurations for $m\in[k]$,
      Lemma~\ref{sec:CheckCon:lem: lower bound of Pe} yields
      $\prod_eP_e\leq\binom{M}{k}^{-1}$, which cancels the factor
      $\binom{M}{k}$. Step $(c)$ follows from separating
      the summation $ \sum_{ \LCTvxsM, \LCTvxlM }(\ \cdots\ ) $. Step $(d)$
      follows from~Definition~\ref{def:alpha:N:constant:1}. Because $R_M \geq 0$, we
      obtain
      \begin{align}
        R_M
        = |R_M|
         & \overset{(a)}{\geq}
        1
        -
        \sum_{k=0}^{M-1} \bigl( \ZBSPA^{*}( \graphN) \bigr)^{\! k-M} \cdot |T_{M,k}| \nonumber \\
         & \overset{(b)}{\geq}
        1-\sum_{k=0}^{M-1}\alpha_{\graphN}^{M-k}
        \geq 1-\frac{\alpha_{\graphN}}{1-\alpha_{\graphN}}
        \overset{(c)}{>}0.
        \label{sec:CheckCon:eqn:75}
      \end{align}%
      Step $(a)$ follows from~\eqref{sec:CheckCon:eqn:R-expansion} and the
      reverse triangle inequality. Step $(b)$ follows
      from~\eqref{sec:CheckCon:eqn:T-def}. Step $(c)$ follows
      from Lemma~\ref{lemma:alpha:N:bounds}. Therefore,%
    \begin{align*}
      \liminf_{M \to \infty}
      \ZBM(\graphN)
       & =
      \liminf_{M \to \infty}
      \Bigl( \ZBSPA^{*}(\graphN)
      \cdot (R_M)^{1/M} \Bigr)
      \\  & \overset{(a)}{\geq}
      \ZBSPA^{*}(\graphN)
      \cdot
      \liminf_{M \to \infty}
      \Biggl(
      1 -
      \frac{ \alpha_{\graphN}
      }{ 1-\alpha_{\graphN} }
      \Biggr)^{\! \!\! 1/M}
      \nonumber                                \\
       & \overset{(b)}{=} \ZBSPA^{*}(\graphN),
    \end{align*}
    where step $(a)$ follows from~\eqref{sec:CheckCon:eqn:75}, and where step
    $(b)$ follows from the fact that $\lim_{M\to\infty}c^{1/M} = 1$ holds for
    any constant $c>0$.
  \end{proof}

  \begin{lemma}\label{sec:CheckCon:lem:limsup Z N0 smaller than ZSPA}
    If the strict inequality in~\eqref{sec:CheckCon:eqn:70} holds, then
    $
      \limsup_{M \to \infty}
      \ZBM(\graphN)
      \leq \ZBSPA^{*}(\graphN).
    $
  \end{lemma}
  \begin{proof}
    Since $R_M\geq0$, we obtain
      \begin{align*}
        R_M
        = |R_M|
         & \overset{(a)}{\leq}
        1
        +
        \sum_{k=0}^{M-1} \bigl( \ZBSPA^{*}( \graphN) \bigr)^{\! k-M} \cdot |T_{M,k}| \\
         & \overset{(b)}{\leq}
        1 + \sum_{k=0}^{M-1}\alpha_{\graphN}^{M-k}
        \overset{(c)}{\leq} \frac{1}{1-\alpha_{\graphN}}.
      \end{align*}
      Step~$(a)$ follows from~\eqref{sec:CheckCon:eqn:R-expansion} and the
      triangle inequality. Step~$(b)$ follows
      from~\eqref{sec:CheckCon:eqn:T-def}. Step~$(c)$ follows from
      Lemma~\ref{lemma:alpha:N:bounds}. Therefore,
    \begin{align*}
      \limsup_{M \to \infty}\ZBM(\graphN)
       & =
      \limsup_{M \to \infty}
      \Bigl( \ZBSPA^{*}(\graphN)
      \cdot (R_M)^{1/M} \Bigr)
      \\& \leq \ZBSPA^{*}(\graphN)
      \cdot \lim_{M\to\infty}
      \Biggl(  \frac{1}{1-\alpha_{\graphN}} \Biggr)^{\!\!1/M}
      \\&=\ZBSPA^{*}(\graphN).\qedhere
    \end{align*}
  \end{proof}

  Lemmas~\ref{sec:CheckCon:lem:liminf Z N0 greater than ZSPA} and~\ref{sec:CheckCon:lem:limsup Z N0 smaller than ZSPA} complete the proof of Theorem~\ref{sec:CheckCon:thm:1}.

  \begin{remark}
    The bounds in Lemmas~\ref{sec:CheckCon:lem:liminf Z N0 greater than
        ZSPA} and~\ref{sec:CheckCon:lem:limsup Z N0 smaller than ZSPA} rest on
      the following cancellation:
    \begin{align*}
       & \hspace{-0.25 cm}
      \binom{M}{k}
      \cdot \!\! \!\! \!\!
      \sum_{ \LCTvxsM, \LCTvxlM }
      \bigl[ \LCTvxsm \! = \! \LCTvxlm \! = \! \tv{0},\, \forall m \in [k] \bigr]
      \\
       & \quad\cdot
      \Biggl(
      \prod_{ m \in [M] \setminus [k] }
      \bigl|\gone( \LCTvxsm, \LCTvxlm )\bigr|
      \Biggr)
      \cdot \prod_{e}
      P_{e}\bigl( \LCTtv{x}_{\efi,[M]}, \LCTtv{x}_{\efj,[M]} \bigr)
      \\
       & \leq
      \!\!\sum_{ \LCTvxsM, \LCTvxlM }
      \!\!\!\!\bigl[ \LCTvxsm \! = \! \LCTvxlm \! = \! \tv{0},\, \forall m \in [k] \bigr]
      \cdot\!\!\!\!\!\!
      \prod_{ m \in [M] \setminus [k] }
      \!\!\!\!\!\!\bigl|\gone( \LCTvxsm, \LCTvxlm )\bigr|,
    \end{align*}
    where $0 \leq k \leq M-1$.  The LCT support identity in
      Lemma~\ref{sec:CheckCon:prop:properties of N(0)} is essential:
    \begin{align*}
       & \bigl|\gone( \LCTvxsm, \LCTvxlm )\bigr| = \bigl|\gone( \LCTvxsm, \LCTvxlm )\bigr| \!\cdot\!
      \bigl[ \wh(\LCTvxsm) \! + \! \wh( \LCTvxlm) \! \geq \! 2 \bigr].
    \end{align*}
    The vanishing of the weight-one terms forces every nonzero factor indexed by $m \in [M] \setminus [k]$ to satisfy the Hamming-weight condition in Lemma~\ref{sec:CheckCon:lem: lower bound of Pe}. Together with the all-zero conditions for $m \in [k]$, this gives
      $\prod_e P_e\leq\binom{M}{k}^{-1}$ and exactly cancels the leading factor $\binom{M}{k}$. Without the LCT, the displayed support identity need not hold. The cover bound would then be unavailable for some nonzero summands, the binomial factor would remain, and this proof would not control the correction terms.
    \eremark
  \end{remark}

  {
{\section{Per-Node Relative Loop Weight for the Class of DE-NFGs with Single-Cycle Topology in Section~\ref{sec:CheckCon}}
    \label{apx:single-cycle-loop-weight}}
}

    For the DE-NFG $\graphN$ in Section~\ref{sec:CheckCon} with
      $\matr{C}_{f_k} = \matr{C}\defeq\matr{I}+\epsilon\cdot\matr{J}$, $k \in [3]$,
      this appendix derives the per-node relative loop weights
      $L_{f_k} = 2 |\epsilon|$, $k \in [3]$, and identifies
      $\LCT{f_k}(\tv{0})$ with the dominant eigenvalue of the per-edge
      transfer operator. The three edges of $\graphN$ are $e_1 = (f_1,f_2)$,
      $e_2 = (f_2,f_3)$, and $e_3 = (f_3,f_1)$. Edge subscripts are
      interpreted modulo $3$; in particular, we have $e_0 = e_3$. We index each
      doubled symbol $\txe = (\xe,\xe')$ by $2\xe+\xe'\in\{0,1,2,3\}$, where
      $\xe,\xe'\in\{0,1\}$, so that $\tv{0}=(0,0)$ has index $0$.

    \emph{Transfer operator.} For $k\in[3]$, node $f_k$ is incident to
      edges $e_{k-1}$ and $e_k$, and the corresponding local function can be
      written as $f_k:\tset{X}_{e_{k-1}}\times\tset{X}_{e_k}\to\sC$. We
    define the complex-valued square matrix
    \begin{align*}
      \matr{T}_{f_k}
       & \defeq
      \bigl( f_k( \tx_{e_{k-1}}, \tx_{e_k} ) \bigr)_{
                                               \tx_{e_{k-1}} \in \tset{X}_{e_{k-1}},\, \tx_{e_k} \in \tset{X}_{e_k} } \in \sC^{ |\tset{X}_{e_{k-1}}| \times |\tset{X}_{e_k}| },
    \end{align*}
    where the row and column indices are
    $\tx_{e_{k-1}}\in\tset{X}_{e_{k-1}}$ and
    $\tx_{e_k}\in\tset{X}_{e_k}$, respectively.\footnote{The Choi matrix $\matr C_{f_k}$ has row index
      $(x_{e_{k-1}},x_{e_k})$ and column index $(x'_{e_{k-1}},x'_{e_k})$.
      In contrast, the transfer matrix $\matr T_{f_k}$ has row index
      $(x_{e_{k-1}},x'_{e_{k-1}})$ and column index $(x_{e_k},x'_{e_k})$.
      Thus, $\matr T_{f_k}$ is obtained by reordering the entries of $\matr C_{f_k}$.} We call $\matr{T}_{f_k}$
    the \emph{transfer operator}, or per-edge Liouville superoperator, of
    $f_k$. The partition function is then given by
    $Z(\graphN) = \operatorname{tr}\bigl(\matr{T}_{f_1}\cdot\matr{T}_{f_2}\cdot\matr{T}_{f_3}\bigr)$.

    Because $\matr{C}_{f_k}=\matr{C}$ for every $k\in[3]$, all three
    function nodes have the same transfer operator. Reordering the entries of
    $\matr{C}=\matr{I}+\epsilon\cdot\matr{J}$ according to the definition above
    gives $\matr{T}_{f_k}=\matr{T}$, where
    \begin{align*}
      \matr{T}
      =
      \begin{pmatrix}
        1 & 0        & 0        & 1 \\
        0 & \epsilon & \epsilon & 0 \\
        0 & \epsilon & \epsilon & 0 \\
        1 & 0        & 0        & 1
      \end{pmatrix},
      \qquad k\in[3].
    \end{align*}
    Note that $\matr{T}$ decouples into the block
    $\bigl( \begin{smallmatrix} 1 & 1 \\ 1 & 1 \end{smallmatrix} \bigr)$ on
    $\{ (0,0), (1,1) \}$ and the block
    $\bigl( \begin{smallmatrix} \epsilon & \epsilon \\ \epsilon &
        \epsilon \end{smallmatrix} \bigr)$ on $\{ (0,1), (1,0) \}$. The
    eigenvalues of $\matr{T}$ are $\{ 2, 2\epsilon, 0, 0 \}$, with dominant
    eigenvalue $2$ for $ |\epsilon| < 1$. Therefore, we have $Z(\graphN) = \operatorname{tr}(\matr{T}^3) = 2^{3} + (2\epsilon)^{3}$.

    \emph{The value of $\LCT{f_k}(\tv{0})$.} By symmetry, all SPA
    fixed-point messages have the same form. Depending on its direction, each
    message is a left or right eigenvector of $\matr{T}$ associated
    with the eigenvalue $2$. Under the normalization
    $\sum_{\txe}\mu(\txe)=1$, this eigenvector is
    $\vmu=(1/2,0,0,1/2)^{\tran}$, supported on $(0,0)$ and $(1,1)$. Thus, for every $k\in[3]$, it holds that
    $Z_{e_k} = \vmu^{\tran} \cdot \vmu = 1/2$ and
    $Z_{f_k}= \vmu^{\tran}\cdot\matr{T}\cdot\vmu = 1$, which implies
    \begin{align*}
      \ZBSPA^{*}(\graphN)
       & = \frac{ \prod_f Z_f }{ \prod_e Z_e }
      = 8.
    \end{align*}
    By~\eqref{sec:CheckCon:eqn:31}, it holds that
    $\ZBSPA^{*}(\graphN) = \prod_f\LCT{f}(\tv{0})$. The symmetries of the
    setup then imply that $\LCT{f_k}(\tv{0}) = 2$, $k\in[3]$.

    \emph{The per-node relative loop weight.} The LCT (see
    Definition~\ref{def:DENFG:LCT:1}) reparameterizes each edge by a
    matrix $\matr{M}$ built from $\vmu$; here
    \begin{align*}
      \matr{M} =
      \frac{1}{\sqrt{2}}
      \cdot
      \begin{pmatrix}
        1 & 0        & 0        & -1 \\
        0 & \sqrt{2} & 0        & 0  \\
        0 & 0        & \sqrt{2} & 0  \\
        1 & 0        & 0        & 1
      \end{pmatrix},
      \qquad
      \matr{M} \cdot \matr{M}^{\tran} = \matr{I} .
    \end{align*}
    For $k \in [3]$, the transformed local function at $f_k$, arranged as a
    matrix, is given by $\matr{M}^{\tran} \cdot \matr{T} \cdot \matr{M}$,
    which equals
    \begin{align}
      \left(
      \begin{array}{c|ccc}
        2 & 0        & 0        & 0 \\
        \hline
        0 & \epsilon & \epsilon & 0 \\
        0 & \epsilon & \epsilon & 0 \\
        0 & 0        & 0        & 0
      \end{array}
      \right) .
      \label{eqn:M matrix special case checkable cond}
    \end{align}
    Here, as elsewhere, the Hamming weight of
    $\LCTtv{x}_{\setpf_k} = (\LCTt{x}_{e_{k-1}},\LCTt{x}_{e_k})$ equals
    $\bigl[ \LCTt{x}_{e_{k-1}} \neq (0,0) \bigr] + \bigl[ \LCTt{x}_{e_k} \neq
        (0,0) \bigr]$. Thus, the index $\LCTtv{x}_{\setpf_k}$ of the upper-left
    entry of the matrix in~\eqref{eqn:M matrix special case checkable cond}
    has Hamming weight zero. In contrast, the index $\LCTtv{x}_{\setpf_k}$ of
    any of the other entries in the first row and first column of the matrix
    in~\eqref{eqn:M matrix special case checkable cond} has Hamming weight
    one, and so the corresponding matrix entry is equal to zero, consistent
    with Property~\ref{sec:LCT:prop:1:item:8} of
    Proposition~\ref{prop:DENFG:LCT:1}. Finally, the index
    $\LCTtv{x}_{\setpf_k}$ of any of the entries of the lower-right block of
    the matrix in~\eqref{eqn:M matrix special case checkable cond} has Hamming
    weight two; four of these entries are equal to $\epsilon$ and five of
    these entries are equal to zero. Therefore,
    \begin{align*}
      L_{f_k}
       & = \sum_{ \LCTtv{x}_{\setpf_{k}} \neq \tv{0} } \
      \Biggl|
      \frac{\LCT{f_k}( \LCTtv{x}_{\setpf_{k}} )}
      {\LCT{f_k}(\tv{0}) }
      \Biggr|
      = 4 \cdot \Bigl| \frac{\epsilon}{2} \Bigr|
      = 2 |\epsilon|,
    \end{align*}
    and condition~\eqref{sec:CheckCon:eqn:70} reduces to
    $(1 + 2|\epsilon|)^{3} < 3/2$, as discussed in
    Section~\ref{sec:CheckCon}.

\end{appendices}

\begin{footnotesize}
  \bibliographystyle{IEEEtran}
  \bibliography{biblio.bib}

\begin{thebibliography}{10}
\providecommand{\url}[1]{#1}
\csname url@samestyle\endcsname
\providecommand{\newblock}{\relax}
\providecommand{\bibinfo}[2]{#2}
\providecommand{\BIBentrySTDinterwordspacing}{\spaceskip=0pt\relax}
\providecommand{\BIBentryALTinterwordstretchfactor}{4}
\providecommand{\BIBentryALTinterwordspacing}{\spaceskip=\fontdimen2\font plus
\BIBentryALTinterwordstretchfactor\fontdimen3\font minus
  \fontdimen4\font\relax}
\providecommand{\BIBforeignlanguage}[2]{{%
\expandafter\ifx\csname l@#1\endcsname\relax
\typeout{** WARNING: IEEEtran.bst: No hyphenation pattern has been}%
\typeout{** loaded for the language `#1'. Using the pattern for}%
\typeout{** the default language instead.}%
\else
\language=\csname l@#1\endcsname
\fi
#2}}
\providecommand{\BIBdecl}{\relax}
\BIBdecl

\bibitem{Kschischang2001}
F.~R. {Kschischang}, B.~J. {Frey}, and H.-A. {Loeliger}, ``Factor graphs and
  the sum-product algorithm,'' \emph{IEEE Trans. Inf. Theory}, vol.~47, no.~2,
  pp. 498--519, Feb. 2001.

\bibitem{Forney2001}
G.~D. {Forney}, ``Codes on graphs: normal realizations,'' \emph{IEEE Trans.
  Inf. Theory}, vol.~47, no.~2, pp. 520--548, Feb. 2001.

\bibitem{Loeliger2004}
H.-A. {Loeliger}, ``An introduction to factor graphs,'' \emph{IEEE Signal
  Process. Mag.}, vol.~21, no.~1, pp. 28--41, Jan. 2004.

\bibitem{Cao2017}
M.~X. {Cao} and P.~O. {Vontobel}, ``Double-edge factor graphs: Definition,
  properties, and examples,'' in \emph{Proc. IEEE Inf. Theory Workshop (ITW)},
  Kaohsiung, Taiwan, Nov. 2017, pp. 136--140.

\bibitem{Alkabetz2021}
R.~Alkabetz and I.~Arad, ``Tensor networks contraction and the belief
  propagation algorithm,'' \emph{Phys. Rev. Research}, vol.~3, p. 023073, Apr
  2021.

\bibitem{Mezard2009}
M.~Mézard and A.~Montanari, \emph{Information, Physics and Computation}.\hskip
  1em plus 0.5em minus 0.4em\relax Oxford, U.K.: Oxford Univ. Press, 2009.

\bibitem{Richardson2008}
T.~Richardson and R.~Urbanke, \emph{Modern Coding Theory}.\hskip 1em plus 0.5em
  minus 0.4em\relax Cambridge, U.K.: Cambridge Univ. Press, 2008.

\bibitem{Wymeersch2007}
H.~Wymeersch, \emph{Iterative Receiver Design}.\hskip 1em plus 0.5em minus
  0.4em\relax Cambridge, U.K.: Cambridge Univ. Press, 2007.

\bibitem{Cao2021}
\BIBentryALTinterwordspacing
M.~X. Cao, ``Factor graphs for quantum information processing,'' Ph.D.
  dissertation, Dept.\ of Information Engineering, Faculty of Engineering, The
  Chinese University of Hong Kong, Apr. 2021. [Online]. Available:
  \url{https://arxiv.org/pdf/2203.12413.pdf}
\BIBentrySTDinterwordspacing

\bibitem{Cirac2009}
J.~I. Cirac and F.~Verstraete, ``Renormalization and tensor product states in
  spin chains and lattices,'' \emph{J. Phys. A: Math. Theor.}, vol.~42, no.~50,
  p. 504004, Dec. 2009.

\bibitem{Coecke2010}
B.~Coecke, ``Quantum picturalism,'' \emph{Contemporary Physics}, vol.~51,
  no.~1, pp. 59--83, 2010.

\bibitem{Robeva2019}
E.~Robeva and A.~Seigal, ``Duality of graphical models and tensor networks,''
  \emph{Information and Inference: A Journal of the IMA}, vol.~8, no.~2, pp.
  273--288, Jun. 2019.

\bibitem{Coecke2008}
B.~Coecke and R.~Duncan, ``Interacting quantum observables,'' in \emph{Proc.
  35th Int. Colloq. Automata, Languages and Programming (ICALP)}, ser. Lecture
  Notes in Computer Science, vol. 5126, Reykjavik, Iceland, Jul. 2008, pp.
  298--310.

\bibitem{Coecke2011}
------, ``Interacting quantum observables: categorical algebra and
  diagrammatics,'' \emph{New J. Phys.}, vol.~13, p. 043016, Apr. 2011.

\bibitem{Murphy1999}
K.~P. Murphy, Y.~Weiss, and M.~I. Jordan, ``Loopy belief propagation for
  approximate inference: An empirical study,'' in \emph{Proc. Uncertainty in
  Artificial Intelligence (UAI)}, Stockholm, Sweden, Jul. 1999, pp. 467--475.

\bibitem{Heskes2003}
T.~Heskes, ``Stable fixed points of loopy belief propagation are minima of the
  {Bethe} free energy,'' in \emph{Advances in Neural Information Processing
  Systems}, vol.~15.\hskip 1em plus 0.5em minus 0.4em\relax MIT Press, 2003,
  pp. 359--366.

\bibitem{Weller2014}
A.~Weller, K.~Tang, D.~Sontag, and T.~Jebara, ``Understanding the {Bethe}
  approximation: when and how can it go wrong?'' in \emph{Proc. Uncertainty in
  Artificial Intelligence (UAI)}, Quebec City, Quebec, Canada, Jul. 2014, pp.
  868--877.

\bibitem{Yedidia2005}
J.~S. {Yedidia}, W.~T. {Freeman}, and Y.~{Weiss}, ``Constructing free-energy
  approximations and generalized belief propagation algorithms,'' \emph{IEEE
  Trans. Inf. Theory}, vol.~51, no.~7, pp. 2282--2312, Jul. 2005.

\bibitem{Bethe1935}
H.~Bethe, ``Statistical theory of superlattices,'' \emph{Proc. Roy. Soc. London
  A}, vol. 150, no. 871, pp. 552--575, Jul. 1935.

\bibitem{Chertkov2006}
M.~{Chertkov} and V.~Y. {Chernyak}, ``Loop series for discrete statistical
  models on graphs,'' \emph{J. Stat. Mech: Theory Exp.}, vol. 2006, no.~06,
  Jun. 2006, art.~no.~P06009.

\bibitem{Chernyak2007}
V.~Y. {Chernyak} and M.~{Chertkov}, ``Loop calculus and belief propagation for
  {q-ary} alphabet: Loop tower,'' in \emph{Proc. IEEE Int. Symp. Information
  Theory}, Nice, France, Jun. 2007, pp. 316--320.

\bibitem{Mori2015}
R.~{Mori}, ``Loop calculus for nonbinary alphabets using concepts from
  information geometry,'' \emph{IEEE Trans. Inf. Theory}, vol.~61, no.~4, pp.
  1887--1904, Apr. 2015.

\bibitem{Koetter2003}
R.~Koetter and P.~O. Vontobel, ``Graph covers and iterative decoding of
  finite-length codes,'' in \emph{Proc. 3rd Int. Symp. Turbo Codes Related
  Top.}, Brest, France, Sep. 2003, pp. 75--82.

\bibitem{Vontobel2013}
P.~O. {Vontobel}, ``Counting in graph covers: A combinatorial characterization
  of the {Bethe} entropy function,'' \emph{IEEE Trans. Inf. Theory}, vol.~59,
  no.~9, pp. 6018--6048, Sep. 2013.

\bibitem{Ruozzi2012}
N.~Ruozzi, ``The {Bethe} partition function of log-supermodular graphical
  models,'' in \emph{Proc.\ Neural Inf.\ Proc.\ Sys.\ Conf.}, Lake Tahoe, NV,
  USA, Dec. 2012, pp. 117--125.

\bibitem{Csikvari2022}
P.~Csikv\'{a}ri, N.~Ruozzi, and S.~Shams, ``{Markov} random fields,
  homomorphism counting, and {Sidorenko’s} conjecture,'' \emph{IEEE Trans.
  Inf. Theory}, vol.~68, no.~9, pp. 6052--6062, Sep. 2022.

\bibitem{E.B.Sudderth2007}
E.~B. Sudderth, M.~J. Wainwright, and A.~S. Willsky, ``Loop series and {B}ethe
  variational bounds in attractive graphical models,'' in \emph{Proc.\ Neural
  Inf.\ Proc.\ Sys.\ Conf.}, Vancouver, Canada, Dec. 2007.

\bibitem{Vontobel2025}
P.~O. Vontobel, ``Understanding the ratio of the partition sum to its {Bethe}
  approximation via double covers,'' in \emph{Proc. 2025 IEEE Information
  Theory Workshop ({ITW})}, Sydney, Australia, Sep.--Oct. 2025, pp. 1--6,
  [Extended version online]. Available: \url{https://arxiv.org/abs/2509.19910}.

\bibitem{Wu2026}
B.~Wu and P.~O. Vontobel, ``Double-cover-based analysis of the {Bethe}
  permanent of block-structured positive matrices,'' in \emph{Proc. 2026 IEEE
  Int. Symp. Information Theory ({ISIT})}, Guangzhou, China, Jul. 2026, pp.
  1--6, [Extended version online]. Available:
  \url{https://arxiv.org/abs/2601.17508}.

\bibitem{Zhou2026}
J.~Zhou and P.~O. Vontobel, ``Complex-valued-matrix permanents: {SPA}-based
  approximations and double-cover analysis,'' in \emph{Proc. 2026 IEEE Int.
  Symp. Information Theory ({ISIT})}, Guangzhou, China, Jul. 2026, pp. 1--6,
  [Extended version online]. Available: \url{https://arxiv.org/abs/2601.18232}.

\bibitem{Altieri2017}
A.~Altieri, M.~C. Angelini, C.~Lucibello, G.~Parisi, F.~Ricci-Tersenghi, and
  T.~Rizzo, ``Loop expansion around the {Bethe} approximation through the
  {$M$}-layer construction,'' \emph{J. Stat. Mech.: Theory Exp.}, vol. 2017,
  no.~11, p. 113303, Nov. 2017.

\bibitem{Angelini2022}
M.~C. Angelini, C.~Lucibello, G.~Parisi, G.~Perrupato, F.~Ricci-Tersenghi, and
  T.~Rizzo, ``Unexpected upper critical dimension for spin glass models in a
  field predicted by the loop expansion around the {Bethe} solution at zero
  temperature,'' \emph{Phys. Rev. Lett.}, vol. 128, p. 075702, Feb 2022.

\bibitem{AngeliniPalazzi2024}
M.~C. Angelini, S.~Palazzi, G.~Parisi, and T.~Rizzo, ``{Bethe} {$M$}-layer
  construction on the {Ising} model,'' \emph{J.\ Stat.\ Mech.: Theory and
  Experiment}, vol. 2024, no.~6, Jun. 2024, art.~no.~063301.

\bibitem{Palazzi2025}
\BIBentryALTinterwordspacing
S.~Palazzi, ``The {$M$}-layer construction: A diagrammatic framework for
  critical behavior in lattice models,'' Ph.D. dissertation, Dept.\ of Physics,
  Sapienza University of Rome, Rome, Italy, Dec. 2025. [Online]. Available:
  \url{https://hdl.handle.net/11573/1756842}
\BIBentrySTDinterwordspacing

\bibitem{Leleu2026}
\BIBentryALTinterwordspacing
T.~Leleu, S.~Reifenstein, A.~Yamamura, and S.~Ganguli, ``Reshaping global loop
  structure to accelerate local optimization by smoothing rugged landscapes,''
  2026. [Online]. Available: \url{https://arxiv.org/abs/2602.01490}
\BIBentrySTDinterwordspacing

\bibitem{Loeliger2012}
H.-A. {Loeliger} and P.~O. {Vontobel}, ``A factor-graph representation of
  probabilities in quantum mechanics,'' in \emph{Proc. IEEE Int Symp.
  Information Theory}, Cambridge, MA, USA, Jul. 2012, pp. 656--660.

\bibitem{Loeliger2017}
------, ``Factor graphs for quantum probabilities,'' \emph{IEEE Trans. Inf.
  Theory}, vol.~63, no.~9, pp. 5642--5665, Sep. 2017.

\bibitem{Loeliger2020}
------, ``Quantum measurement as marginalization and nested quantum systems,''
  \emph{IEEE Trans. Inf. Theory}, vol.~66, no.~6, pp. 3485--3499, Jun. 2020.

\bibitem{Mori2015a}
R.~{Mori}, ``Holographic transformation, belief propagation and loop calculus
  for generalized probabilistic theories,'' in \emph{Proc. IEEE Int Symp.
  Information Theory}, Hong Kong, China, Jun. 2015, pp. 1099--1103.

\bibitem{LohJr1990}
E.~Y. Loh, Jr., J.~E. Gubernatis, R.~T. Scalettar, S.~R. White, D.~J.
  Scalapino, and R.~L. Sugar, ``Sign problem in the numerical simulation of
  many-electron systems,'' \emph{Phys. Rev. B, Condens. Matter}, vol.~41,
  no.~13, pp. 9301--9307, May 1990.

\bibitem{Guo2023}
C.~Guo, D.~Poletti, and I.~Arad, ``Block belief propagation algorithm for
  two-dimensional tensor networks,'' \emph{Phys. Rev. B}, vol. 108, p. 125111,
  Sep 2023.

\bibitem{Tindall2023}
J.~Tindall and M.~Fishman, ``{Gauging tensor networks with belief
  propagation},'' \emph{SciPost Phys}, vol.~15, p. 222, 2023.

\bibitem{Tindall2024}
J.~Tindall, M.~Fishman, E.~M. Stoudenmire, and D.~Sels, ``Efficient tensor
  network simulation of {IBM's} {Eagle} kicked {Ising} experiment,'' \emph{PRX
  Quantum}, vol.~5, p. 010308, Jan 2024.

\bibitem{Kim2023}
Y.~Kim, A.~Eddins, S.~Anand, K.~X. Wei, E.~van~den Berg, S.~Rosenblatt,
  H.~Nayfeh, Y.~Wu, M.~Zaletel, K.~Temme, and A.~Kandala, ``Evidence for the
  utility of quantum computing before fault tolerance,'' \emph{Nature}, vol.
  618, no. 7965, pp. 500--505, Jun. 2023.

\bibitem{Tindall2026}
J.~Tindall, G.~M. Sommers, and H.~Kappen, ``Contracting tensor networks with
  generalized belief propagation,'' {Apr.} 2026, [Preprint online]. Available:
  \url{https://arxiv.org/abs/2604.24760}.

\bibitem{Gutman2026}
\BIBentryALTinterwordspacing
N.~Gutman, ``A block belief-propagation algorithm for the contraction of
  tensor-networks,'' {M.Sc.\ thesis}, Faculty of Electrical and Computer
  Engineering, Technion -- Israel Institute of Technology, Aug. 2024. [Online].
  Available: \url{https://arxiv.org/pdf/2603.13304.pdf}
\BIBentrySTDinterwordspacing

\bibitem{AlBashabsheh2011}
A.~{Al-Bashabsheh} and Y.~{Mao}, ``Normal factor graphs and holographic
  transformations,'' \emph{IEEE Trans. Inf. Theory}, vol.~57, no.~2, pp.
  752--763, Feb. 2011.

\bibitem{Huang2024}
\BIBentryALTinterwordspacing
Y.~Huang, ``Finite-graph-cover-based analysis of factor graphs in classical and
  quantum information processing systems,'' Ph.D. dissertation, Dept.\ of
  Information Engineering, Faculty of Engineering, The Chinese University of
  Hong Kong, Sep. 2024. [Online]. Available:
  \url{https://arxiv.org/pdf/2412.05942.pdf}
\BIBentrySTDinterwordspacing

\bibitem{Evenbly2024}
G.~Evenbly, N.~Pancotti, A.~Milsted, J.~Gray, and G.~K.-L. Chan, ``Loop series
  expansions for tensor networks,'' \emph{Phys. Rev. Research}, vol.~8, no.~1,
  p. 013245, Mar. 2026.

\bibitem{Ruozzi2013}
N.~Ruozzi, ``Beyond log-supermodularity: lower bounds and the {Bethe} partition
  function,'' in \emph{Proc. Uncertainty Artificial Intelligence (UAI)},
  Bellevue, WA, USA, Jul. 2013, pp. 546--555.

\bibitem{Vontobel2013a}
P.~O. {Vontobel}, ``The {Bethe} permanent of a nonnegative matrix,'' \emph{IEEE
  Trans. Inf. Theory}, vol.~59, no.~3, pp. 1866--1901, Mar. 2013.

\bibitem{Gurvits2011}
L.~Gurvits, ``Unleashing the power of {Schrijver's} permanental inequality with
  the help of the {Bethe} approximation,'' \emph{Elec. Coll. Comp. Compl.},
  Dec. 2011, report TR11-169.

\bibitem{Straszak2019}
D.~Straszak and N.~K. Vishnoi, ``Belief propagation, {Bethe} approximation, and
  polynomials,'' \emph{IEEE Trans. Inf. Theory}, vol.~65, no.~7, pp.
  4353--4363, Jul. 2019.

\bibitem{Anari2021}
N.~Anari and S.~Oveis~Gharan, ``A generalization of permanent inequalities and
  applications in counting and optimization,'' \emph{Adv. Math.}, vol. 383, p.
  107657, Jun. 2021.

\bibitem{Huang2024a}
Y.~Huang and P.~O. Vontobel, ``The {B}ethe partition function and the {SPA} for
  factor graphs based on homogeneous real stable polynomials,'' in \emph{Proc.
  IEEE Int. Symp. Inf. Theory}, Athens, Greece, Jul. 2024, pp. 3029--3034.

\bibitem{Huang2023}
Y.~Huang, N.~Kashyap, and P.~O. Vontobel, ``Degree-{$M$} {B}ethe and {S}inkhorn
  permanent based bounds on the permanent of a non-negative matrix,''
  \emph{IEEE Trans. Inf. Theory}, vol.~70, no.~7, pp. 5289--5308, Jul. 2024.

\bibitem{Wood2015}
C.~J. Wood, J.~D. Biamonte, and D.~G. Cory, ``Tensor networks and graphical
  calculus for open quantum systems,'' \emph{Quantum Info. Comput.}, vol.~15,
  no. 9-10, pp. 759--811, Jul. 2015.

\bibitem{J.Wainwright2008}
M.~J. {Wainwright} and M.~I. {Jordan}, ``Graphical models, exponential
  families, and variational inference,'' \emph{Foundations and Trends in
  Machine Learning}, vol.~1, no. 1--2, pp. 1--305, 2008.

\bibitem{Koetter2007}
R.~Koetter, W.-C.~W. Li, P.~O. Vontobel, and J.~L. Walker, ``Characterizations
  of pseudo-codewords of (low-density) parity-check codes,'' \emph{Adv. Math.},
  vol. 213, no.~1, pp. 205--229, 2007.

\bibitem{Stark:Terras:96:1}
H.~M. Stark and A.~A. Terras, ``Zeta functions of finite graphs and
  coverings,'' \emph{Adv.\ in Math.}, vol. 121, no.~1, pp. 124--165, Jul. 1996.

\bibitem{Vontobel:16:1}
P.~O. Vontobel, ``Analysis of double covers of factor graphs,'' in \emph{Proc.\
  Int.\ Conf.\ Sig.\ Proc.\ and Comm.}, Bangalore, India, Jun. 2016, pp. 1--5.

\bibitem{Cover:Thomas:06:1}
T.~M. Cover and J.~A. Thomas, \emph{Elements of Information Theory},
  2nd~ed.\hskip 1em plus 0.5em minus 0.4em\relax Hoboken, NJ, USA: John Wiley
  \& Sons Inc., 2006.

\bibitem{Harrow2013}
\BIBentryALTinterwordspacing
A.~W. Harrow, ``The church of the symmetric subspace,'' Aug. 2013. [Online].
  Available: \url{https://arxiv.org/pdf/1308.6595v1.pdf}
\BIBentrySTDinterwordspacing

\bibitem{Dobrushin1968}
R.~L. Dobrushin, ``The problem of uniqueness of a {G}ibbsian random field and
  the problem of phase transitions,'' \emph{Funct. Anal. Appl.}, vol.~2, no.~4,
  pp. 302--312, 1968.

\bibitem{Kotecky1986}
R.~Koteck\'{y} and D.~Preiss, ``Cluster expansion for abstract polymer
  models,'' \emph{Commun. Math. Phys.}, vol. 103, no.~3, pp. 491--498, 1986.

\bibitem{Friedli2017}
S.~Friedli and Y.~Velenik, \emph{Statistical Mechanics of Lattice Systems: A
  Concrete Mathematical Introduction}.\hskip 1em plus 0.5em minus 0.4em\relax
  Cambridge, UK: Cambridge University Press, 2017.

\end{thebibliography}
\end{footnotesize}

\end{document}